\documentclass[journal,draftcls,onecolumn,twoside]{IEEEtran}
\normalsize
\usepackage{amsmath,amssymb,amsthm}
\usepackage{enumerate}
\usepackage{stfloats}
\usepackage{comment}
\usepackage{subcaption}
\usepackage{siunitx}
\usepackage{mathtools}
\usepackage{accents,latexsym}

\usepackage[dvipsnames]{xcolor}
\definecolor{myPink}{RGB}{255,105,183}

\usepackage[T1]{fontenc}
\usepackage{graphics} 
\usepackage{epsfig} 
\usepackage[mathscr]{euscript}
\usepackage{algorithm}
\usepackage[noend]{algpseudocode}
\usepackage{bbm}
\makeatletter
\def\BState{\State\hskip-\ALG@thistlm}
\makeatother

\usepackage{hyperref}

\usepackage{tikz}
\usetikzlibrary{arrows,shapes,chains,matrix,positioning,scopes,patterns,calc,
decorations.markings,
decorations.pathmorphing,
}

\usepackage{pgfplots}
\pgfplotsset{compat=1.3}
\usepgflibrary{shapes}

\newtheorem{theorem}{Theorem}
\newtheorem{lemma}[theorem]{Lemma}
\newtheorem{proposition}[theorem]{Proposition}
\newtheorem{definition}[theorem]{Definition}

\newtheorem{remark}[theorem]{Remark}
\newtheorem{corollary}[theorem]{Corollary}
\newtheorem{assumption}[theorem]{Assumption}

\renewcommand{\epsilon}{\varepsilon}

\newcommand{\mc}[1]{\mathcal{#1}}

\newcommand{\RNum}[1]{\uppercase\expandafter{\romannumeral #1\relax}}

\newcommand{\av}{\ensuremath{\underline{a}}}
\newcommand{\bv}{\ensuremath{\underline{b}}}
\newcommand{\bve}{\ensuremath{\hat{\underline{b}}}}
\newcommand{\cv}{\ensuremath{\underline{c}}}
\newcommand{\pv}{\ensuremath{\underline{p}}}
\newcommand{\sv}{\ensuremath{\underline{s}}}
\newcommand{\vv}{\ensuremath{\underline{v}}}

\newcommand{\wv}{\ensuremath{\underline{w}}}
\newcommand{\xv}{\ensuremath{\underline{x}}}
\newcommand{\yv}{\ensuremath{\underline{y}}}
\newcommand{\zv}{\ensuremath{\underline{z}}}
\newcommand{\ptree}{\ensuremath{p_{\mathrm{tree}}}}

\newcommand{\Ktot}{\ensuremath{K_{\mathrm{tot}}}}
\newcommand{\Ka}{\ensuremath{K_{\mathrm{a}}}}

\def \pcs{p_{\mathrm{cs}}}

\def\Pr{\mathrm{Pr}}

\DeclareMathAlphabet{\mcl}{OMS}{cmsy}{m}{n}

\newlength\tikzwidth
\newlength\tikzheight

\definecolor{mycolor1}{rgb}{0.63529,0.07843,0.18431}%
\definecolor{mycolor2}{rgb}{0.00000,0.44706,0.74118}%
\definecolor{mycolor3}{rgb}{0.00000,0.49804,0.00000}%
\definecolor{mycolor4}{rgb}{0.87059,0.49020,0.00000}%
\definecolor{mycolor5}{rgb}{0.00000,0.44700,0.74100}%
\definecolor{mycolor6}{rgb}{0.74902,0.00000,0.74902}%

\newif\ifproof
\prooftrue

\def\fig_path{./}
\begin{document}
\title{A Coded Compressed Sensing Scheme for Uncoordinated Multiple Access}

%
\author{\IEEEauthorblockN{Vamsi K. Amalladinne, \emph{Student Member, IEEE},
Jean-Francois Chamberland,  \emph{Senior Member, IEEE},
Krishna R. Narayanan, \emph{Fellow, IEEE}}
\thanks{
This material is based upon work supported by the National Science Foundation (NSF) under Grant No.~CCF-1619085.
This work was presented in part at the International Conference on Acoustics, Speech, and Signal Processing, 2018.

The authors are with the Department of Electrical and Computer Engineering, Texas A\&M University, College Station, TX 77843, USA (emails: \{vamsi128,krn,chmbrlnd\}@tamu.edu).}
}

\maketitle

\begin{abstract}
This article introduces a novel communication scheme, termed coded compressed sensing, for unsourced multiple-access communication.
The proposed divide-and-conquer approach leverages recent advances in compressed sensing and forward error correction to produce a novel uncoordinated access paradigm, along with a computationally efficient decoding algorithm.
Within this framework, every active device partitions its data into several sub-blocks and, subsequently, adds redundancy using a systematic linear block code.
Compressed sensing techniques are then employed to recover sub-blocks up to a permutation of their order, and the original messages are obtained by stitching fragments together using a tree-based algorithm.
The error probability and computational complexity of this access paradigm are characterized.
An optimization framework, which exploits the tradeoff between performance and computational complexity, is developed to assign parity-check bits to each sub-block.
In addition, two emblematic parity bit allocation strategies are examined and their performances are analyzed in the limit as the number of active users and their corresponding payloads tend to infinity.
The number of channel uses needed and the computational complexity associated with these allocation strategies are established for various scaling regimes.
Numerical results demonstrate that coded compressed sensing outperforms other existing practical access strategies over a range of operational scenarios.
\end{abstract}

\begin{IEEEkeywords}
Communication, forward error correction, unsourced multiple-access, compressed sensing.
\end{IEEEkeywords}

\section{Introduction}
\label{sec:intro}

Uncoordinated and unsourced multiple access communication (MAC) is a novel formulation for non-orthogonal multiple access.
This framework, which was introduced by Polyanskiy in~\cite{polyanskiy2017perspective}, is particularly relevant in the context of the Internet of Things (IoT).
It is closely related to coded random access~\cite{paolini2015coded} and the many access channel~\cite{chen2017capacity}.
In this new paradigm, a wireless network is composed of $\Ktot$ users, out of which a smaller group of $\Ka$ users are active at any given time.
These $\Ka$ active users each wish to transmit a $B$-bit message to the access point in an uncoordinated fashion. The access point is tasked with recovering the set of transmitted messages, without regard for the identities of the corresponding sources.
The total number of users $\Ktot$, can be very large, whereas parameters $\Ka$ and $B$ are envisioned to be orders of magnitude smaller than $\Ktot$.

For typical IoT applications, the message length $B$ is envisioned to remain small and, in this regime, asymptotic information-theoretic results only offer limited insight.
Rather, finite-length bounds are often more meaningful.
Along these lines, Polyanskiy~\cite{polyanskiy2017perspective} computes finite block-length (FBL) achievability bounds based on random Gaussian codebooks and maximum-likelihood (ML) decoding.
While this work provides a benchmark to evaluate coding schemes, random coding with ML decoding is computationally infeasible in most practical situations and, hence, there is an important need for computationally-efficient coding and decoding schemes aimed at the unsourced MAC.

In ~\cite{ordentlich2017low}, Ordentlich and Polyanskiy show that many existing multiple access strategies, including treating inference as noise (TIN) and ALOHA perform poorly in this context, especially when $\Ka$ exceeds $100$.
They also propose the first low-complexity coding scheme tailored to this setting.
In their scheme, the transmission period is divided into sub-blocks, or slots, and the system operates in a synchronous fashion.
Specifically, all the users are aware of slot boundaries.
Within this framework, every active user transmits a codeword during a randomly chosen slot.
A data block is formed with a concatenated code that is designed for a $T$-user, real-addition Gaussian multiple access channel ($T$-GMAC); values for $T$ range from 2 to 5.
Although this scheme performs significantly better than ALOHA and TIN, there remains an important gap of approximately $20$~dB between this realized performance and the aforementioned achievability limit associated with the unsourced MAC~\cite{polyanskiy2017perspective}.
In subsequent work~\cite{vem2017user}, Vem et al.\ introduce a low-complexity coding scheme that relies on a similar slotted structure.
This latter framework consists of an improved, close-to-optimal coding strategy for the $T$-GMAC; coupled to the application of successive interference cancellation across slots to reduce the performance degradation caused by overcrowded slots.
The combination of these two features constitutes a significant improvement over \cite{ordentlich2017low}, with a performance curve that lies only approximately $6$~dB away from the above mentioned achievability limit.
Both schemes discussed above adopt a channel coding viewpoint, wherein the $\Ka$-user GMAC is reduced to multiple smaller $T$-GMAC channel problems.

In contrast, this article promotes an alternate compressed sensing (CS) approach tailored to the uncoordinated and unsourced MAC problem.
Enabling uncoordinated multiple access to a massive number of users has a strong connection to the problem of support recovery in noisy compressed sensing~\cite{polyanskiy2017perspective}.
Conceptually, decoding an instance of the uncoordinated MAC entails finding the support of an unknown vector of length~$2^B$. A naive CS solution requires operations on sensing matrices with $2^B$ columns, which is computationally intractable for values of $B$ on the order of 100. Consequently, any pragmatic solution to this problem needs to have a computational cost that is sub-linear in the dimension of the problem.

To this end, we devise a novel CS algorithm, called coded compressed sensing (CCS), for the unsourced MAC problem.
This algorithm achieves sub-linear complexity in the dimension of the CS problem using a divide-and-conquer approach.
Information messages at the users are partitioned into smaller sub-blocks such that each sub-problem is amenable to CS recovery.
Before transmission, redundancy is added to individual fragments using a systematic linear code.
The collection of sub-blocks transmitted within a slot are recovered using a standard CS algorithm.
Once this is achieved, the individual fragments of the original messages need to be pieced together.
This is accomplished via a low-complexity tree-based algorithm.
The overall structure of this communication architecture yields better performance compared to previously proposed algorithms with comparable computational complexity.
The main contributions of this article are summarized below.
\begin{enumerate}
    \item A novel low-complexity compressed sensing algorithm is proposed to solve the unsourced, uncoordinated multiple access problem.
    \item Explicit closed-form expressions are provided to characterize the error probability and the average computational complexity of this scheme.
    \item We show how the parameters of this algorithm can be tuned to gracefully trade off complexity and performance.
    An optimization framework is developed to exploit this tradeoff.
    \item In the finite block length regime, the algorithm is shown to perform close to the FBL achievability bounds.
    \item In the asymptotic regime where $\Ka$ and $B$ approach infinity, bounds are provided for the average computational complexity and number of channel uses needed for this scheme to be asymptotically reliable.
    The algorithm has an average complexity that is sub-linear in the dimension of the problem.
\end{enumerate}
In the section below, we review the connection between compressed sensing and multiple access, and we survey pertinent recent developments in the field of compressed sensing.

\subsection{Compressed Sensing and Multiple Access Schemes}

The connection between compressed sensing and multiple access has been explored in the literature \cite{jin2011limits, hong2015sparsity,zhang2013neighbor,wunder2015sparse}.
The work in \cite{jin2011limits} provides necessary and sufficient conditions for exact support recovery by interpreting the sparse recovery problem as a Gaussian MAC coding problem.
In~\cite{hong2015sparsity}, the LASSO algorithm is employed for user identification in a random multiple access scenario.
Yet, the complexity of this algorithm does not scale well in many regimes of interest.
Another closely related area where compressed sensing techniques are applied for random access is the problem of discovering the access points within the range of a wireless device in a network, also known as the neighbor discovery problem~\cite{zhang2013neighbor}.
This setting is characterized by devices attempting to identify the network interface addresses of nodes within a single hop.
Therein, Zhang et al.\ propose two compressed sensing schemes based on group testing and second-order Reed Muller codes followed by chirp decoding.
The second scheme is shown to have sub-linear computational cost.
In \cite{chen2017sparse}, a low complexity neighbor discovery scheme, referred to as sparse-orthogonal frequency division multiplexing (sparse-OFDM), which is based on the recent developments in sparse Fourier transform \cite{pawar2013computing}, is proposed for the asynchronous neighbor discovery problem.
Still, there does not appear to exist a straightforward way to extrapolate these schemes to the uncoordinated MAC problem and, hence, their connection to the FBL bounds for this latter problem remains unclear.
In a recent contribution \cite{amalladinne2019asynchronous}, we make a performance comparison between CCS and the sparse-OFDM scheme found in \cite{chen2017sparse}; we demonstrate that CCS significantly outperforms sparse-OFDM scheme for asynchronous neighbor discovery.

Schemes that combine random access and compressed sensing have been proposed in \cite{wunder2015compressive} and a nice survey is presented in \cite{wunder2015sparse}. A multiple access scheme that uses compressed sensing in sensor networks is presented in \cite{fazel2013random}. These papers do not use a divide-and-conquer strategy and also do not use a tree code. As such, our proposed coding scheme is substantially different from these works.

Exact support recovery in noisy compressed sensing has been studied extensively in the literature \cite{wainwright2009sharp,wainwright2009information,wang2010information}.
In \cite{wainwright2009sharp}, the authors establish a connection between problem dimension, sparsity index, and the number of observations needed for exact support recovery using the LASSO algorithm.
Furthermore, \cite{wainwright2009information} and \cite{wang2010information} offer necessary and sufficient conditions for exact support recovery in the presence of noise.
A key result common among these studies is that $\mathcal{O} \left( k \log \left( {p}/{k} \right) \right)$ measurements are sufficient to recover a $k$-sparse vector of dimension $p$ in the presence of noise.
Conventional compressed sensing solvers like LASSO \cite{tibshirani1996regression} and iterative hard thresholding~\cite{blumensath2009iterative} are known to achieve this scaling when $k$ scales sub-linearly with $p$.
However, most of these algorithms admit a computational complexity that scales as $\operatorname{poly} (p)$, which precludes the application of these schemes to the uncoordinated MAC problem.

Several works in the field of data stream computations, e.g., \cite{gilbert2007one,gilbert2010sparse,indyk2008explicit}, aim to recover a $k$ out of $p$ sparse signal from a low-dimensional sketch.
The algorithms therein feature measurement costs of $\mathcal{O} \left( k \log \left( {p}/{k} \right) \right)$ and computational costs of $\mathcal{O} \left( k \operatorname{polylog} (p) \right)$.
However, these algorithms admit a failure probability that is bounded away from zero and, consequently, are not asymptotically reliable.

\subsection{Organization and Notation}

The remainder of this article is organized as follows.
In Section~\ref{sec:systemmodel}, we describe the system model and introduce the various parameters used in this context.
Section~\ref{sec:proposedscheme} provides detailed descriptions of the encoding and decoding operations employed within our proposed scheme.
The performance of this system is analyzed in terms of error probability and average complexity in Section~\ref{perr}.
In the process, an optimization framework is established to exploit the tension between these two considerations.
Bounds on the average computational cost and the number of channel uses needed for our scheme to work reliably in the asymptotic regime where $\Ka, B \rightarrow \infty$ are derived in Section~\ref{section:AsymptoticAnalysis}.
Simulation results are given in Section~\ref{sec:simresults} to illustrate the performance of our scheme in the finite-block length regime.
Finally, we conclude the paper in Section~\ref{sec:conc}.
Intricate proofs are relegated to the appendix.

At this stage, it is worth reviewing the notation we adopt throughout.
We employ $\mathbb{R}_{+}$, $\mathbb{Z}_{+}$, and $\mathbb{N}$ to represent the non-negative real numbers, non-negative integers, and the natural numbers.
For $a,b \in \mathbb{Z}_{+}$ with $a \le b$, we use $[a:b]$ to denote $\{c \in \mathbb{Z}_{+}: a \le c \le b \}$.
For any $a,b \in \mathbb{R}_+$ with $a \le b$, we use $[a,b]$ to denote $\{c \in \mathbb{R}_+: a \le c \le b \}$.
We employ $|A|$ for the cardinality of set $A$, and we use $[x]$ to designate the closest integer to $x$.
We write $f(n)=\mathcal{O}(g(n))$ when there are constants $c$, $n_0$ such that $f(n) \le cg(n)$ for all $n \ge n_0$.

\section{System Model}
\label{sec:systemmodel}

Let $\mathbf{S}_\mathrm{tot}$ be the collection of devices within a network, and let $\mathbf{S}_\mathrm{a}$ denote the subset of active devices within a communication round, $\mathbf{S}_\mathrm{a} \subset \mathbf{S}_\mathrm{tot}$.
We label the size of these sets by $|\mathbf{S}_\mathrm{tot}| = \Ktot$ and $|\mathbf{S}_\mathrm{a}| = \Ka$.
Every active device wishes to communicate $B$ bits of information to a base station and, collectively, these data transfers must take place through an uncoordinated uplink transmission scheme.
That is, transmissions are not scheduled centrally and, consequently, active devices must act independently of one another.
The number of channel uses dedicated to this process is $N$, and we employ $W = \left\{ \wv_i : i \in \mathbf{S}_\mathrm{a} \right\}$ to represent the collection of $B$-bit message vectors associated with these active devices.
In our proposed scheme, we assume that active devices pick their information message independently and uniformly at random from the set of binary sequences $\{0, 1\}^B$.

The base station facilitates a slotted structure for multiple access on the uplink through coarse synchronization.
As such, the signal available at the receiver assumes the form
\begin{equation} \label{syseq}
\yv= \sum_{i \in \mathbf{S}_\mathrm{a}} \xv_i + \zv ,
\end{equation}
where $\xv_i$ is the $N$-dimensional vector transmitted by device~$i$, and $\zv$ represents additive white Gaussian noise with covariance $\sigma^2 \mathbf{I}$.
The signal sent by a device is power constrained, i.e., $\| \xv_i \|_2^2 \le N E_{\mathrm{s}}$ for $i \in \mathbf{S}_\mathrm{a}$, a scenario akin to~\cite{polyanskiy2017perspective}.
The energy-per-bit is then equal to $E_{\mathrm{b}} = \frac{N E_{\mathrm{s}}}{B}$.
If $N_0/2$ is two-sided power spectral density of the underlying noise process, then $\sigma^2 = N_0 / 2$ and, upon setting the noise variance to one, we get $\frac{E_{\mathrm{b}}}{N_0} = \frac{N E_{\mathrm{s}}}{2B}$. 
Based on the observed signal, the receiver produces an estimate $\widehat{W}(\yv)$ for the list of transmitted binary vectors $W$ with $| \widehat{W}(\yv) | \le \Ka$.
As in \cite{polyanskiy2017perspective}, the per-user error probability defined by
\begin{equation}
P_{\mathrm{e}}
= \frac{1}{\Ka} \sum_{i \in \mathbf{S}_\mathrm{a}}
\Pr \left( \wv_i \notin \widehat{W}(\yv) \right)
\end{equation}
serves as a performance objective.
In words, there is a penalty when a sent message is missed by the base station.
Moreover, the total number of vectors in $\widehat{W}(\yv)$ is subject to a hard constraint, which prevents the base station from admitting an excessive number of guesses.
The following key features distinguish this model from the traditional multiple access channel.
\begin{enumerate}
\item Transmissions are unsourced and, hence, all the active users share a same codebook.
\item The decoder is tasked with providing an unordered list of messages.
\item Error probability is defined on a per user basis, as opposed to a success being contingent on identifying all the sent messages correctly.
\end{enumerate}

We note that \eqref{syseq} can  be expressed in matrix form as
\begin{equation} \label{cseq}
\yv= \mathbf{X} \bv + \zv ,
\end{equation}
where $\mathbf{X} \in  \mathbb{R}^{N\times2^B}$ denotes the common codebook and $\bv \in \{0,1\}^{2^B}$ is a binary vector that contains the indices of all the transmitted codewords.
Under this viewpoint, the $i$th column $\mathbf{X}[:,i]$ represents the signal associated with message index~$i$, and $\| \bv \|_0 = \Ka$.
The above formulation highlights a close connection between this viewpoint and compressed sensing problem, with $\mathbf{X}$ taking the role of a sensing matrix and $\bv$ being an unknown $\Ka$-sparse vector.
As pointed out in the previous section, the matrix $\mathbf{X}$ has $2^B$ columns with $B \approx 100$ and, hence, a naive CS algorithm cannot be employed to recover sparse vector $\bv$.
The dimensionality of this problem precludes the application of commodity algorithms without further modifications.
Our goal is to devise an encoding and decoding scheme that achieves $P_\mathrm{e} \le \varepsilon$, where $\varepsilon$ is a target error probability, and does so with manageable computational complexity.
Table~\ref{table:notation} summarizes the important parameters encountered in this article.

\begin{table}[tbh]
\centerline{
\begin{tabular}{||c|l||}
\hline
Notation & Parameter Description \tabularnewline
\hline
$\Ktot$ & Total number of users in the system \tabularnewline
$\Ka$ & Number of active users \tabularnewline
$B$ & Message length in bits \tabularnewline
$N$ & Number of channel uses per round \tabularnewline
$\epsilon$ & Maximum tolerable probability of error per user \tabularnewline
$n$ & Number of coded sub-blocks per round \tabularnewline
$J$ & Length of coded sub-blocks \tabularnewline
$M$ & Length of coded $B$-bit message, $M=nJ$ \tabularnewline
$\epsilon_{\mathrm{tree}}$ & Maximum probability of error for tree decoding \tabularnewline
$C_{\mathrm{cs}}$ & Computational complexity of CS sub-problem \tabularnewline
$C_{\mathrm{tree}}$ & Computational complexity of tree decoding \tabularnewline
$m_j$ & Number of information bits in $j$th sub-block \tabularnewline
$l_j$ & Number of parity bits in $j$th sub-block \tabularnewline
$K$ & Size of output list for CS sub-problem \tabularnewline
\hline
\end{tabular}}
\caption{This list contains key parameters encountered in the treatment of coded compressed sensing.}
\label{table:notation}
\end{table}

\section{Proposed Scheme}
\label{sec:proposedscheme}

\begin{figure}[tbh]
\centering
\begin{tikzpicture}[
  font=\small, >=stealth', line width=0.75pt,
  block/.style={rectangle, draw, minimum height=7mm, minimum width=10mm}
]

\node (te) at (0.125,0.75) {Tree Code};
\node[block] (te1) at (0.125,0) {$G$}
  edge[<-] node[auto,xshift=-1mm]{$\wv_1$} (-1,0);
\node[block] (te2) at (0.125,-1.25) {$G$}
  edge[<-] node[auto,xshift=-1mm]{$\wv_2$} (-1,-1.25);
\node (tei) at (0.125,-2) {$\vdots$};
\node[block] (teKa) at (0.125,-3) {$G$}
  edge[<-] node[auto,xshift=-1mm]{$\wv_{\Ka}$} (-1,-3);

\node (cs) at (2,0.75) {CS Encoder};
\node[block] (cs1) at (2,0) {CS}
  edge[<-,densely dashed] node[auto]{$\vv_1$} (te1);
\node[block] (cs2) at (2,-1.25) {CS}
  edge[<-,densely dashed] node[auto]{$\vv_2$} (te2);
\node (csi) at (2,-2) {$\vdots$};
\node[block] (csKa) at (2,-3) {CS}
  edge[<-,densely dashed] node[auto]{$\vv_{\Ka}$} (teKa);

\node[draw,circle] (MAC) at (3.5,-1.5) {$\sum$};
\draw[->, densely dashed] (cs1.east) -- (2.75,0) -- (MAC);
\draw[->, densely dashed] (cs2.east) -- (2.75,-1.25) -- (MAC);
\draw[->, densely dashed] (csKa.east) -- (2.75,-3) -- (MAC);
\node (noise) at (3.5,-3.25) {Noise}
  edge[->] node[auto]{$\zv$} (MAC);

\node[block,rotate=90,minimum width=25mm] (csDecoder) at (4.75,-1.5) {CS Algorithm}
  edge[<-, densely dashed] (MAC);
\node[block,rotate=90,minimum width=25mm] (treeDecoder) at (5.75,-1.5) {Tree Decoder}
  edge[->] node[auto,xshift=0.5mm]{$\widehat{W}$} (6.75,-1.5);

\draw (5.1,-0.5) -- (5.4,-0.5);
\draw (5.1,-1) -- (5.4,-1);
\draw (5.1,-1.5) -- (5.4,-1.5);
\draw (5.1,-2) -- (5.4,-2);
\draw (5.1,-2.5) -- (5.4,-2.5);

\end{tikzpicture}
\caption{This schematic diagram captures the overall architecture of the proposed scheme.
The information bits are split into sub-blocks, and redundancy is added to individual components.
Transmitted signals are then determined via a CS matrix, and sent over a MAC channel.
A CS algorithm recovers the lists of sub-blocks, and a tree decoder reconstructs the original messages.}
\label{fig:1}
\end{figure}
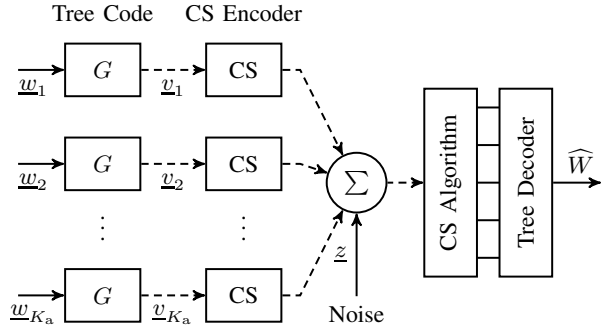

The main idea to limit complexity consists in dividing the data stream generated by active devices into several sub-blocks.
These sub-blocks, with their very short packet fragments, are amenable to computationally efficient compressed sensing (CS) algorithms.
The transmission of sub-blocks takes place sequentially, with every collection of fragments becoming an instance of unsourced multiple access, albeit one with a much reduced number of information bits.
Upon completion of these consecutive CS instances, the original messages are recovered by stitching together compatible fragments.
Technically, this latter step is accomplished by preemptively adding redundancy to the codewords, and then leveraging this redundancy while combining sub-blocks via a low-complexity tree-based algorithm.
The specifics of this envisioned scheme are detailed below. 
A notional diagram for the proposed system appears in Fig.~\ref{fig:1}.

\subsection{Encoding Process}
\label{encoder}
\begin{figure}[tbh]
\centering
\begin{tikzpicture}[
  font=\small, >=stealth', line width = 0.75pt,
  wblock/.style={rectangle, minimum height=7mm, draw=black, fill=gray!10, rounded corners},
  pblock/.style={rectangle, minimum height=7mm, draw=black, fill=gray!40, rounded corners}
]

\node[wblock,minimum width=20mm] (w0) at (0.5,2) {$\wv(0)$};
\node[wblock,minimum width=10mm] (w1) at (2,2) {$\wv(1)$};
\node (wdot) at (3,2) {$\cdots$};
\node[wblock,minimum width=10mm] (wn) at (4,2) {$\wv(4)$};

\node (bits) at (-0.4,0) {$\vv$};
\node[wblock,minimum width=20mm] (vw0) at (-1,0) {$\wv(0)$};
\node[wblock,minimum width=10mm] (vw1) at (0.75,0) {$\wv(1)$};
\node[pblock,minimum width=10mm] (vp1) at (1.75,0) {$\pv(1)$};
\node (vdot) at (2.75,0) {$\cdots$};
\node[wblock,minimum width=10mm] (vwn) at (3.75,0) {$\wv(4)$};
\node[pblock,minimum width=10mm] (vpn) at (4.75,0) {$\pv(4)$};

\draw[|-|] (-2,-0.5) to node[midway,below] {$m_0$} (0,-0.5);
\draw[|-|] (0.25,-0.5) to node[midway,below] {$m_1$} (1.25,-0.5);
\draw[-|] (1.25,-0.5) to node[midway,below] {$l_1$} (2.25,-0.5);
\draw[|-|] (3.25,-0.5) to node[midway,below] {$m_4$} (4.25,-0.5);
\draw[-|] (4.25,-0.5) to node[midway,below] {$l_4$} (5.25,-0.5);

\draw[->] (w0.south) to (vw0.north);
\draw[->] (w1.south) to (vw1.north);
\draw[->] (wn.south) to (vwn.north);
\draw[loosely dashed] (w0.south) to (vp1.north);
\draw[loosely dashed] (w0.south) to (vpn.north);
\draw[loosely dashed] (w1.south) to (vpn.north);
\draw[loosely dashed] (2.75,1.65) to (vpn.north);
\draw[loosely dashed] (3.25,1.65) to (vpn.north);

\foreach \v in {-1.00,1.25,4.25} {
  \draw[->] (\v,-1) to (\v,-1.5);
  \node at (\v-0.0625,-2) {$\mathbf{A}$};
  \draw[line width=1pt] (\v-0.375,-1.75) -- (\v-0.5,-1.75) -- (\v-0.5,-2.25) -- (\v-0.375,-2.25);
  \draw[line width=1pt] (\v+0.25,-1.75) -- (\v+0.375,-1.75) -- (\v+0.375,-2.25) -- (\v+0.25,-2.25);
  \draw[line width=1pt] (\v+0.45,-1.75) -- (\v+0.45,-2.625) -- (\v+0.57,-2.625) -- (\v+0.575,-1.75) -- (\v+0.45,-1.75);
}

\foreach \v in {-1.00,1.25,4.25} {
  \draw[->]  (\v,-2.75) -- (\v,-3.25);
  \draw[densely dotted, draw=gray]  (\v-0.25,-2.85) -- (\v,-3.25);
  \draw[densely dotted, draw=gray]  (\v-0.125,-2.8) -- (\v,-3.25);
  \draw[densely dotted, draw=gray]  (\v+0.125,-2.8) -- (\v,-3.25);
  \draw[densely dotted, draw=gray]  (\v+0.25,-2.85) -- (\v,-3.25);
}

\node (slot0) at (-1,-3.75) {Slot 0};
\node (slot0) at (1.25,-3.75) {Slot 1};
\node (slot0) at (4.25,-3.75) {Slot 4};

\end{tikzpicture}
\caption{Encoding for CCS proceeds as follows.
Information bits are partitioned into $n$ fragments.
These fragments are enhanced with redundancy in the form of parity bits.
Each sub-block is converted into a signal via a CS matrix, and subsequently transmitted over a time slot.}
\label{fig:EncodingProcess}
\end{figure}
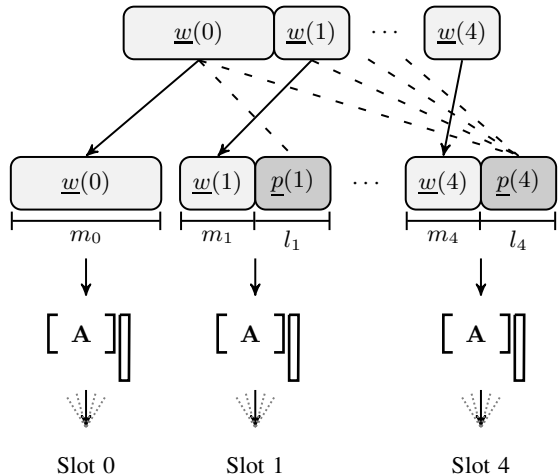

The transmission strategy features two complementary components: a systematic linear block code based on random parity checks, which we call tree code, and a signaling scheme rooted in compressed sensing.
These building blocks are detailed below, and a top level overview of the encoding process is illustrated in Fig.~\ref{fig:EncodingProcess}.


\subsubsection{Tree Encoding}

Every $B$-bit binary message $\wv$ is partitioned into $n$ sub-blocks, where the $j$th sub-block consisting of $m_j$ message bits, with $\sum_{j=0}^{n-1} m_j = B$.
Under this fragmentation, we can express a message as $\wv = \wv(0) \wv(1) \cdots \wv(n-1)$.
The tree encoder appends $l_j$ parity check bits to sub-block~$j$, bringing the total length of every sub-block to $m_j + l_j = J = {M}/{n}$ bits.
The first block is chosen to have $m_0 = J$ information bits and, necessarily, $l_0 = 0$ parity bits.
For subsequent sub-blocks, the parity bits are constructed as follows.
Let $\pv(j)$ denote the parity bits in sub-block~$j$.
These $l_j$ checks are selected to satisfy random parity constraints associated with the message bits that are contained in preceding sub-blocks.
Mathematically,
\begin{equation*}
\pv(j) = \sum_{\ell=0}^{j-1} \wv(\ell) G_{\ell,j-1}
\end{equation*}
where $G_{\ell,j-1}$ is a $m_{\ell} \times l_j$ Rademacher matrix.
That is, the entries in $G_{\ell,j-1}$ are independent Bernoulli trials, each with parameter half.
Parity bits are computed using modulo-2 arithmetic and, as such, they remain binary.
These parity bits are subsequently injected into a codeword $\vv$, which assumes the form
\begin{equation}
\vv = \underbrace{\wv(0)}_{\vv(0)} \underbrace{\wv(1) \pv(1)}_{\vv(1)} \cdots \underbrace{\wv(n-1) \pv(n-1)}_{\vv(n-1)} .
\end{equation}
In effect, $\pv(j)$ acts as parity constraints for random collections of information bits preceding fragment~$j$.
In Section~\ref{parityvector}, we describe an optimization framework for the selection of parity lengths $l_1, l_2, \dots, l_{n-1}$.

\subsubsection{CS Encoding}

Let $\mathbf{A}=[\av_1,\ldots,\av_{2^J}] \in \{\pm \sqrt{E_{\mathrm{s}}}\}^{\frac{N}{n} \times 2^J}$ be a compressed sensing matrix that is designed to recover $\Ka$-sparse binary vectors from $N/n$ noisy observations with a low probability of failure.
This matrix is meant to operate on tree-coded sub-blocks, which explains its size.
Conceptually, every possible tree-coded $J$-bit vector is mapped to a column of $\mathbf{A}$.
Thus, we can view the columns of $\mathbf{A}$ as the set of potentially transmitted signals by a user over the duration of a CS sub-block.
In Fig.~\ref{fig:EncodingProcess}, the CS encoding process is portrayed as the product of an abstract sensing matrix and a tall index vector of length $2^J$.
Under this representation, the index vector $\bv_i(j)$ associated with $\vv_i(j)$, the codeword fragment of user~$i$, has zeros everywhere except at location $\operatorname{decimal}(\wv_i(j))$, where the entry is one.
The figure also emphasizes the divide-and-conquer approach where the processing of one message leads to the creation of $n$ CS instances.
During every transmission round, the multiple access channel sums up the contributions of all the active users.
This is equivalent to sending aggregate signal $\mathbf{A} \bv(j)$ over a noisy channel, where $\Ka$-sparse vector $\bv(j) = \sum_{i \in \mathbf{S}_\mathrm{a}} \bv_i(j)$.


\subsection{Decoding Process}
\label{decoder}
\begin{figure}[tbh]
\centering
\begin{tikzpicture}[
  font=\small, >=stealth', line width = 0.75pt,
  wblock/.style={rectangle, minimum height=2mm, draw=black, fill=gray!10},
  pblock/.style={rectangle, minimum height=2mm, draw=black, fill=gray!40}
]

\foreach \v in {-1.00,1.25,4.25} {
  \draw[->] (\v,1.5) -- (\v,1);
  \draw (\v-0.25,1.4) -- (\v,1);
  \draw (\v-0.125,1.45) -- (\v,1);
  \draw (\v+0.125,1.45) -- (\v,1);
  \draw (\v+0.25,1.4) -- (\v,1);

  \node at (\v-0.0625,0.5) {$\mathbf{A}$};
  \draw[line width=1pt] (\v-0.375,0.75) -- (\v-0.5,0.75) -- (\v-0.5,0.25) -- (\v-0.375,0.25);
  \draw[line width=1pt] (\v+0.25,0.75) -- (\v+0.375,0.75) -- (\v+0.375,0.25) -- (\v+0.25,0.25);
  \draw[line width=1pt] (\v+0.45,0.75) -- (\v+0.45,-0.125) -- (\v+0.57,-0.125) -- (\v+0.575,0.75) -- (\v+0.45,0.75);
  \draw[->] (\v,-0.125) to (\v,-0.625);

  \node (vdot-\v) at (\v,-2.2) {$\vdots$};
}
\node (hdot) at (2.75,0.375) {$\cdots$};

\foreach \w in {-1.0,-1.4,-1.8, -2.8} {
  \node[wblock,minimum width=20mm] (w0-\w) at (-1,\w) {};
  \node[wblock,minimum width=10mm] (w1-\w) at (0.75,\w) {};
  \node[pblock,minimum width=10mm] (p1-\w) at (1.75,\w) {};
  \node[wblock,minimum width=10mm] (wn-\w) at (3.75,\w) {};
  \node[pblock,minimum width=10mm] (pn-\w) at (4.75,\w) {};
}

\node (L0) at (-1,-3.25) {$\mathcal{L}_0$};
\node (L1) at (1.25,-3.25) {$\mathcal{L}_1$};
\node (Ln) at (4.25,-3.25) {$\mathcal{L}_{n-1}$};

\draw[line width=1.5pt,color=black,line cap=round] plot[smooth, tension=.55] coordinates {(-1.75,-1.0) (-0.25,-1) (0.5,-1.8) (2, -1.8) (3.5,-1.4) (5,-1.4)};

\end{tikzpicture}
\caption{The decoding process at the destination starts by running a CS recovery algorithm on the noisy signal aggregate corresponding to each time slot.
This yields lists of coded fragments $\mathcal{L}_0, \mathcal{L}_1, \ldots, \mathcal{L}_{n-1}$, one list per time slot.
Message fragments are then stitched together using the redundant structure of the tree code, as depicted in this illustration.
}
\label{fig:DecodingProcess}
\end{figure}
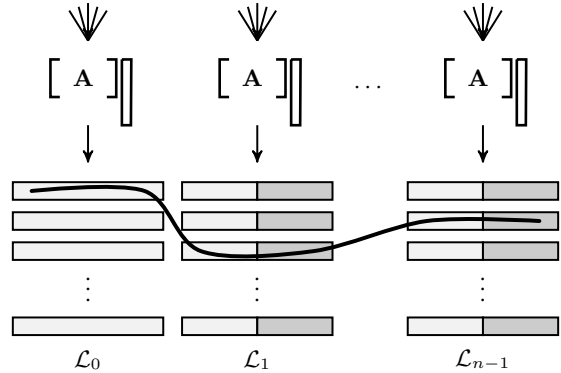

As mentioned above, the input to the decoding process is the sum of the signals transmitted by the active users plus noise.
This received signal is segmented along the boundaries of the fragments, with the portion corresponding to each slot viewed as a sub-block.
Not too surprisingly, the decoding strategy also consists of two components: a CS recovery algorithm that operates over every sub-block, and a tree decoder operating across sub-blocks.
These steps are illustrated in Fig.~\ref{fig:DecodingProcess}, and they are detailed below.


\subsubsection{CS Decoding}
\label{csdec}
The aggregate signal received at the base station during the $j$th sub-block can be expressed as $\yv(j) = \mathbf{A}\bv(j) + \zv(j)$, where $\bv(j) \in \{0,1\}^{2^J}$ is a $\Ka$-sparse binary vector that incorporates the indices of $j$th sub-blocks transmitted by the active users.
The task of the CS decoder is to provide an estimate of the sparse vector $\bv(j)$ based on the received signal $\yv(j)$ during the corresponding time slot.
This is accomplished by first applying a conventional CS recovery algorithm, e.g., non-negative least squares (NNLS) or LASSO, to get an estimate $\bve^{(\mathrm{cs})}(j)$ of vector $\bv(j)$.
Yet, this does not ensure that the entries of vector $\bve^{(\mathrm{cs})}(j)$ are binary.
The desired binary estimate $\bve(j)$ is obtained by setting the $K$ largest entries of vector $\bve^{(\mathrm{cs})}(j)$ to one and the remaining $2^J-K$ entries to zero.
Sparsity parameter $K$ is chosen as $K=\Ka+K_{\delta}$, where $K_{\delta}$ is a small positive integer.
The fragments corresponding to index vector $\bve(j)$ are aggregated on a list $\mathcal{L}_j$, which acts as the output of the CS algorithm for this sub-problem.
Although the list output by the CS recovery algorithm is larger than $\Ka$, the quantity $K_{\delta}$ is carefully chosen such that erroneous fragments remain very unlikely to satisfy the parity check constraints associated with the tree code.

\subsubsection{Tree Decoding}
\label{Treedec}
The tree decoder seeks to recover the original messages transmitted by all the users by piecing together valid sequences of elements drawn from the various CS lists.
Towards this end, the access point constructs a decoding tree for \textit{every candidate message} as follows.
We fix a sub-block from $\mathcal{L}_0$, the list of all possible initial sub-blocks supplied by the CS decoder.
We view this fragment as the root node of a tree.
Once this sub-block is selected, there are $K$ possible choices for the subsequent sub-block, and these are the nodes which appear on $\mathcal{L}_1$.
Similarly, there are $K$ possible choices for the second sub-block for each partial path identified thus far and, hence, there are $K^2$ partial paths at stage two.
This process continues until stage~$(n-1)$ is reached.
At this point, the tree features $K^{n-1}$ leafs.
Every path connecting the root node to a leaf becomes a candidate message, but invalid paths that do not meet their own parity requirements can be eliminated.
If there remains a single valid path that meets its parity checks, then the decoder outputs the corresponding message; otherwise, it reports a failure.

\begin{figure}
        \centering
        \begin{subfigure}[t]{0.4\textwidth}
            \centering
            \resizebox{0.9\textwidth}{!}{\begin{tikzpicture}[
  font=\small, >=stealth', line width = 0.75pt,
  node/.style={circle, minimum width=5mm, inner sep=0pt, draw=black},
  pnode/.style={circle, minimum width=5mm, inner sep=0pt, draw=black, fill=gray!40},
  snode/.style={circle, minimum width=4.5mm, inner sep=0pt, draw=black}
]

\node[pnode] (root) at (0,0) {$1$};

\node[node] (level1-1) at (-3,-1.5) {$1$}
  edge[<-] (root);
\node[node] (level1-2) at (-0.5,-1.5) {$2$}
  edge[<-] (root);
\node at (1,-1.5) {$\cdots$};
\node[node] (level1-4) at (2.5,-1.5) {$K$}
  edge[<-] (root);

\node[node] (level21-1) at (-3-0.9,-3) {$1$}
  edge[<-] (level1-1);
\node[node] (level21-2) at (-3-0.3,-3) {$2$}
  edge[<-] (level1-1);
\node  at (-3+0.3,-3) {$\cdots$};
\node[node] (level21-4) at (-3+0.9,-3) {$K$}
  edge[<-] (level1-1);

\node[node] (level22-1) at (-0.5-0.9,-3) {$1$}
  edge[<-] (level1-2);
\node[node] (level22-2) at (-0.5-0.3,-3) {$2$}
  edge[<-] (level1-2);
\node at (-0.5+0.3,-3) {$\cdots$};
\node[node] (level22-4) at (-0.5+0.9,-3) {$K$}
  edge[<-] (level1-2);

\node[node] (level24-1) at (2.5-0.9,-3) {$1$}
  edge[<-] (level1-4);
\node[node] (level24-2) at (2.5-0.3,-3) {$2$}
  edge[<-] (level1-4);
\node at (2.5+0.3,-3) {$\cdots$};
\node[node] (level24-4) at (2.5+0.9,-3) {$K$}
  edge[<-] (level1-4);

\foreach \ell in {-3, -0.5, 2.5} {
  \node at (\ell-0.9,-3.65) {$\vdots$};
  \node at (\ell-0.3,-3.65) {$\vdots$};
  \node at (\ell+0.9,-3.65) {$\vdots$};

  \node[snode] at (\ell-0.95,-4.5) {};
  \node[snode] at (\ell-0.475,-4.5) {};
  \node[snode] at (\ell,-4.5) {};
  \node[snode] at (\ell+0.475,-4.5) {};
  \node[snode] at (\ell+0.95,-4.5) {};
}
\end{tikzpicture}}
            \caption[]%
            {Stage $0$: Processing one element from $\mathcal{L}_0$ at a time, a fragment is selected as the root node of a tree.}
        \end{subfigure}
        \quad
        \begin{subfigure}[t]{0.4\textwidth}  
            \centering 
            \resizebox{0.9\textwidth}{!}{\begin{tikzpicture}[
  font=\small, >=stealth', line width = 0.75pt,
  node/.style={circle, minimum width=5mm, inner sep=0pt, draw=black},
  pnode/.style={circle, minimum width=5mm, inner sep=0pt, draw=black, fill=gray!40},
  snode/.style={circle, minimum width=4.5mm, inner sep=0pt, draw=black}
]

\node[pnode] (root) at (0,0) {$1$};

\node[pnode] (level1-1) at (-3,-1.5) {$1$}
  edge[<-] (root);
\node[pnode] (level1-2) at (-0.5,-1.5) {$2$}
  edge[<-] (root);
\node at (1,-1.5) {$\cdots$};
\node[node] (level1-4) at (2.5,-1.5) {$K$}
  edge[dashed] (root);

\node[node] (level21-1) at (-3-0.9,-3) {$1$}
  edge[<-] (level1-1);
\node[node] (level21-2) at (-3-0.3,-3) {$2$}
  edge[<-] (level1-1);
\node  at (-3+0.3,-3) {$\cdots$};
\node[node] (level21-4) at (-3+0.9,-3) {$K$}
  edge[<-] (level1-1);

\node[node] (level22-1) at (-0.5-0.9,-3) {$1$}
  edge[<-] (level1-2);
\node[node] (level22-2) at (-0.5-0.3,-3) {$2$}
  edge[<-] (level1-2);
\node at (-0.5+0.3,-3) {$\cdots$};
\node[node] (level22-4) at (-0.5+0.9,-3) {$K$}
  edge[<-] (level1-2);

\node[node] (level24-1) at (2.5-0.9,-3) {$1$}
  edge[dashed] (level1-4);
\node[node] (level24-2) at (2.5-0.3,-3) {$2$}
  edge[dashed] (level1-4);
\node at (2.5+0.3,-3) {$\cdots$};
\node[node] (level24-4) at (2.5+0.9,-3) {$K$}
  edge[dashed] (level1-4);

\foreach \ell in {-3, -0.5, 2.5} {
  \node at (\ell-0.9,-3.65) {$\vdots$};
  \node at (\ell-0.3,-3.65) {$\vdots$};
  \node at (\ell+0.9,-3.65) {$\vdots$};

  \node[snode] at (\ell-0.95,-4.5) {};
  \node[snode] at (\ell-0.475,-4.5) {};
  \node[snode] at (\ell,-4.5) {};
  \node[snode] at (\ell+0.475,-4.5) {};
  \node[snode] at (\ell+0.95,-4.5) {};
}
\end{tikzpicture}}
            \caption[]%
            {Stage $1$: Fragments in $\mathcal{L}_1$ act as the children of the root node. Parity requirements are checked and only complying nodes, nodes~$1$ and~$2$ highlighted in the figure, are retained.}
        \end{subfigure}
        \vskip\baselineskip
        \begin{subfigure}[t]{0.4\textwidth}   
            \centering 
            \resizebox{0.9\textwidth}{!}{\begin{tikzpicture}[
  font=\small, >=stealth', line width = 0.75pt,
  node/.style={circle, minimum width=5mm, inner sep=0pt, draw=black},
  pnode/.style={circle, minimum width=5mm, inner sep=0pt, draw=black, fill=gray!40},
  snode/.style={circle, minimum width=4.5mm, inner sep=0pt, draw=black}
]

\node[pnode] (root) at (0,0) {$1$};

\node[pnode] (level1-1) at (-3,-1.5) {$1$}
  edge[<-] (root);
\node[pnode] (level1-2) at (-0.5,-1.5) {$2$}
  edge[<-] (root);
\node at (1,-1.5) {$\cdots$};
\node[node] (level1-4) at (2.5,-1.5) {$K$}
  edge[dashed] (root);

\node[node] (level21-1) at (-3-0.9,-3) {$1$}
  edge[dashed] (level1-1);
\node[pnode] (level21-2) at (-3-0.3,-3) {$2$}
  edge[<-] (level1-1);
\node  at (-3+0.3,-3) {$\cdots$};
\node[node] (level21-4) at (-3+0.9,-3) {$K$}
  edge[dashed] (level1-1);

\node[pnode] (level22-1) at (-0.5-0.9,-3) {$1$}
  edge[<-] (level1-2);
\node[node] (level22-2) at (-0.5-0.3,-3) {$2$}
  edge[dashed] (level1-2);
\node at (-0.5+0.3,-3) {$\cdots$};
\node[pnode] (level22-4) at (-0.5+0.9,-3) {$K$}
  edge[<-] (level1-2);

\node[node] (level24-1) at (2.5-0.9,-3) {$1$}
  edge[dashed] (level1-4);
\node[node] (level24-2) at (2.5-0.3,-3) {$2$}
  edge[dashed] (level1-4);
\node at (2.5+0.3,-3) {$\cdots$};
\node[node] (level24-4) at (2.5+0.9,-3) {$K$}
  edge[dashed] (level1-4);

\foreach \ell in {-3, -0.5, 2.5} {
  \node at (\ell-0.9,-3.65) {$\vdots$};
  \node at (\ell-0.3,-3.65) {$\vdots$};
  \node at (\ell+0.9,-3.65) {$\vdots$};

  \node[snode] at (\ell-0.95,-4.5) {};
  \node[snode] at (\ell-0.475,-4.5) {};
  \node[snode] at (\ell,-4.5) {};
  \node[snode] at (\ell+0.475,-4.5) {};
  \node[snode] at (\ell+0.95,-4.5) {};
}
\end{tikzpicture}}
            \caption[]%
            {Subsequent Stages: Candidate fragments from a subsequent stage become the children of complying nodes.
            Again, parity constraints are verified for every child and the tree is pruned accordingly, as highlighted above.}
        \end{subfigure}
        \quad
        \begin{subfigure}[t]{0.4\textwidth}   
            \centering 
            \resizebox{0.9\textwidth}{!}{\begin{tikzpicture}[
  font=\small, >=stealth', line width = 0.75pt,
  node/.style={circle, minimum width=5mm, inner sep=0pt, draw=black},
  pnode/.style={circle, minimum width=5mm, inner sep=0pt, draw=black, fill=gray!40},
  snode/.style={circle, minimum width=4.5mm, inner sep=0pt, draw=black}
]

\node[pnode] (root) at (0,0) {$1$};

\node[pnode] (level1-1) at (-3,-1.5) {$1$}
  edge[<-] (root);
\node[pnode] (level1-2) at (-0.5,-1.5) {$2$}
  edge[<-] (root);
\node at (1,-1.5) {$\cdots$};
\node[node] (level1-4) at (2.5,-1.5) {$K$}
  edge[dashed] (root);

\node[node] (level21-1) at (-3-0.9,-3) {$1$}
  edge[dashed] (level1-1);
\node[node] (level21-2) at (-3-0.3,-3) {$2$}
  edge[<-] (level1-1);
\node  at (-3+0.3,-3) {$\cdots$};
\node[node] (level21-4) at (-3+0.9,-3) {$K$}
  edge[dashed] (level1-1);

\node[pnode] (level22-1) at (-0.5-0.9,-3) {$1$}
  edge[<-] (level1-2);
\node[node] (level22-2) at (-0.5-0.3,-3) {$2$}
  edge[dashed] (level1-2);
\node at (-0.5+0.3,-3) {$\cdots$};
\node[pnode] (level22-4) at (-0.5+0.9,-3) {$K$}
  edge[<-] (level1-2);

\node[node] (level24-1) at (2.5-0.9,-3) {$1$}
  edge[dashed] (level1-4);
\node[node] (level24-2) at (2.5-0.3,-3) {$2$}
  edge[dashed] (level1-4);
\node at (2.5+0.3,-3) {$\cdots$};
\node[node] (level24-4) at (2.5+0.9,-3) {$K$}
  edge[dashed] (level1-4);

\foreach \ell in {-3, -0.5, 2.5} {
  \node at (\ell-0.9,-3.65) {$\vdots$};
  \node at (\ell-0.3,-3.65) {$\vdots$};
  \node at (\ell+0.9,-3.65) {$\vdots$};

  \node[snode] at (\ell-0.95,-4.5) {};
  \node[snode] at (\ell-0.475,-4.5) {};
  \node[snode] at (\ell,-4.5) {};
  \node[snode] at (\ell+0.475,-4.5) {};
  \node[snode] at (\ell+0.95,-4.5) {};
}
\node[snode,fill=gray!40] at (-0.5-0.95,-4.5) {};

\draw[line width=1.5pt,color=black,line cap=round] plot[smooth, tension=.55] coordinates {(0.15,0) (-0.65,-1.5) (-0.35+0.9,-3) (-0.6-0.95, -4.5)};
\end{tikzpicture}}
            \caption[]%
            {Last Stage: Parity constraints are verified at the leafs. A valid message on the CS tree will survives, but decoding is successful only if no other paths meet its parity requirements.
            We highlight the legitimate path in black above.}
        \end{subfigure}
        \caption[]
        {These images illustrate the operation of the tree decoding algorithm across sub-blocks.} 
        \label{fig:Tree Dec}
    \end{figure}
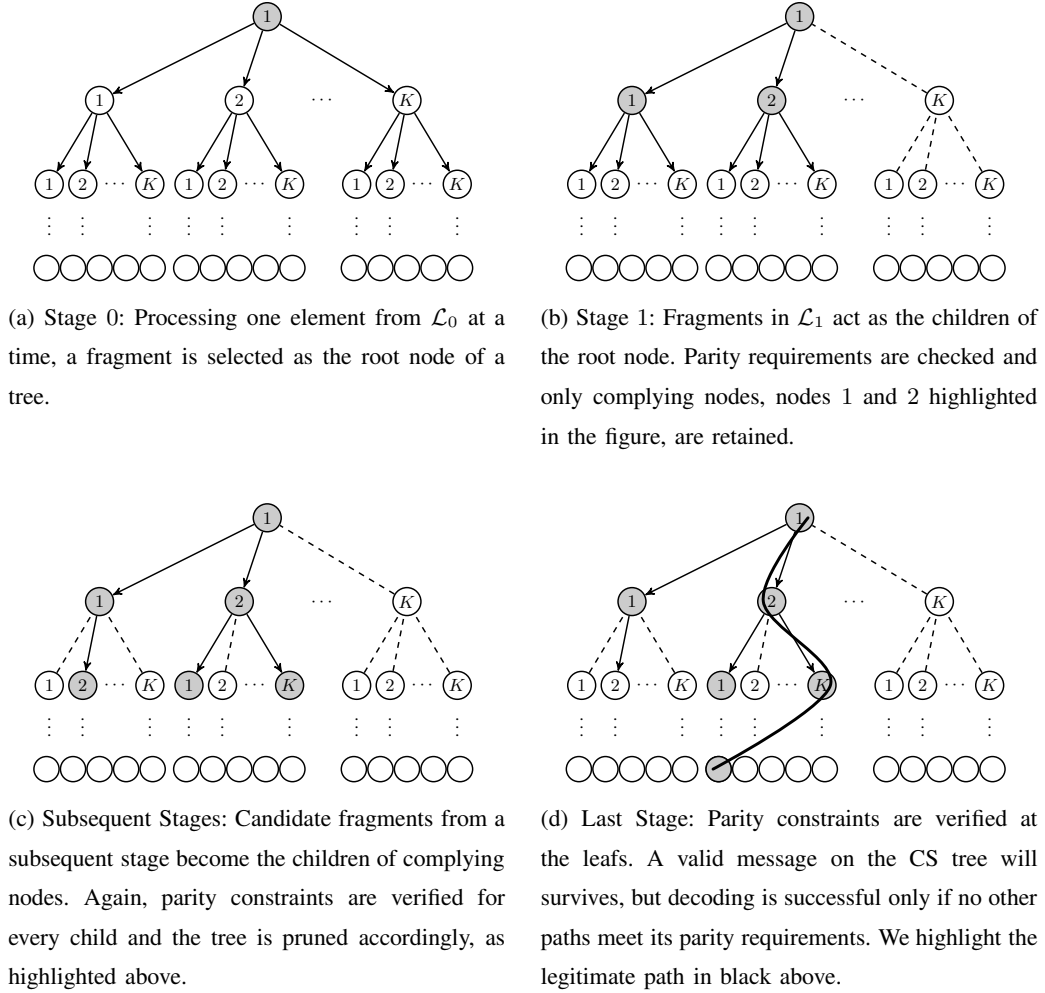

We emphasize that the number of possible paths increases exponentially with the stages of the tree and, hence, a naive search through all the leaf nodes is infeasible.
In practice, invalid paths can be pruned iteratively through the parity check constraints.
Specifically, at stage $j \ge 1$, the decoder retains only nodes that satisfy the $l_j$ parity constraints on all the message bits preceding that stage.
This iterative procedure continues until the last stage is reached.
The complexity of this decoding scheme depends on the number of nodes surviving each stage because parity checks only need to be enforced on the children of surviving nodes in subsequent stages of the tree decoding process.
Figure~\ref{fig:Tree Dec} gives a step-by-step description of the various stages involved in the tree decoding algorithm. 

\subsubsection{Iterative Extension}
\label{section:IterativeExtension}

The CCS framework admits an iterative extension based on successive cancellation.
At the end of the tree decoding process, the receiver has identified a collection of candidate messages $\widehat{W}(\yv)$.
The indices of the $j$th fragments of these messages necessarily appear on $\bve(j) \in \{0,1\}^{2^B}$.
However, when a sub-block is mistakenly discarded by the CS recovery algorithm or if the tree decoder fails to stitch a message without including extraneous paths, the number of used fragments within $\bve(j)$ may not be equal to $\Ka$.
In particular, on a first pass, $|\widehat{W}(\yv)|$ may not be equal to $\| \bv(j) \|_0 = \Ka$.
This situation invites an iterative procedure to recover the missing fragments and messages.
The contributions of the fragments associated with $\widehat{W}(\yv)$ can be subtracted from their corresponding CS sub-signal $\yv(j)$ to yield
\begin{equation}
\begin{split}
\underbrace{\yv(j) - \mathbf{A} \bve_{\mathrm{tree}} (j)}_{\yv_{\mathrm{residual}}(j)}
&= \underbrace{\xv(j) - \mathbf{A} \bve_{\mathrm{tree}} (j)}_{\xv_{\mathrm{residual}}(j)} + \zv(j)
\end{split}
\end{equation}
where $\bve_{\mathrm{tree}}$ only contains the indices within $\bve(j)$ that are associated with fragments found on the output of the tree decoder $\widehat{W}(\yv)$.
The above equation resembles the standard form of the noisy compressed sensing problem, where $\mathbf{A}$ is the sensing matrix and $\bv(j) - \bve_{\mathrm{tree}}(j)$ is the unknown binary vector with sparsity index $\Ka-|\widehat{W}(\yv)|$.
An improved estimate for vector $\bv(j)$ can then be obtained by solving the above residual problem using the divide-and-conquer approach where the CS recovery algorithm operates at sub-block level and the tree decoder across sub-blocks.
The list size output by the CS algorithm at the sub-block level for this iteration is reduced accordingly, based on the number of unidentified messages $\Ka-|\widehat{W}(\yv)|$.
This successive interference cancellation method can be repeated iteratively, potentially leading to significant gains in performance, especially for the first few iterations.

\section{Performance Analysis}
\label{perr}

In this section, we study the performance of the CCS multiple access scheme.
We rely mostly on established results to characterize the CS component of the scheme.
On the other hand, the analysis of tree coding is new and, as such, it is our primary focus.

Let $E_i$ denote the event that the message sent by user~$i$ does not appear on the list output by the tree decoder.
Similarly, let $C_i$ be the event that all the sub-blocks generated by this user are present on the lists output by the CS decoder.
With this notation, we can write probability $\Pr (E_i)$ as
\begin{equation} \label{Totalprob}
\Pr (E_i) = \Pr (E_i\lvert C_i) \Pr (C_i) + \Pr (E_i \lvert \overline{C_i}) \Pr (\overline{C_i}) .
\end{equation}
Consider the event where sub-block~$i_j$ transmitted by user~$i$ during slot~$j$ fails to appear on the list output by the CS algorithm at that stage.
We note that $\pcs$, the probability of this event, is uniformly the same among users and across time slots due to the symmetric structure of the decoding process.
Equivalently, the probability that the list output by the CS recovery algorithm contains sub-block~$i_j$ is equal to $1-\pcs$. When the CS recovery algorithm fails to identify at least one of the sub-blocks belonging to a particular user, then the output of the tree decoder may not contain the original codeword sent by that user.
Given that the CS sub-blocks are decoded independently, $\Pr (C_i)$ can be computed as $\Pr (C_i) = (1 - \pcs)^n$.

We write the conditional event that the tree decoder declares a failure because of more than one path surviving the tree decoding process by $E_i \lvert C_i$, and we represent the  probability corresponding to this event as $\ptree$.
In the base form of the algorithm, without the optional iteration process discussed in Section~\ref{section:IterativeExtension}, the quantity $P_e$ is the same as $\Pr (E_i)$.
They can be computed using \eqref{Totalprob}, along with the aforementioned observations,
\begin{equation} \label{petotal}
\begin{split}
P_e &= \ptree (1 - \pcs)^n + \left( 1 - (1 - \pcs)^n \right) \\
&= 1-(1-\ptree)(1-\pcs)^n .
\end{split}
\end{equation}
In view of this expression, we turn our attention to better understanding $\ptree$.

\subsection{Tree Code Analysis}
\label{subsection:TreeCodeAnalysis}

The exact analysis of the tree decoding process, although possible, is quite intricate.
Herein, we present the key results, but relegate details to Appendix~\ref{Section:PerformanceTreeCodeAppendix}. 
The main idea is to compute the expected number of surviving erroneous paths at various levels of the tree decoding process, and then bound the probability that erroneous paths survive altogether using the Markov inequality.
Let $L_j$ denote the random variable corresponding to the number of erroneous paths that survive stage $j \in[1:n-1]$ of the tree decoding process, assuming that all the fragments corresponding to the root sub-blocks appear on the CS output lists and that the list size is $K = \Ka$.
The expected value of this random variable is characterized in Theorem~\ref{Theorem:ExpectedNumberErroneouspaths}.

\begin{theorem} \label{Theorem:ExpectedNumberErroneouspaths}
The expected number of erroneous surviving paths at level $j-1$ is given by
\begin{equation*}
\mathbb{E}[L_{j-1}] = \sum_{\sv \in \mc{P}_j} n(\sv) \Phi_{\sv} \left( \frac{1}{2} \right) - 1,
\quad j \in [1:n].
\end{equation*}
where $\mc{P}_j$ represents the collection of $j$-patterns (OEIS A008277),
\begin{equation*}
n(\sv) = (K-1)(K-2) \cdots (K-(d(\sv)-1))
\end{equation*}
with $d(\sv)$ denoting the number of distinct integers in $\sv$.
Furthermore, $\Phi_{\sv}(x)$ is the probability generating function of $T(\sv)$, the number of statistically discriminating parity bits for a path associated with pattern $\sv$.
It takes the form
\begin{equation*}
\Phi_{\sv}(x) = \mathbb{E} \left[ x^{T(\sv)} \right]
= \sum_{t=0}^{l_1 + \cdots + l_{j-1}} \Pr (T(\sv)=t)x^t
\end{equation*}
where $\Pr (T(\sv)=t)$ is the probability that exactly $t$ parity bits are statistically discriminating for $j$-pattern sequence $\sv$.
\end{theorem}
\begin{IEEEproof}
A proof for this theorem is available in Appendix~\ref{Section:PerformanceTreeCodeAppendix}.
Section~\ref{subsection:FragmentedCodewrds} presents conditions for erroneous paths to survive.
A complexity reduction scheme based on $j$-patterns and the Bell numbers is found in Section~\ref{subsection:PatternSequence}.
The notion of statistically discriminating parity bits appears in Section~\ref{subsection:StatisticallyDiscriminatingParityBits}, along with a characterization of their probability distributions.
Collectively, these components provide a computationally tractable means to compute $\mathbb{E}[L_{j-1}]$ for the parameters of interest.
\end{IEEEproof}

\begin{figure}[htb]
\centerline{
\begin{tikzpicture}[
  font=\scriptsize, >=stealth', line width=1.25pt, line cap=round,
  node/.style={circle, minimum size=3.25mm, inner sep=0pt, draw=black}
]

\node[node] (s1) at (0,3.1) {1};
  \node[node] (s1-1) at (1,3.1) {1};
    \node[node] (s1-1-1) at (2.5,3.1) {1};
      \node[node,label=right:{$(1,1,1,1)$}] (s1-1-1-1) at (5,3.1) {1};
      \node[node,label=right:{$(1,1,1,4)$}] (s1-1-1-4) at (5,-1.5) {3};
    \node[node] (s1-1-3) at (2.5,-0.5) {3};
      \node[node,label=right:{$(1,1,3,1)$}] (s1-1-3-1) at (5,1.9) {3};
      \node[node,label=right:{$(1,1,3,3)$}] (s1-1-3-3) at (5,-0.5) {3};
      \node[node,label=right:{$(1,1,3,4)$}] (s1-1-3-4) at (5,-2.7) {6};
  \node[node] (s1-2) at (1,0.9) {3};
    \node[node] (s1-2-1) at (2.5,2.7) {3};
      \node[node,label=right:{$(1,2,1,1)$}] (s1-2-1-1) at (5,2.7) {3};
      \node[node,label=right:{$(1,2,1,2)$}] (s1-2-1-2) at (5,0.9) {3};
      \node[node,label=right:{$(1,2,1,4)$}] (s1-2-1-4) at (5,-1.9) {6};
    \node[node] (s1-2-2) at (2.5,0.9) {3};
      \node[node,label=right:{$(1,2,2,1)$}] (s1-2-2-1) at (5,2.3) {3};
      \node[node,label=right:{$(1,2,2,2)$}] (s1-2-2-2) at (5,0.5) {3};
      \node[node,label=right:{$(1,2,2,4)$}] (s1-2-2-4) at (5,-2.3) {6};
    \node[node] (s1-2-3) at (2.5,-0.9) {6};
      \node[node,label=right:{$(1,2,3,1)$}] (s1-2-3-1) at (5,1.5) {6};
      \node[node,label=right:{$(1,2,3,2)$}] (s1-2-3-2) at (5,0.1) {6};
      \node[node,label=right:{$(1,2,3,3)$}] (s1-2-3-3) at (5,-0.9) {6};
      \node[node,label=right:{$(1,2,3,4)$}] (s1-2-3-4) at (5,-3.1) {6};

\draw[line width=0.5pt] (s1) -- (s1-1);
  \draw[line width=0.5pt] (s1-1) -- (s1-1-1);
    \draw[line width=0.5pt] (s1-1-1) -- (s1-1-1-1);
    \draw[line width=1.5pt] (s1-1-1) -- (s1-1-1-4);
  \draw[line width=1.5pt] (s1-1) -- (s1-1-3);
    \draw[line width=1.5pt] (s1-1-3) -- (s1-1-3-1);
    \draw[line width=1.5pt] (s1-1-3) -- (s1-1-3-3);
    \draw[line width=3.0pt] (s1-1-3) -- (s1-1-3-4);
\draw[line width=1.5pt] (s1) -- (s1-2);
  \draw[line width=1.5pt] (s1-2) -- (s1-2-1);
    \draw[line width=1.5pt] (s1-2-1) -- (s1-2-1-1);
    \draw[line width=1.5pt] (s1-2-1) -- (s1-2-1-2);
    \draw[line width=3.0pt] (s1-2-1) -- (s1-2-1-4);
  \draw[line width=1.5pt] (s1-2) -- (s1-2-2);
    \draw[line width=1.5pt] (s1-2-2) -- (s1-2-2-1);
    \draw[line width=1.5pt] (s1-2-2) -- (s1-2-2-2);
    \draw[line width=3.0pt] (s1-2-2) -- (s1-2-2-4);
  \draw[line width=3.0pt] (s1-2) -- (s1-2-3);
    \draw[line width=3.0pt] (s1-2-3) -- (s1-2-3-1);
    \draw[line width=3.0pt] (s1-2-3) -- (s1-2-3-2);
    \draw[line width=3.0pt] (s1-2-3) -- (s1-2-3-3);
    \draw[line width=3.0pt] (s1-2-3) -- (s1-2-3-4);

\draw[dashed] (-0.4,3.4) -- (6.8,3.4);
\draw[dashed] (-0.1,1.2) -- (6.5,1.2);
\draw[dashed] (-0.1,-0.2) -- (6.5,-0.2);
\draw[dashed] (-0.1,-1.2) -- (6.5,-1.2);
\draw[dashed] (-0.4,-3.4) -- (6.8,-3.4);

\node[rotate=90] at (-0.4,2.3) {\small level~$0$};
\node[rotate=90] at (-0.4,0.5) {\small level~$1$};
\node[rotate=90] at (-0.4,-0.7) {\small level~$2$};
\node[rotate=90] at (-0.4,-2.3) {\small level~$3$};

\node[rotate=-90] at (6.8,0) {\small $j$-patterns $\sv$};
\end{tikzpicture}}
\caption{This diagram illustrates the level structure of tree coding and the reduction afforded by $j$-patterns.
In this particular scenario, the total number of paths to consider decreases from 64 to 15.
The number of (partial) paths $n(\sv)$ in each equivalence class $\sv$ appears inside the corresponding node, and it is also highlighted through the thickness of the connection.}
\label{figure:Bell_diagram}
\end{figure}
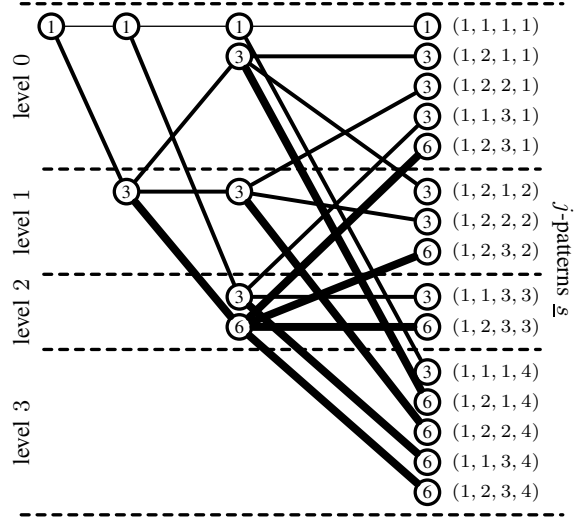
Figure~\ref{figure:Bell_diagram} illustrates the reduction in complexity afforded by $j$-patterns for $n=4$ and $K=4$.
A candidate path enters a new level whenever its next fragment comes from a message that has not been visited before.
As a corollary to this Theorem~\ref{Theorem:ExpectedNumberErroneouspaths}, we get a bound on $\ptree$.
This bound, presented in Corollary~\ref{corollary:PtreeBound}, is particularly close when $\ptree$ is small.

\begin{corollary} \label{corollary:PtreeBound}
Assuming that all the fragments of a message appear on their respective CS lists, the probability that the tree decoder declares a failure due to more than one path surviving is bounded by
\begin{equation*}
\ptree \leq \mathbb{E}[L_{n-1}] .
\end{equation*}
\end{corollary}
\begin{IEEEproof}
We can bound the probability of at least one extraneous path surviving by
\begin{equation*}
\begin{split}
\ptree &= \Pr ( E_i \lvert C_i )
= \Pr (L_{n-1} \geq 1) \\
&\leq \mathbb{E}[L_{n-1}] .
\end{split}
\end{equation*}
The inequality follows from an application of the Markov inequality.
\end{IEEEproof}

This bound can be substituted in \eqref{petotal} to provide guarantees on the performance of tree coding.
We illustrate its use in our upcoming numerical results.
Still, the amount of computations involved in applying Theorem~\ref{Theorem:ExpectedNumberErroneouspaths} remains significant.
Above, the derivations are presented for a given parity allocation $l_1, l_2, \ldots, l_{n-1}$.
Moving forward, we need to find suitable values for these parameters.
In principle, this latter task involves repetitive computations of $\ptree$ for numerous candidate assignments.
To circumvent the computational load this would entail, we develop an easily computable approximation for $\ptree$, which enables the selection of appropriate values for $l_1, \ldots, l_{n-1}$.
This is accomplished below.

\subsection{Approximate Tree Code Analysis}
\label{subsection:ApproximateTreeCodeAnalysis}

The complexity associated with the exact characterization of tree codes stems from the fact that many cases must be considered separately.
These cases correspond to equivalent classes associated with $j$-patterns, as discussed above.
From an abstract point of view, three confounding factors make the performance analysis of tree coding challenging.
Several fragments within an erroneous candidate codeword may come from a same message.
When $\wv_i(j)$ and $\wv_i(q)$ are both part of an erroneous candidate, with $j < q$, the probability that the parity bits of the later fragment remain statistically discriminating is reduced.
Moreover, although messages must be distinct overall, two different messages can share a same fragment.
Mathematically, $\wv_i \neq \wv_k$ does not imply $\wv_i(j) \neq \wv_k(j)$ for all $j \in [0 : n-1]$.
When fragments across messages are identical, they produce a loss in discrimination power in the corresponding parity bits.
This also introduces correlation in error events across time slots.

As mentioned above, exact analysis may be suitable for performance characterization, but an approximate analysis is desirable for parameter selection.
To facilitate the latter, we make a simplifying assumption about the nature of erroneous paths.
A partial candidate vector
\begin{equation*}
\wv_{i_0}(0) \wv_{i_1}(1) \cdots \wv_{i_j}(j)
\end{equation*}
is termed erroneous if not all indices $i_0, i_1, \ldots, i_j$ come from a same message.
That is, there exists at least one $q < j$ such that $i_q \neq i_j$.
Identifying erroneous vectors is the main purpose of the tree decoder.
We state our simplifying assumption that render analysis more tractable formally below for future reference.

\begin{assumption}[Simplifying Structure] \label{assumption:DistinctFragments}
The collection of messages sent to the destination is such that, for any $i \neq k$,
\begin{equation} \label{equation:DistinctFragments}
\wv_i(j) \neq \wv_k(j)
\end{equation}
for all $j \in [0:n-1]$ where $m_j > 0$.
\end{assumption}

We point out that the condition in \eqref{equation:DistinctFragments} is met with high probability at every stage~$j$ where $m_j$ is large.
Also, we emphasize that it has a greater impact in earlier stages where $m_j$ tends to be large.
Given \eqref{equation:DistinctFragments}, the events that erroneous partial candidate vectors fulfill their parity constraints are pairwise independent.
A stronger notion of independence is obtained when the collection of erroneous partial candidate vectors form a linearly independent set.
For parameters of interest, this is nearly achieved in that the dimensionality of the span of the erroneous partial candidate vectors is typically close to the number of such vectors.
In any case, we conduct a mean analysis below, which circumvents technical issues related to independence.
As we will see shortly, the level of performance predicted under this approximation appears accurate for parameters of interest.
However, we also note that one can create idiosyncratic cases where the predicted performance under Assumption~\ref{assumption:DistinctFragments} is overly optimistic.
We employ $\tilde{L}_j$ to denote the approximate number of erroneous paths that survive stage~$j$, computed under Assumption~\ref{assumption:DistinctFragments}.
This makes the approximate results easy to distinguish from the true $L_j$ introduced in Section~\ref{subsection:TreeCodeAnalysis}.

\begin{proposition} \label{proposition:ExpectedValuesDistinctFragments}
Under Assumption~\ref{assumption:DistinctFragments}, the expected values of $\tilde{L}_j$ can be computed as
\begin{equation} \label{expLi}
\mathbb{E} \left[ \tilde{L}_j \right]
= \sum_{q=1}^{j} \left( K^{j-q}(K-1) \prod_{\ell=q}^j p_{\ell} \right)
\end{equation}
where $p_{\ell} = 2^{-l_{\ell}}$ and $j \in [1:n-1]$.
\end{proposition}
\begin{IEEEproof}
A proof is available in Appendix~\ref{Section:ApproximatePerformanceTreeCodeAppendix}.
\end{IEEEproof}

Having gained a handle on the expected growth in the number of erroneous paths, we are in a position to analyze the complexity of the tree decoder.
We define the computational complexity $C_{\mathrm{tree}}$ of the tree decoder as the total number of parity check constraints that need to be verified.
The expected value of this quantity is captured in Proposition~\ref{proposition:ExpectedParityChecks}.

\begin{proposition} \label{proposition:ExpectedParityChecks}
Under Assumption~\ref{assumption:DistinctFragments}, a closed-form expression for computing the expected computational complexity $\mathbb{E}[C_{\mathrm{tree}}]$ is given by
\begin{equation} \label{expcomp}
\mathbb{E}[C_{\mathrm{tree}}]
= K \left( \sum_{j=1}^{n-1} l_j
+ \sum_{j=1}^{n-2} l_{j+1} \sum_{q=1}^j K^{j-q}(K-1) \prod_{\ell=q}^j p_{\ell} \right)
\end{equation}
where $p_{\ell} = 2^{-l_{\ell}}$.
\end{proposition}
\begin{IEEEproof}
Each non-leaf node that survives stage~$j$ engenders $K$ children.
For every such child, a total of $l_{j+1}$ parity checks must be verified.
Hence, the computational complexity and its expected value can be expressed as
\begin{align}
C_{\mathrm{tree}} &= K l_1 + \sum_{j=1}^{n-2} ( L_j+1 ) K l_{j+1} \\
\mathbb{E}[C_{\mathrm{tree}}] &= K l_1 + \sum_{j=1}^{n-2} \left( \mathbb{E}[L_j] + 1 \right) K l_{j+1}. \label{comput}
\end{align}
Using the approximate expression for $L_j$ found in \eqref{expLi} and substituting it into \eqref{comput}, we get the closed-form expression of \eqref{expcomp}.
\end{IEEEproof}

A related quantity, which we call approximate tree complexity, is the number of tree nodes on which parity check constraints must be verified.
This latter quantity is less accurate, as it does not take into consideration that distinct nodes may have a different number of parity bits.
However, this looser definition facilitates mathematical tractability in certain cases and, as such, we characterize its expected value below.

\begin{corollary} \label{corollary:ExpectedParityNodes}
Under Assumption~\ref{assumption:DistinctFragments}, the expected number of nodes for which parity checks must be computed is given by
\begin{equation} \label{equation:ExpectedParityNodes}
\mathbb{E} \left[ \tilde{C}_{\mathrm{tree}} \right]
= (n-1)K + K\sum_{j=1}^{n-2} \sum_{q=1}^j K^{j-q}(K-1) \prod_{\ell=q}^j p_{\ell}
\end{equation}
where $p_{\ell} = 2^{-l_{\ell}}$.
\end{corollary}
\begin{IEEEproof}
As before, every non-leaf node that survives stage~$j$ will produce $K$ children.
Thus, the number of nodes on which parity checks must be performed, and its expected value are
\begin{align}
\tilde{C}_{\mathrm{tree}} &= (n-1)K + \sum_{j=1}^{n-2} L_j K \\
\mathbb{E} \left[ \tilde{C}_{\mathrm{tree}} \right]
&= (n-1)K + \sum_{j=1}^{n-2} \mathbb{E}[L_j] K . \label{computNodes}
\end{align}
Substituting the approximate expression for $L_j$ found in \eqref{expLi} into \eqref{computNodes}, we get \eqref{equation:ExpectedParityNodes}.
\end{IEEEproof}

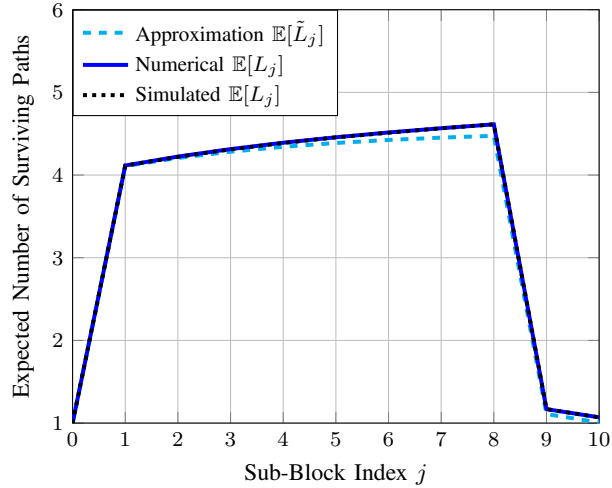
\begin{figure}[ht]
\centerline{
\begin{tikzpicture}

\begin{axis}[%
font=\footnotesize,
width=7cm,
height=5.5cm,
scale only axis,
xmin=0,
xmax=10,
xtick = {0,1,...,10},
xlabel={\small Sub-Block Index $j$},
xmajorgrids,
ymin=1,
ymax=6,
ytick = {1,2,...,7},
ylabel={\small Expected Number of Surviving Paths},
ymajorgrids,
legend style={at={(0,1)},anchor=north west,draw=black,fill=white,legend cell align=left}
]

\addplot [color=cyan,dashed,line width=1.5pt]
  table[row sep=crcr]{
0 1.000\\
1 4.1094\\
2 4.2065\\
3 4.2825\\
4 4.3418\\
5 4.3881\\
6 4.4243\\
7 4.4526\\
8 4.4747\\
9  1.1091\\
10 1.0067\\
};
\addlegendentry{Approximation $\mathbb{E}[\tilde{L}_j]$};

\addplot [color=blue,solid,line width=1.5pt]
  table[row sep=crcr]{
0 1.000\\
1 4.1154\\
2 4.2234\\
3 4.3139\\
4 4.3908\\
5 4.4570\\
6 4.5150\\
7 4.5666\\
8 4.6132\\
9  1.1685\\
10 1.0695\\
};
\addlegendentry{Numerical $\mathbb{E}[L_j]$};

\addplot [color=black,dotted,line width=1.5pt]
  table[row sep=crcr]{
0 1.000\\
1 4.1146\\
2 4.2197\\
3 4.3105\\
4 4.3880\\
5 4.4559\\
6 4.5119\\
7 4.5656\\
8 4.6135\\
9  1.1689\\
10 1.0693\\
};
\addlegendentry{Simulated $\mathbb{E}[L_j]$};

\end{axis}

\end{tikzpicture}%
}
\caption{This plot shows the expected number of surviving paths at various stages during the tree decoding process.
The approximate values derived under simplifying Assumptions~\ref{assumption:DistinctFragments} remain close to the exact values over a range of system parameters.}
\label{fig:numerical_results}
\end{figure}

Figure~\ref{fig:numerical_results} offers supportive evidence to the fact that the expected number of surviving paths predicted under Assumption~\ref{assumption:DistinctFragments} is reasonably accurate for parameters of interest.
This plot shows the exact values computed using Theorem~\ref{Theorem:ExpectedNumberErroneouspaths}, the optimistic approximation of Proposition~\ref{proposition:ExpectedValuesDistinctFragments}, and the results obtained via numerical simulations.
Reported findings are obtained under the following system parameters: $K=200$, $n=11$, $J=15$.
The parity allocation is $(l_1,l_2,l_3,\ldots,l_{10}) = (6,8,8,\ldots,8,13,15)$.
Similar findings can be found for a range of system parameters.

\subsection{Allocating Parity Checks}
\label{parityvector}

The rate of the tree code is intrinsically determined by the number of parity constraints added to the information bits.
Once the rate is fixed, one must decide where to position these parity bits. 
Allocating more parity bits towards the initial stages of the tree will limit expected complexity, as these parity checks will prune the tree early.
However, this comes at the expense of a higher probability of error.
On the other hand, pushing parity bits towards later stages will have the dual effect of reducing the probability of error and increasing computational complexity.
Hence, the location of parity checks should be selected judiciously as a means to tradeoff performance and complexity.

To this end, we formulate the constrained optimization problem of minimizing the expected complexity subject to the probability of failure being less than a carefully chosen threshold $\varepsilon_{\mathrm{tree}}$:
\begin{alignat}{2}
&\min_{(l_1, \ldots, l_{n-1})} & \quad & \mathbb{E}[C_{\mathrm{tree}}] \\
&\text{subject to} && \ptree \leq \varepsilon_{\mathrm{tree}} \\
&&& \sum_{j=1}^{n-1} l_j = M-B \\
&&& l_j \in [0:J] \quad \forall j \in [1:n-1] .
\end{alignat}
The last condition ensures that $l_j$ is a non-negative integer that does not exceed the length of a sub-block.

This optimization problem is non-linear and it is, in general, difficult to solve.
As such, we alter the objective function and relax constraints.
Drawing inspiration from the Markov inequality, we replace $\ptree \le \varepsilon_{\mathrm{tree}}$ with $\mathbb{E}[L_{n-1}] \le \varepsilon_{\mathrm{tree}}$.
We also relax the integer requirements to $(l_1, \dots, l_{n-1}) \in \mathbb{R}^{n-1}_+$.
A first attempt at finding a tractable objective function consists in adopting the expected complexity $\mathbb{E}[C_{\mathrm{tree}}]$ found in \eqref{expcomp}, which is derived under Assumption~\ref{assumption:DistinctFragments}.
Still, the ensuing problem remains challenging because of the non-convex nature of the objective function.
Hence, instead of minimizing the average number of parity check constraints that need to be verified, we minimize the average number of nodes on which parity check evaluations must be performed, as in Corollary~\ref{corollary:ExpectedParityNodes}.
After these modifications, the optimization problem for allocating parity lengths can be posed as follows:
\begin{alignat}{2}
&\min_{(p_1, \ldots, p_{n-1})} & \quad & \mathbb{E} \left[ \tilde{C}_{\mathrm{tree}} \right] \\
&\text{subject to} && \mathbb{E} \left[ \tilde{L}_{n-1} \right] \leq \varepsilon_{\mathrm{tree}} \\
&&& \sum_{j=1}^{n-1} \log_2 \left( \frac{1}{p_j} \right) = M-B \\
&&& p_j \in \left[ \frac{1}{2^J}, 1 \right] \quad \forall j \in [1:n-1] . \label{optprob3}
\end{alignat}
Above, we express the optimization problem in terms of $p_1, \ldots, p_{n-1}$ to emphasize the fact that this problem is a geometric program~\cite{boyd2004convex}.
Given this structure, an optimal solution can be attained using standard convex solvers.
In implementing the proposed scheme, we adopt a parity allocation based on a quantized version of the unconstrained solution to this relaxed optimization.
Specifically, if $p_1^*, \dots, p_{n-1}^*$ is the optimal solution to the aforementioned problem, then we allocate parity check bits as $l_j = \left[ \log_2 \left({1}/{p_j^*}\right) \right]$ for $j \in [1:n-1]$.

Empirically, the solutions obtained using this technique perform well.
For a specific scenario, one could seek to improve performance by running a local search around this suggested operating point using the exact expressions derived in Section~\ref{subsection:TreeCodeAnalysis} as a performance criterion.
Still, in view of the adequate performance achieved using the relaxed optimization, this more involved approach seems unnecessary to capture most of the benefits of tree coding.

\section{Asymptotic Analysis}
\label{section:AsymptoticAnalysis}

As is often the case with problems of inference and communication systems, we seek additional insights by studying the CCS algorithm in the context of large settings.
In particular, we explore the average computational complexity and the number of channel uses needed for CCS to be asymptotically reliable.
Throughout this section, we assume that the number of active users $\Ka$ and the message length $B$ jointly tend to infinity.
We examine two relevant regimes for parameters $\Ka$ and $B$, which will be referred henceforth as the logarithmic regime and the linear regime, respectively:
\begin{enumerate}[(i)]
\item $B=\alpha \log_2\Ka$ for some fixed constant $\alpha>1$,
\item $B=\varrho \Ka$ for some fixed constant $\varrho >0$.
\end{enumerate}
The scaling in the logarithmic regime applies, for instance, to scenarios akin to \cite{zhang2013neighbor} where every active user sends its identity as data.
When the number of active users $\Ka$ scales sub-linearly with the total number of users $\Ktot$, i.e., $\Ka = \Ktot^{{1}/{\alpha}}$ for $\alpha>1$, each active user needs to send $B=\log_2 \Ktot = \alpha \log_2\Ka$ bits to uniquely identify themselves.
On the other hand, the linear regime corresponds to situations where the payload is proportional to the number of active users.
For ease of exposition, we focus on the special case $B = \Ka$ in regime~(ii).
Nevertheless, the results we derived are valid for any linear factor $\varrho > 0$.
We remark that these regimes are selected, partly, to enable the comparison of our algorithm with prior art.

From Section~\ref{parityvector}, we already know that the positioning of parity bits can act as a mechanism to tradeoff performance and computational complexity.
Again, allocating parity bits towards the later stages of CCS improves performance at the expense of complexity; whereas placing check bits in early stages has the reverse effect.
In view of this principle, we consider two distinct parity bit allocation strategies for these asymptotic regimes, one for each of the extreme behaviors outlined above.
Specifically, we consider a scenario where all the parity bits are located towards the end of the codewords and we study the situation where the parity bits are equally distributed among all the non-root sub-blocks.
In this context, we provide bounds for the number of parity bits, the number of channel uses, and the average computational complexity needed to achieve asymptotic reliable communication ($P_e \rightarrow 0$ as $\Ka \rightarrow \infty$).
This is accomplished for the trailing and uniform parity allocations in both regime~(i) and regime~(ii), leading to four distinct cases.

For the two parity profiles mentioned above, we show that the number of channel uses (i.e., the sample complexity in the compressed sensing literature) scales in an order-optimal fashion with respect to $\Ka$ and $B$.
The average computational complexity scales sub-linearly with the dimension of the original compressed sensing problem, $2^B$, in both cases.
We also demonstrate that, although the number of channel uses is order optimal in both scenarios, the pre-constant in the Big-$\mathcal{O}$ terms is larger when parity bits are uniformly distributed across fragments than when they are placed at the end.
This advantage in terms of channel uses in the latter case comes at the expense of a higher average computational complexity.
These findings are stated formally in Theorem~\ref{theorem:AsymptoticBehaviorParityTrailing} and Theorem~\ref{theorem:AsymptoticBehaviorParityUniform}.

\begin{remark}
\label{rmk:convCS}
It may be useful to reiterate that the application of conventional CS solvers (e.g., LASSO, OMP) to the global uncoordinated MAC problem of dimension $2^B$ entails a computational complexity that scales polynomially (at least linearly) with $2^B$.
Specifically, the complexity for this approach is on the order of $\mathcal{O}(2^{\gamma B})$ for some $\gamma \ge 1$.
Moreover, the number of channel uses needed for this scheme to work is equivalent to the optimal number of samples needed to solve a $\Ka$-sparse problem of dimension $2^B$, which is $\mathcal{O}(\Ka \log_2 2^B)$.
Thus, the optimal number of samples scales as $\mathcal{O}(\Ka \log_2 \Ka)$ in regime~(i), and $\mathcal{O}(\Ka^2)$ in regime~(ii).
While the conventional approach is impractical for unsourced multiple access, these theoretical scaling behaviors serve as benchmarks for the asymptotic analysis of CCS.
\end{remark}

Proceeding forward, the various parameters employed in our asymptotic analysis are chosen as follows.
The list size output by the CS algorithm at the sub-block level is set to $K=\Ka$.
The length of each coded sub-block is selected to be $J = c_1 \log_2 \Ka$ for fixed constant $c_1 > 1$ (with $c_1 < \alpha$ in regime (i)).
We note that $c_1$ can be chosen arbitrarily close to one from the right. With this scaling, the sparsity index admits a sub-linear scaling with the dimension of the compressed sensing problems at the sub-block level;
that is, ${\Ka}/{2^J} = \Ka^{-(c_1-1)} \rightarrow 0$ as $\Ka \rightarrow \infty$.
We take the outer code to have a fixed rate, with $n J=\Theta(B)$.
Hence, the number of sub-blocks $n$ is of the form $n=c_2$ and $n = c_2 \frac{\Ka}{\log_2 \Ka}$ for regimes (i) and (ii), respectively.
We stress that constant $c_2$ depends on $c_1$ and the total number of parity bits.
This information is summarized in Table~\ref{table:AsymptoticParameters}.
\begin{table}[tbh]
\centerline{
\begin{tabular}{||c|l|l||}
\hline
& Regime (i) & Regime (ii) \tabularnewline
\hline
$B$ & $\alpha \log_2 \Ka$ & $\Ka$ \tabularnewline
\hline
$J$ & $c_1 \log_2 \Ka$ & $c_1 \log_2 \Ka$ \tabularnewline
\hline
$n$ & $c_2$ & $c_2 \frac{\Ka}{\log_2 \Ka}$ \tabularnewline
\hline
$P$ & $(c_1 c_2 - \alpha) \log_2 \Ka$ & $(c_1 c_2 - 1) \Ka$ \tabularnewline
\hline
\end{tabular}
}
\caption{Parameters of scaling regimes.}
\label{table:AsymptoticParameters}
\end{table}

From \eqref{petotal}, we get an upper bound on the overall error probability,
\begin{equation} \label{peub}
P_e \le n \pcs + \ptree.
\end{equation}
Thence, asymptotic reliability is guaranteed whenever $n \pcs$ and $\ptree$ both decay to zero as $\Ka \rightarrow \infty$.
In the upcoming analysis, we show that the former condition is obtained by allocating the required number of channel uses needed to solve a $\Ka$ out of $2^J$ sparse problem for the CS component of the scheme. 
Once this is accomplished, it suffices to identify conditions under which $\ptree \rightarrow 0$ asymptotically to get the desired results.
This proof strategy is applied to the two scenarios of interest below.

\subsection[i]{Parity Bits Located at the End of Codewords}
\label{subsection:TailParityBits}

Let $C_{\mathrm{cs}}$, $\mathbb{E}[C_{\mathrm{tree}}]$, $\mathbb{E}[C]$, and $N$ denote the computational complexity of solving the $n$ CS sub-problems, the average tree decoding complexity, the average overall decoding complexity, and the total number of channel uses, respectively. Then, the following theorem holds.

\begin{theorem} \label{theorem:AsymptoticBehaviorParityTrailing}
Consider a coded compressed sensing scenario, without successive cancellation, where all the parity bits are located in the trailing sub-blocks.
The CCS algorithm can decode a user with vanishing error probability $P_e \rightarrow 0$, asymptotically in $\Ka$, whenever the total number of parity bits is $P=(n + \delta - 1) \log_2 \Ka$ for some fixed constant $\delta>0$.
Complexity orders for this allocation strategy appear in Table~\ref{table:AsymptoticBehaviorParityTrailing}.
\end{theorem}
\begin{table}[h!]
\centerline{
\begin{tabular}{||c|l|l||}
\hline
& Regime (i) & Regime (ii) \tabularnewline
\hline
$N$ & $\mathcal{O}(\Ka \log_2 \Ka)$ & $\mathcal{O} \left( \Ka^2 \right)$ \tabularnewline
\hline
$C_{\mathrm{cs}}$ & $\mathcal{O} \left( \Ka^{c_1} \right)$ & $\mathcal{O}\left(\frac{\Ka^{c_1+1}}{\log_2\Ka}\right)$ \tabularnewline
\hline
$\mathbb{E}[C_{\mathrm{tree}}]$ & $\mathcal{O}\left(\Ka^{\alpha/c_1} \log_2\Ka\right)$ & $\mathcal{O}\left(2^{\Ka/c_1} \log_2\Ka\right)$ \tabularnewline
\hline
$\mathbb{E}[C]$ & 
$\mathcal{O}\left(\max\{\Ka^{c_1},\Ka^{\alpha/c_1}\log_2\Ka\}\right)$ &
$\mathcal{O}\left(2^{\Ka/c_1} \log_2\Ka\right)$ \tabularnewline
\hline
\end{tabular}
}
\caption{Order performance of the CCS algorithm when all the parity bits are located towards the end of codewords}
\label{table:AsymptoticBehaviorParityTrailing}
\end{table}
\begin{IEEEproof}
A proof is available in Appendix~\ref{section:AsymptoticPerformanceAppendix}.
\end{IEEEproof}

It can be seen from Theorem~\ref{theorem:AsymptoticBehaviorParityTrailing} and Table~\ref{table:AsymptoticBehaviorParityTrailing} that, in regime (ii), even thought the average complexity scales sub-linearly in $2^{\Ka}$, it remains sizable as the gain in complexity only appears in the exponent, with $2^{\Ka/c_1}$.
The decoding complexity for this scenario is high because the tree is not pruned during the initial stages of the process.
The number of parity bits allocated to each node is zero for a constant fraction of the tree depth and, consequently, the tree grows exponentially until the first parity bit appears.
To avoid such an exponential growth, the tree should be pruned during the initial stages via the early placement of parity bits.

Throughout the paper, we use order notation to express the dependence of the number of channel uses and decoding complexity of the compressed sensing part on the number of active users.
We omit the pre-constant which depends on the signal to noise ratio.

\subsection{Parity Bits Equally Distributed among Sub-Blocks}

Again, we use $C_{\mathrm{cs}}$, $\mathbb{E}[C_{\mathrm{tree}}]$, $\mathbb{E}[C]$, and $N$ to denote the computational complexity of solving the $n$ CS sub-problems, the average tree decoding complexity, the average overall decoding complexity, and the total number of channel uses, respectively.
Recall that having an equal number of parity-check bits per sub-block, except for the root block, limits complexity at the expense of performance.
We present our findings below in the form of a theorem.

%

\begin{theorem} \label{theorem:AsymptoticBehaviorParityUniform}
Consider a coded compressed sensing scenario, without successive cancellation, where parity bits are assigned uniformly across sub-blocks, except for the root fragment.
Suppose that the non-negative parameters $\alpha$, $c_1$, and $c_2$ in Table~\ref{table:AsymptoticParameters} are chosen such that
\begin{equation} \label{equation:AsymptoticBehaviorParityUniformRegime1}
2 < \frac{\alpha - c_1}{c_2 - 1} < c_1 - 1
\end{equation}
in regime~(i), and
\begin{equation} \label{equation:AsymptoticBehaviorParityUniformRegime2}
3 < \frac{1}{c_2} < c_1 - 1
\end{equation}
in regime~(ii).
Then, the CCS algorithm can decode users with vanishing error probability, $P_e \rightarrow 0$, asymptotically in $\Ka$.
Furthermore, the corresponding complexity orders for this allocation strategy appear in Table~\ref{table:AsymptoticBehaviorParityUniform}.
\end{theorem}
\begin{table}[h!]
\centerline{
\begin{tabular}{||c|l|l||}
\hline
& Regime (i) & Regime (ii) \tabularnewline
\hline
$N$ & $\mathcal{O}(\Ka \log_2 \Ka)$ & $\mathcal{O} \left( \Ka^2 \right)$ \tabularnewline
\hline
$C_{\mathrm{cs}}$ & $\mathcal{O} \left( \Ka^{c_1} \right)$ & $\mathcal{O}\left(\frac{\Ka^{c_1+1}}{\log_2\Ka}\right)$ \tabularnewline
\hline
$\mathbb{E}[C_{\mathrm{tree}}]$ & $\mathcal{O}(\Ka\log_2 \Ka)$ & $\mathcal{O} \left( \Ka^2 \right)$ \tabularnewline
\hline
$\mathbb{E}[C]$ & $\mathcal{O} \left( \Ka^{c_1} \right)$ & $\mathcal{O}\left(\frac{\Ka^{c_1+1}}{\log_2\Ka}\right)$ \tabularnewline
\hline
\end{tabular}
}
\caption{Order performance of the CCS algorithm when parity bits are assigned uniformly across sub-blocks.}
\label{table:AsymptoticBehaviorParityUniform}
\end{table}
\begin{IEEEproof}
A proof is available in Appendix~\ref{section:AsymptoticPerformanceAppendix}.
\end{IEEEproof}

The ratio in \eqref{equation:AsymptoticBehaviorParityUniformRegime1} is related to the number of information bits per non-root sub-block, which is equal to $m = \frac{\alpha-c_1}{c_2-1} \log_2 \Ka$ in regime~(i).
These bounds ensure that all the information bits are accounted for, while also leaving enough slack for parity bits to grow rapidly.
As an example, the requirements for regime~(i) are met by $\alpha = 5.25$, $c_1 = 4$, $c_2 = 1.5$.
A similar statement applies to the ratio in \eqref{equation:AsymptoticBehaviorParityUniformRegime2}.
The number of information bits per non-root sub-block in regime~(ii) is $m = \frac{1}{c_2} \log_2 \Ka - o(1)$.
In this case, ${1}/{c_2}$ should be large enough to make room for all the information bits, while also leaving enough space for parity bits.
We note that, for instance, parameters $c_1 = 6$, $c_2 = 0.25$ are admissible in regime~(ii).
Finally, in both cases, the lower bound guarantees that the conditions of Assumption~\ref{assumption:DistinctFragments} are met with probability approaching one as $\Ka \rightarrow \infty$.

\subsection{Comparison of Parity Allocations}

It is interesting to note that the number of channel uses for reliable message recovery is of the same order when all the parity bits are located towards the end of the codewords and when the parity bits are equally distributed among all the non-root sub-blocks.
This statement is valid for both regime~(i) and regime~(ii).
The complexity order on the other hand is more severe for the scenario where all the parity bits are at the end.
This suggests that, for large systems, it is beneficial to distribute parity bits across sub-blocks.
The per step, multiplicative growth of the tree is $\Ka$.
Reliable performance is achieved by bringing the expected number of surviving erroneous candidates very close to zero at every step.
In the large scale setting, the tradeoff between performance and complexity appears to be favoring aggressive pruning early on, with parity bits being distributed equally among non-root sub-blocks.
This parity allocation scheme is a particular case of what we observe for finite systems through numerical simulations.
In the latter case, parity bits are assigned sparsely until the decoding tree reaches a certain expected girth, and they are assigned regularly thereafter.
These findings are reported below.

\section{Simulation Results}
\label{sec:simresults}

In this section, we study the performance of the proposed scheme for realistic parameters attuned to practical applications.
We also provide a comparison with existing results found in the literature.
We consider a system with $\Ka \in [25:300]$ active users, each having $B=75$~bits of information to transmit.
These bits are divided into $n=11$ sub-blocks, each of length $J$; the specific value for $J$ depends on $\Ka$.
Design attributes are summarized in Table~\ref{tableofparams}.
In a manner akin to \cite{vem2017user}, we adopt sensing matrices that are constructed based on BCH codes.
These matrices are employed for signaling in the compressed sensing sub-problems.
In particular, we select a subset $\mathcal{C}^0$ of codewords of size $|\mathcal{C}^0|=2^J$ from the (2047,23) BCH codebook $\mathcal{C}$.
This subset is created to possess the following properties:
$\underline{0} \in \mathcal{C}^0$;
$\cv \in \mathcal{C}^0$ implies $\underline{1} \oplus \cv \in \mathcal{C} \setminus \mathcal{C}^0$;
if $\cv_1, \cv_2 \in \mathcal{C}^0$ then $\cv_1 \oplus \cv_2 \in \mathcal{C}^0$.
Above, $\underline{0}$ and $\underline{1}$ are the all-zero and all-one vectors, respectively.
The $\oplus$ operator represents binary addition and, as such, 
$\underline{1} \oplus \cv$ denotes the ones' complement of $\cv$.
We then choose sensing matrix $\mathbf{A}$ as $\mathbf{A} = [\av_0, \av_1, \ldots, \av_{2^J-1}]$ with dimension $2047 \times 2^J$, where $\av_j=\sqrt{E_{\mathrm{s}}}(2\cv_j-1)$ and $\cv_j \in \mathcal{C}^0$ for $j \in [0:2^J-1]$.
With this construction, the total number of channel uses is given by $N = 11 \times 2047 = 22,517$.
The target error probability for the system is fixed at $\varepsilon=0.05$.

\begin{table}[tbh]
\centerline{
\begin{tabular}{||c|| c| c| c| c| c| c| c| c| c| c| c| c||}
\hline
$\Ka$ & 25 & 50 & 75 & 100 & 125 & 150 \tabularnewline
\hline
$J$ & 14 & 14 & 14 & 14 & 14 & 15 \tabularnewline
\hline
$\varepsilon_{\mathrm{tree}}$ & 0.0025 & 0.0045 & 0.006 & 0.01 & 0.0125 & 0.0055 \tabularnewline
\hline
\hline
$\Ka$ & 175 & 200 & 225 & 250 & 275 & 300 \tabularnewline
\hline
$J$ & 15 & 15 & 15 & 15 & 15 & 15 \tabularnewline
\hline
$\varepsilon_{\mathrm{tree}}$ & 0.0065 & 0.007 & 0.008 & 0.01 & 0.0125 & 0.0175 \tabularnewline
\hline
\end{tabular}}
\caption{Summary of the operational parameters used in simulating the CCS algorithm.}
\label{tableofparams}
\end{table}

We set list size $K$ for the NNLS CS sub-problems to $K= \Ka + 10$.
For selected values $\Ka \in [25:300]$, we solve the optimization problem~\eqref{optprob3} with CVX~\cite{grant2008cvx}.
The solution then dictates the assignment of parity bits, as described in Section~\ref{parityvector}.
Prescribed values for $\varepsilon_{\mathrm{tree}}$ appear in Table~\ref{tableofparams} as a function of $\Ka$.
Parameters $B$ and $N$ are chosen such that the rate $\frac{B}{N}=\frac{75}{22,517}$ is approximately equal to the rate resulting from the choice of parameters $B=100$ and $N=30,000$ in \cite{ordentlich2017low,vem2017user}.
This enables a fair comparison between the schemes contained therein and the present approach.
We emphasize that the choice of $B$ and $N$ for our simulations is motivated by the existence of good compressed sensing matrices based on BCH codes.
When these parameters are proportionally scaled up, the operation of the system improves, as the finite block length effects become less severe.

\begin{table}[htb]
\centering
\begin{tabular}{|c|c|c|}
\hline
$\varepsilon_{\mathrm{tree}}$ & $\mathbb{E} [ \tilde{C}_{\mathrm{tree}} ]$ & Parity Length Vector \\ [0.5ex]
\hline\hline
0.006 & Infeasible & Infeasible \tabularnewline
\hline
0.0061930 & 3.2357$\times$ \num{e11} & [0 ,0, 0, 0, 15, 15, 15, 15, 15, 15] \tabularnewline
\hline
0.0061931 & 3357300 & [0, 3, 8, 8, 8, 8, 10, 15, 15, 15]  \tabularnewline
\hline
0.0061932 & 1737000 & [0, 4, 8, 8, 8, 8, 9, 15, 15, 15] \tabularnewline
\hline
0.0061933 & 926990 & [0, 5, 8, 8, 8, 8, 8, 15, 15, 15] \tabularnewline
\hline
0.0061935  & 467060 & [1, 8, 8, 8, 8, 8, 8, 11, 15, 15]  \tabularnewline
\hline
0.0062 & 79634 & [1, 8, 8, 8, 8, 8, 8, 11, 15, 15] \tabularnewline
\hline
0.007 & 7358 & [6, 8, 8, 8, 8, 8, 8, 8, 13, 15] \tabularnewline
\hline
0.008 & 6153 & [7,  8, 8, 8, 8, 8, 8, 8, 12, 15] \tabularnewline
\hline
0.02 & 5023 & [6, 8, 8, 9, 9, 9, 9, 9, 9, 14] \tabularnewline
\hline
0.04 & 4158 & [7, 8, 8, 9, 9, 9, 9, 9, 9, 13] \tabularnewline
\hline
0.6378 & 3066 & [9, 9, 9, 9, 9, 9, 9, 9, 9, 9] \tabularnewline
\hline
\end{tabular}
\caption{Results from the optimization framework developed in the paper. Error probability is least when parity check bits are pushed to end and average computational complexity is least when equal parity-check bits are allocated per sub-block.}
\label{table:optimization demonstration}
\end{table}

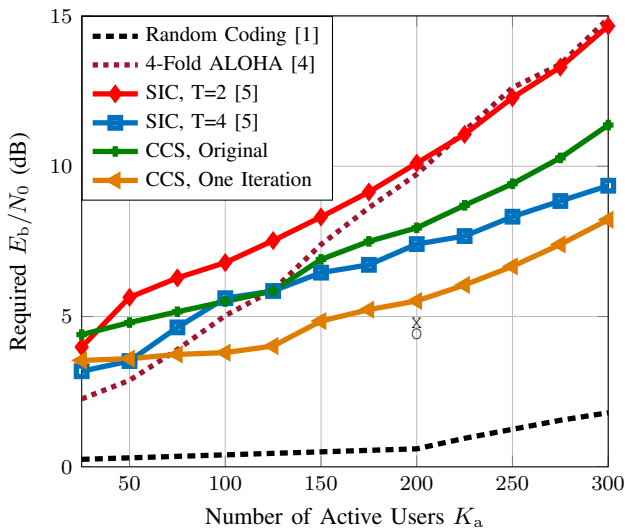
\begin{figure}[!ht]
\centerline{
\begin{tikzpicture}

\begin{axis}[%
font=\footnotesize,
width=7cm,
height=6cm,
scale only axis,
xmin=25,
xmax=300,
xtick = {50,100,...,300},
xlabel={\small Number of Active Users $\Ka$},
xmajorgrids,
ymin=0,
ymax=15,
ytick = {0,5,...,15},
ylabel={\small Required $E_{\mathrm{b}}/N_0$ (dB)},
ymajorgrids,
legend style={at={(0,1)},anchor=north west,draw=black,fill=white,legend cell align=left}
]

\addplot [color=black,densely dashed,line width=2.0pt]
  table[row sep=crcr]{
 25	0.25\\
50	0.3\\
75	0.35\\
100	0.4\\
125	0.45\\
150	0.5\\
175	0.55\\
200	0.6\\
225	0.95\\
250	1.25\\
275	1.55\\
300	1.8\\
};
\addlegendentry{Random Coding~\cite{polyanskiy2017perspective}};

\addplot [color=mycolor1,dotted,line width=2.0pt]
  table[row sep=crcr]{25	2.26\\
50	2.88\\
75	3.9\\
100	5.03\\
125	5.8798\\
150	7.3954\\
175	8.6199\\
200	9.7328\\
225	11.1761\\
250	12.6127\\
275	13.3907\\
300	14.9116\\
};
\addlegendentry{4-Fold ALOHA~\cite{ordentlich2017low}};
\addplot [color=red,solid,line width=2.0pt,mark=diamond,mark options={solid}]
  table[row sep=crcr]{25	3.9849\\
50	5.6396\\
75	6.2855\\
100	6.7952\\
125	7.5262\\
150	8.3122\\
175	9.1418\\
200	10.103\\
225	11.062\\
250	12.279\\
275	13.296\\
300	14.6648\\
};
\addlegendentry{SIC, T=2~\cite{vem2017user}};
\addplot [color=mycolor2,solid,line width=2.0pt,mark=square,mark options={solid}]
  table[row sep=crcr]{25	3.18\\
50	3.52\\
75	4.64\\
100	5.61\\
125	5.85\\
150	6.46\\
175	6.72\\
200	7.41\\
225	7.6772\\
250	8.3217\\
275	8.8428\\
300	9.352\\
};
\addlegendentry{SIC, T=4~\cite{vem2017user}};
\addplot [color=mycolor3,solid,line width=2.0pt,mark=+,mark options={solid}]
  table[row sep=crcr]{25	4.4\\
50	4.8\\
75	5.16\\
100	5.5\\
125	5.85\\
150	6.9\\
175	7.5\\
200	7.95\\
225	8.7\\
250	9.4167\\
275	10.2773\\
300	11.37\\
};
\addlegendentry{CCS, Original};
\addplot [color=mycolor4,solid,line width=2.0pt,mark=triangle,mark options={solid,rotate=90}]
  table[row sep=crcr]{25	3.54\\
50	3.6\\
75	3.74\\
100	3.8\\
125	4.02\\
150	4.85\\
175	5.23\\
200	5.52\\
225	6.05\\
250	6.67\\
275	7.4\\
300	8.22\\
};
\addlegendentry{CCS, One Iteration};

\node[] at (axis cs: 200,4.8) {\scriptsize \texttt{X}};
\node[] at (axis cs: 200,4.43) {\scriptsize \texttt{O}};

%

\end{axis}

\end{tikzpicture}%
}
\caption{This graph plots the minimum $E_{\mathrm{b}}/N_0$ required to achieve $P_e \le 0.05$ as a function of $\Ka$, the number of active users.
The proposed scheme and its enhanced version, with one iteration, have their own curves.
The more demanding versions with $2$ and $3$ iterations are computed for a single operating point;
they are represented by \texttt{x} and \texttt{o}, respectively.}
\label{fig:sim_results}
\end{figure}

Table~\ref{table:optimization demonstration} captures the tradeoff between performance and average computational cost as a function of the parity assignment vector for $\Ka=200$.
These results are consistent with our treatment of the CCS algorithm.
The early allocation of parity bits decreases complexity at the expense of performance.
Suitable allocations tend to have a ramp-up phase where bits are allocated sparsely while the expected girth of the decoding tree grows, followed by a stable phase where the girth of the tree is maintained by a stead allocation of parity bits, and finally there is a heavy pruning phase where sub-blocks are filled almost exclusively with parity bits.
In Fig.~\ref{fig:sim_results}, the $E_{\mathrm{b}}/N_0$ required to achieve a target error probability of 0.05 is plotted as a function of $\Ka$ for various schemes.
The bottom most curve corresponds to Polyanskiy's acheivability bound \cite{polyanskiy2017perspective} on the performance of a finite block length code for the communication problem at hand.
The curves labelled $T=2$, $T=4$, and 4-fold ALOHA, which correspond to the performance curves in \cite{vem2017user} and \cite{ordentlich2017low}, assume the existance of a code for the $T$-user MAC channel which achieves the bound in~\cite{polyanskiy2017perspective}. 
It can be seen from Fig.~\ref{fig:sim_results} that our proposed approach with just one extended round of iteration outperforms previously published schemes for $\Ka \in [75:300]$.

\section{Conclusion}
\label{sec:conc}
In this article, we propose a novel compressed sensing architecture for the uncoordinated massive random access problem.
This scheme, called coded compressed sensing, innovatively combines techniques from compressed sensing and forward error correction to yield an algorithm with a manageable computational complexity.
This conceptual structure, when applied to unsourced multiple access, offers significant performance improvements compared to older, low-complexity algorithms.
After the publication of the conference version of this work \cite{amalladinne2018couple}, the tree code framework has been applied to design novel random access schemes in \cite{Giuseppe} and \cite{calderbank2018chirrup}. In \cite{Giuseppe}, a tree code is used in conjunction with a compressed sensing scheme based on sparse regression codes and approximate message passing. This scheme shows performance improvement over the proposed scheme in certain regimes. In \cite{calderbank2018chirrup}, the tree code framework is used in conjunction with a Hadamard matrix based compressing scheme which permits very low complexity decoding even for large number of users.

The framework can be leveraged with a variety of sensing matrices, beyond those featured in this article.
In fact, CCS can serve as a conceptual blueprint for the design of algorithms in the context of very high-dimensional problems.
In this sense, this technical approach may be relevant to many compressed sensing applications where the unknown sparse vector is very large.
This may include, for instance, IoT neighbor discovery and heavy hitters detection in wired networks \cite{chen2017sparse,gilbert2007one}.
Still, the treatment of such problems would be a significant departure from the primary focus of this article and, as such, the aforementioned opportunities are left as potential avenues for future research.

\appendices



\section{Performance of Tree Code}
\label{Section:PerformanceTreeCodeAppendix}

In this section, we characterize the performance of tree codes, as they pertain to coded compressive sensing.
To do so, we briefly review relevant results related to random linear codes.
We then extend these results to the desired setting.
As we will see shortly, the exact analysis of the probability of failure under tree decoding, although tractable, is cumbersome, especially for deep trees.
Nevertheless, a precise expression can be derived for performance characterization, and it is employed to compute probabilities of error for the parameters of interest.

\subsection{Random Linear Codes}
\label{sectionRandomLInearCodes}

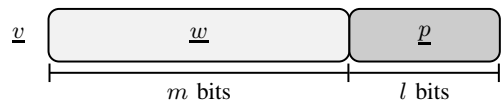
\begin{figure}[htb]
  \centering
  \begin{tikzpicture}[
  font=\small, >=stealth', line width=0.75pt,
  infobits/.style={rectangle, minimum height=7mm, minimum width=40mm, draw=black, fill=gray!10, rounded corners},
  paritybits/.style={rectangle, minimum height=7mm, minimum width=20mm, draw=black, fill=gray!40, rounded corners}
]

\node[infobits] (vb) at (2,0) {$\wv$};
\node[paritybits] (vp) at (5,0) {$\pv$};
\node (bits) at (-0.4,0) {$\vv$};
\draw[|-|] (0,-0.5) to node[midway,below] {$m$ bits} (4,-0.5);
\draw[-|] (4,-0.5) to node[midway,below] {$l$ bits} (6,-0.5);

\end{tikzpicture}
  \caption{Message $\vv$ is obtained through systematic encoding.
  It is composed of $m$ information bits and $l$ parity-check bits.}
  \label{figure:subvector}
\end{figure}

To begin, we inspect the type of (random) linear codes we are interested in.
Consider a binary information vector $\wv$ of length $m$.
This vector is systematically encoded using parity generator matrix $G$ of size $m \times l$.
The resulting codeword, $\vv = \wv \pv$, is then obtained by taking message $\wv$ and appending parity vector $\pv$ to it.
Specifically, the parity bits are generated via linear equation
\begin{equation*}
\pv = \wv G ,
\end{equation*}
where operations are taken over binary field $\mathbb{F}_2$.
Thus, $\vv$ has length $m + l$, as depicted in Fig.~\ref{figure:subvector}.

Suppose that an alternate information vector $\wv_{\mathrm{r}}$ is selected at random from $\{ 0, 1 \}^m$.
This vector is encoded using the same generator matrix $G$, yielding concatenated vector $\vv_{\mathrm{r}} = \wv_{\mathrm{r}} \pv_{\mathrm{r}}$ with $\pv_{\mathrm{r}} = \wv_{\mathrm{r}} G$.
We wish to compute the probability that $\vv$ and $\vv_{\mathrm{r}}$ share a same parity sub-component, i.e., $\pv = \pv_{\mathrm{r}}$.

\begin{lemma}
Fix information vector $\wv$ and parity generating matrix $G$.
The probability that a randomly selected information vector $\wv_{\mathrm{r}} \in \{ 0, 1 \}^m$ produces the same parity sub-component as $\wv$ under $G$ is given by
\begin{equation*}
\Pr (\pv = \pv_{\mathrm{r}}) = 2^{- \operatorname{rank} (G)} .
\end{equation*}
\end{lemma}
\begin{IEEEproof}
Information vector $\wv_{\mathrm{r}}$ is drawn at random, uniformly over $\{ 0, 1 \}^m$.
Codeword $\vv_{\mathrm{r}}$ is then created by computing parity bits $\pv_{\mathrm{r}} = \wv_{\mathrm{r}} G$ and, subsequently, appending these bits to the information vector.
This process yields codeword $\vv_{\mathrm{r}} = \wv_{\mathrm{r}} \pv_{\mathrm{r}}$.
The event $\{ \pv = \pv_{\mathrm{r}} \}$ can therefore be expressed as
\begin{equation*}
\begin{split}
\{ \pv = \pv_{\mathrm{r}} \}
&= \{ \wv G = \wv_{\mathrm{r}} G \}
= \{ (\wv + \wv_{\mathrm{r}}) G = \underline{0} \} \\
&= \{ \wv + \wv_{\mathrm{r}} \in \operatorname{nullspace}(G) \} .
\end{split}
\end{equation*}
Note that subtraction and addition are equivalent under operations in $\mathbb{F}_2$.
Moreover, vector $\wv + \wv_{\mathrm{r}}$ is uniformly distributed over $\{ 0, 1 \}^m$.
Since the number of vectors in the nullspace of $G$ is
$2^{\operatorname{nullity}(G)} = 2^{m - \operatorname{rank} (G)}$,
it follows that
\begin{equation*}
\Pr ( \pv = \pv_{\mathrm{r}} )
= \frac{2^{m - \operatorname{rank} (G)}}{2^m}
= \frac{1}{2^{\operatorname{rank}(G)}} .
\end{equation*}
This completes the proof.
\end{IEEEproof}

The rank of parity generator matrix $G$ is of fundamental importance in assessing performance.
In many scenarios, including our current treatment of unsourced multiple access, matrix $G$ is drawn at random from a large ensemble.
In such cases, it becomes appropriate to compute the probability that an erroneous vector fulfills the parity constraints associated with $\wv$.

\begin{lemma} \label{lemma:RandomGenerator}
Fix erroneous vector $\wv_{\mathrm{e}} \neq \wv$.
Let parity generator matrix $G$ be a Rademacher matrix of size $m \times l$.
That is, the entries in $G$ are drawn at random from a uniform Bernoulli distribution, independently of one another.
Under such circumstances, the probability of event $\{ \pv = \pv_{\mathrm{e}} \}$ is given by
\begin{equation*}
\Pr (\pv = \pv_{\mathrm{e}}) = 2^{-l} .
\end{equation*}
\end{lemma}
\begin{IEEEproof}
The event $\{ \pv = \pv_{\mathrm{e}} \}$ is equivalent to $\{ (\wv + \wv_{\mathrm{e}})G = \underline{0} \}$.
Since $\wv \neq \wv_{\mathrm{e}}$, there exists at least one pair of vector entries, say at location~$q$, such that $\wv(q) + \wv_{\mathrm{e}}(q) = 1$.
Then,
\begin{equation*}
(\wv + \wv_{\mathrm{e}})G
= G[q,:] + \sum_{\ell \neq q} (\wv(\ell) + \wv_{\mathrm{e}}(\ell))G[\ell,:] ,
\end{equation*}
where $G[q,:]$ denotes the $q$th row of $G$.
By constructions, $G[q,:]$ is a sequence of independent, uniform Bernoulli trials.
Therefore, $(\wv + \wv_{\mathrm{e}})G$ also forms a sequence of independent, uniform Bernoulli bits and, hence, the probability that this sequence is the all-zero sequence is equal to $2^{-l}$.
\end{IEEEproof}

Conceptually, the results presented in this section form the cornerstones of our upcoming analysis.
However, when codewords are fragmented and redundancy is added at different stages, assessing performance becomes more complicated.
We initiate our treatment of this more elaborate scenario next.

\subsection{Fragmented Codewords}
\label{subsection:FragmentedCodewrds}

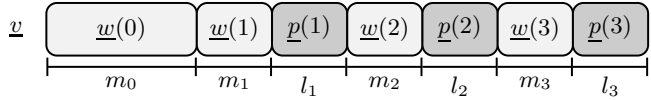
\begin{figure}[htb]
  \centering
  \begin{tikzpicture}[
  font=\small, >=stealth', line width=0.75pt,
  infobits/.style={rectangle, minimum height=7mm, minimum width=10mm, draw=black, fill=gray!10, rounded corners},
  paritybits/.style={rectangle, minimum height=7mm, minimum width=10mm, draw=black, fill=gray!40, rounded corners}
]

\node[infobits, minimum width=20mm] (vb0) at (1,0) {$\wv(0)$};
\node[infobits] (vb1) at (2.5,0) {$\wv(1)$};
\node[paritybits] (vp1) at (3.5,0) {$\pv(1)$};
\node[infobits] (vb2) at (4.5,0) {$\wv(2)$};
\node[paritybits] (vp2) at (5.5,0) {$\pv(2)$};
\node[infobits] (vb3) at (6.5,0) {$\wv(3)$};
\node[paritybits] (vp3) at (7.5,0) {$\pv(3)$};
\node (bits) at (-0.4,0) {$\vv$};
\draw[|-|] (0,-0.5) to node[midway,below] {$m_0$} (2,-0.5);
\draw[-|] (2,-0.5) to node[midway,below] {$m_1$} (3,-0.5);
\draw[-|] (3,-0.5) to node[midway,below] {$l_1$} (4,-0.5);
\draw[-|] (4,-0.5) to node[midway,below] {$m_2$} (5,-0.5);
\draw[-|] (5,-0.5) to node[midway,below] {$l_2$} (6,-0.5);
\draw[-|] (6,-0.5) to node[midway,below] {$m_3$} (7,-0.5);
\draw[-|] (7,-0.5) to node[midway,below] {$l_3$} (8,-0.5);

\end{tikzpicture}
  \caption{The fragmented nature of tree coding leads to peculiarities in the performance analysis of this scheme.
  This situation arises, partly, from the fact that two different information sequences can share identical fragments.}
  \label{figure:subvector3}
\end{figure}

An important distinction that arises in coded compressive sensing when compared to standard random linear codes stems from the fragmented structure of tree codes depicted in Fig.~\ref{figure:subvector3}.
A second related factor ensues from the fact that erroneous candidates are formed by piecing together fragments from valid codewords.
Typically, when studying error events under random linear coding, the starting point is an erroneous message of the form $\wv_{\mathrm{e}} \neq \wv$, as in Lemma~\ref{lemma:RandomGenerator}.
However, the situation is more intricate for tree coding and meticulous accounting must be performed.

There are essentially three confounding factors in the analysis of erroneous messages in tree coding.
First, several fragments within an erroneous candidate codeword may come from a same message.
Such occurrences can significantly reduce the discriminating power of parity bits.
Second, two different messages may have overlapping information fragments, as they only need to differ at one location overall.
Thence, when comparing a valid codeword to an erroneous candidate, the two messages are necessarily distinct, yet some of their information fragments may be identical.
Third, the loss of discriminating power from parity bits is correlated across fragments in certain cases, which exacerbates the probability of detection failure.
We elaborate on these idiosyncrasies below.

Suppose $\wv = \wv(0) \wv(1) \cdots \wv(n-1)$ is a partitioned information vector.
Let
\begin{equation} \label{equation:CodewordStructure}
\vv = \wv(0) \wv(1) \pv(1) \cdots \wv(n-1) \pv(n-1)
\end{equation}
be the corresponding codeword produced by tree encoding.
Since parity bits only operate on information bits contained in fragments that precede a slot, the generator matrix admits an upper triangular block structure of the form
\begin{equation*}
G = \begin{bmatrix} G_{0,0} & G_{0,1} & G_{0,2} & \cdots \\
\mathbf{0} & G_{1,1} & G_{1,2} & \cdots \\
\mathbf{0} & \mathbf{0} & G_{2,2} & \cdots \\
\vdots & \vdots & \vdots & \ddots \end{bmatrix} .
\end{equation*}
The parity bits $\pv(j)$ are then produced by column group~$j$,
\begin{equation*}
\pv(j) = \sum_{\ell=0}^{j-1} \wv(\ell) G_{\ell,j-1} .
\end{equation*}
These parity bits are subsequently injected into the codeword as in \eqref{equation:CodewordStructure}.
We wish to capture the performance of the tree code when active blocks within $G$ are Rademacher sub-matrices.
That is, the entries of every sub-matrices are uniform Bernoulli trials, independent of one another and across blocks.

To examine the behavior of the tree code, consider a collection of codewords $\mathcal{L} = \{ \vv_1, \vv_2, \ldots, \vv_K \}$.
Under tree coding, every codeword has the form
\begin{equation*}
\vv_i = \wv_i(0) \wv_i(1) \pv_i(1) \cdots \wv_i(n-1) \pv_i(n-1)
\end{equation*}
where $\pv_i(j) = \sum_{\ell=0}^{j-1} \wv_i(\ell) G_{\ell,j-1}$.
The collection of fragment candidates at stage~$0$ is
\begin{equation*}
\mathcal{L}_0 = \{ \wv_i (0) : i = 1, \ldots, K \} .
\end{equation*}
Similarly, the list of fragment candidates at stage~$j$, where $j \in [1:n-1]$, can be expressed as
\begin{equation*}
\mathcal{L}_j = \{ \wv_i (j) \pv_i (j) : i = 1, \ldots, K \} .
\end{equation*}
The goal of the tree decoder is to produce an unordered list that matches the original set of messages $\mathcal{M} = \{ \wv_1, \wv_2, \ldots, \wv_K \}$.
We emphasize that valid messages invariably appear on the final output list because they necessarily fulfill their own parity constraints.
The probability of error is then dictated by instances where invalid codewords also meet their own parity constraints.
In such cases, the decoder fails to identify these extraneous candidates as erroneous and, therefore, they make it to the final output list as well.
As mentioned above, one of the challenges in analyzing error events is the fact that distinct candidate codewords are not guaranteed to have different fragments at any particular stage.
That is, information fragments can appear more than once in $\mathcal{M}_j = \{ \wv_1(j), \wv_2(j), \ldots, \wv_K(j) \}$, possibly paired with different parity bits.
This situation precludes the naive application of Lemma~\ref{lemma:RandomGenerator} in a divide-and-conquer fashion.

Consider the situation where the tree decoder seeks to validate codewords that start with root fragment $\wv_{i_0}(0)$.
For a given collection $\mathcal{L}$ of transmitted codewords, the list of candidate codewords visited during this portion of the tree decoding process is composed of elements of the form
\begin{equation} \label{equation:CandidateCodewords}
\begin{split}
&\vv_{\mathrm{c}} = \vv_{i_0}(0) \vv_{i_1}(1) \vv_{i_2} (2) \cdots \vv_{i_{n-1}}(n-1) \\
&= \wv_{i_0}(0) \wv_{i_1}(1) \pv_{i_1}(1) \cdots \wv_{i_{n-1}}(n-1) \pv_{i_{n-1}}(n-1) ,
\end{split}
\end{equation}
where $i_j \in [1:K]$ for all slots $j \in [1:n-1]$.
The ability of the tree code to discriminate between valid and erroneous sequences hinges on the structure of the candidate vectors in \eqref{equation:CandidateCodewords} and their connection to parity generator matrix $G$.
We turn our attention to the role of $G$ in identifying invalid codewords.

To survive stage~1, candidate vector $\vv_{\mathrm{c}}$ must fulfill $\pv_{i_1}(1) = \wv_{i_1}(0) G_{0,0} = \wv_{i_0}(0) G_{0,0}$.
Equivalently, we can write $(\wv_{i_0}(0) + \wv_{i_1}(0)) G_{0,0} = \underline{0}$.
To outlast the first two stages, a candidate vector must fulfill two conditions,
\begin{gather*}
\pv_{i_1}(1) = \wv_{i_1}(0) G_{0,0} = \wv_{i_0}(0) G_{0,0} \\
\begin{split}
\pv_{i_2}(2) &= \left( \wv_{i_2}(0), \wv_{i_2}(1) \right)
\begin{bmatrix} G_{0,1} \\ G_{1,1} \end{bmatrix} \\
&= \left( \wv_{i_0}(0), \wv_{i_1}(1) \right)
\begin{bmatrix} G_{0,1} \\ G_{1,1} \end{bmatrix} .
\end{split}
\end{gather*}
This translates into linear equations
\begin{gather*}
\left( \wv_{i_0}(0) + \wv_{i_1}(0) \right) G_{0,0} = \underline{0} \\
\left( \wv_{i_0}(0) + \wv_{i_2}(0), \wv_{i_1}(1) + \wv_{i_2}(1) \right)
\begin{bmatrix} G_{0,1} \\ G_{1,1} \end{bmatrix}
= \underline{0} .
\end{gather*}
In general, to survive the first $j$ stages, a candidate vector must meet the constraints
\begin{equation} \label{equation:GeneralG}
\sum_{\ell = 0}^{q-1} \left( \wv_{i_{\ell}}(\ell) + \wv_{i_q}(\ell) \right) G_{\ell,q-1}
= \underline{0}
\end{equation}
for $q = 1, \ldots, j$.
Keeping these requirements in mind and factoring in the randomness in the problem, we can characterize the probability that an arbitrary candidate vector $\vv_{\mathrm{c}}$ survives the first $j$ stages by analyzing the number of nontrivial random linear equations it must satisfy.
In addressing this question, it is meaningful to study the effective number of parity constraints this candidate sequence must fulfill on a per slot basis.
The insight afforded by Lemma~\ref{lemma:RandomGenerator} is that the $l_j$ parity bits associated with stage~$j$ either act as a statistically discriminating sequence of independent Bernoulli samples, each with probability half, or they are fulfilled trivially based on the combined vector in \eqref{equation:GeneralG} vanishing.
More specifically, the criterion for the former condition to apply is that at least one of the entries in
\begin{equation} \label{equation:InputVector}
\left( \wv_{i_{0}}(0) + \wv_{i_j}(0), \ldots, \wv_{i_{j-1}}(j-1) + \wv_{i_j}(j-1) \right)
\end{equation}
is not equal to zero.
As mentioned above, there are two distinct phenomena that can drive individual entries in \eqref{equation:InputVector} to zero.
Matching occurs in a block fashion whenever $i_{\ell} = i_{j}$ for some $\ell < j$.
Under such circumstances, multiple portions of the candidate vector comes from valid codeword $\wv_{i_j}$ itself and, consequently, $\wv_{i_{\ell}}$ and $\wv_{i_j}$ match at the corresponding locations.
The second possibility arises from two distinct valid messages randomly sharing a common section.
At stage~$j$, the probability that two independent sub-blocks match one another is given by
\begin{equation*}
\Pr \left( \wv_{i_{\ell}}(\ell) = \wv_{i_j}(\ell) | i_{\ell} \neq i_j \right) = 2^{-m_{\ell}} ,
\end{equation*}
where $\ell \in \{ 0, \ldots, j-1 \}$.
Not surprisingly, this probability is intrinsically linked to the number of information bits contained in the sub-block.
We already know from Lemma~\ref{lemma:RandomGenerator} that there is a dichotomy between parity bits associated with column group~$j$.
A collection $\pv(j)$ of parity bits is either statistically discriminating or, as a block, it becomes uninformative.
The distinction between these two situations stems from the character of the input vectors in \eqref{equation:InputVector}, whether they are identical or different.
One subtlety related to the number of active parity bits for a particular candidate codeword can be seen by stacking the combined vectors,
\begin{gather*}
\left( \wv_{i_0}(0) + \wv_{i_1}(0) \right) \\
\left( \wv_{i_0}(0) + \wv_{i_2}(0), \wv_{i_1}(1) + \wv_{i_2}(1) \right) \\
\vdots \\
\left( \wv_{i_0}(0) + \wv_{i_j}(0), \ldots, \wv_{i_{j-1}}(j-1) + \wv_{i_j}(j-1) \right) .
\end{gather*}
Whenever a candidate codeword draws multiple fragments from a same valid message, the events where the combined vectors vanish and the corresponding parity bits lose their discriminatory power become more likely within a block.
These events are also correlated across blocks.
For example, if $i_2 = i_3$ and $\wv_{i_0}(0) + \wv_{i_2}(0) = \underline{0}$, then $\wv_{i_0}(0) + \wv_{i_3}(0) = \underline{0}$.
Consequently, for certain index patterns, the probability that an erroneous candidate meets its parity constraints at different stages cannot be treated separately.
Altogether, the probability that candidate codeword $\vv_{\mathrm{c}}$ is consistent with its own parity bits depends on its index sequence, which can be taken as visits to distinct levels over the progression of the candidate message.

Let index sequence $i_0, i_1, \ldots, i_{n-1}$ be given.
We are interested in conditional probabilities of the form
\begin{equation*}
\Pr (T = t | i_0, i_1, \ldots, i_{n-1})
\end{equation*}
where $T$ is the number of statistically discriminating parity bits.
Since $i_j \in [1:K]$, there are $K^n$ possible conditions.
A careful examination of the problem and, in particular, of \eqref{equation:GeneralG} reveals that this probability is permutation invariant.
That is, the state labeling is irrelevant, only the order in which previously visited states are reentered matters.
Mathematically, the structure of the problem enables us to write
\begin{equation*}
\Pr (T = t | i_0, i_1, \ldots, i_{n-1})
= \Pr (T = t | \pi(i_0), \pi(i_1), \ldots, \pi(i_{n-1}))
\end{equation*}
for any permutation $\pi$ of the integers $1, \ldots, K$.
When the number of active devices exceeds the slot count, this symmetric property reduces the number of possible cases from $K^n$ to the $n$th Bell number (OEIS A000110) given by the formula
\begin{equation*}
B_n=\sum_{j=0}^n S_n^{(j)} .
\end{equation*}
Above, $S_n^{(j)}$ represents the Stirling numbers of the second kind (OEIS A008277), which can be computed recursively using the formula $S_n^{(j)} = S_{n-1}^{(j-1)} + j S_{n-1}^{(j)}$.

\subsection{Pattern Sequences}
\label{subsection:PatternSequence}

To simplify the problem and reduce the total number of cases to be considered, we leverage equivalence classes of erroneous patterns wherein every sequence within a class share the same probability of going undetected.
The structure of the equivalence classes is informed by $j$-patterns (OEIS A008277).
For completeness, we include an appropriate definition below.

\begin{definition}[$j$-Pattern Sequences] \label{definition:jPattern}
A $j$-pattern sequence is an integer sequence
\begin{equation*}
\sv = (\sv(0), \sv(1), \ldots, \sv(j-1))
\end{equation*}
such that $\sv(\ell) = \ell+1$ or $\sv(\ell) = \sv(q)$ for some $q \in [0:\ell-1]$.
\end{definition}

We denote the collection of all $j$-pattern sequences by $\mathcal{P}_j$, where $1 \leq j \leq n$.
There is a single $1$-pattern sequence, $\mathcal{P}_1 = \{ (1) \}$.
Similarly,
\begin{gather*}
\mathcal{P}_2 = \{(1,1), (1,2)\} \\
\mathcal{P}_3 = \{ (1,1,1), (1,1,3), (1,2,1), (1,2,2), (1,2,3) \}
\end{gather*}
indicate the collection of all $2$-pattern and $3$-pattern sequences, respectively.
To be consistent with our established notation, we index the entries of $j$-pattern sequence $\sv$ using integers $[0:j-1]$.
We  emphasize that a $j$-pattern sequence can be created iteratively starting with a shorter $(j-1)$-pattern sequence $\sv$ and appending to it either integer $j$ or entry $\sv(k)$ for some $k < j-1$.
With this construction, it can be verified that the total number of $j$-pattern sequences equal to the $j$th Bell number (OEIS A000110); that is, $|\mathcal{P}_j| = B_j$.

Under this viewpoint, each $j$-pattern sequence $\sv \in \mc{P}_j$ acts as a representative element to a collection of vectors of the form $(i_0, i_1, \ldots, i_{j-1})$.
The mapping from an arbitrary integer sequences in $[1:K]^n$ to the representative of its equivalence class can be defined recursively.
The initial entry of the representative is $\sv(0) = 1$, irrespective of $i_0$.
Then, for any $q > 1$,
\begin{equation*}
\sv(q) = \begin{cases} q+1 & i_{q} \notin \{ i_0, \ldots, i_{q-1} \} \\
\min  \{ \ell \in [1:q-1] \mid i_{\ell} = i_q \} & \text{otherwise} . \end{cases}
\end{equation*}
For instance, any vector of the form $\{(i_0,i_1,i_2):i_0=i_1=i_2\}$ maps to the $3$-pattern sequence $(1,1,1)$.
Likewise, any vector in $\{(i_0,i_1,i_2):i_0=i_1,i_1 \neq i_2\}$ map to $(1,1,3)$.
The number of admissible integer sequences within an equivalence class is characterized below.
Without loss of generality, we focus on a single rooted tree with $i_0 = 1$ to simplify our upcoming discussion.

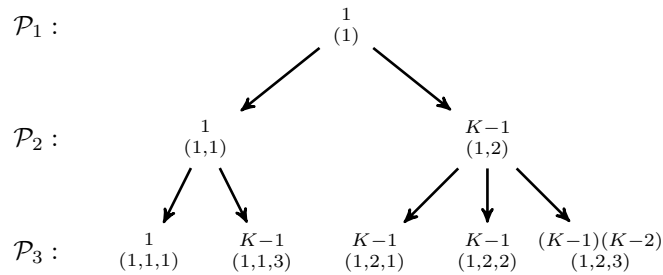
\begin{figure}[htb]
  \centering
  \begin{tikzpicture}
[font=\normalsize, >=stealth', line width=1pt]

\node (v1) at (2.625,3) {${\genfrac..{0pt}{1}{1}{(1)}}$};
\node (v2) at (0.75,1.5) {${\genfrac..{0pt}{1}{1}{(1,1)}}$};
\node (v3) at (4.5,1.5) {${\genfrac..{0pt}{1}{K-1}{(1,2)}}$};
\node (v4) at (0,0) {${\genfrac..{0pt}{1}{1}{(1,1,1)}}$};
\node (v5) at (1.5,0) {${\genfrac..{0pt}{1}{K-1}{(1,1,3)}}$};
\node (v6) at (3,0) {${\genfrac..{0pt}{1}{K-1}{(1,2,1)}}$};
\node (v7) at (4.5,0) {${\genfrac..{0pt}{1}{K-1}{(1,2,2)}}$};
\node (v8) at (6,0) {${\genfrac..{0pt}{1}{(K-1)(K-2)}{(1,2,3)}}$};
\node (p1) at (-1.5, 3) {$\mathcal{P}_1:$};
\node (p2) at (-1.5, 1.5) {$\mathcal{P}_2:$};
\node (p3) at (-1.5, 0) {$\mathcal{P}_3:$};
\draw[->] (v1) edge (v2);
\draw[->] (v1) edge (v3);
\draw[->] (v2) edge (v4);
\draw[->] (v2) edge (v5);
\draw[->] (v3) edge (v6);
\draw[->] (v3) edge (v7);
\draw[->] (v3) edge (v8);

\end{tikzpicture}
  \caption{This illustration showcases $j$-pattern sequences for stages $0$, $1$, and $2$, along with the sizes of their equivalence classes.
  The figure adopts the stacked format ${\genfrac..{0pt}{1}{n(\sv)}{\sv}}$.}
  \label{figure:Bellnumbers}
\end{figure}

\begin{lemma}\label{Number of states}
Let $\sv \in \mc{P}_j$ denote a $j$-pattern sequence.
The number of integer sequences $i_0, \ldots, i_{j-1}$ that map to $\sv$ is given by
\begin{equation} \label{equation:SizeEquivalenceClass}
n(\sv) = (K-1)(K-2) \cdots (K-(d(\sv)-1)),
\end{equation}
where $d(\sv)$ denotes the number of distinct integers in $\sv$.
\end{lemma}
\begin{IEEEproof}
By assumption, we have $i_0 = 1$.
For $\ell > 0$, integer $i_{\ell}$ is unambiguously determined by its predecessors whenever a previous entry is repeated in sequence $\sv$.
On the other hand, when $s(\ell) = \ell + 1$, integer $i_{\ell}$ can be any element within $[1:K]$ that has not appeared in the sequence thus far.
Combining these observations with a straightforward application of the counting principle, we get \eqref{equation:SizeEquivalenceClass}.
\end{IEEEproof}
The total number of candidate paths in the tree at stage $j-1$ is $\sum_{\sv \in \mc{P}_j}n(\sv)=K^{j-1}$.
Figure~\ref{figure:Bellnumbers} shows $j$-pattern sequences and the cardinality of their equivalence classes for stages $0$, $1$, and $2$ in the tree.

Our next task is to compute the probability that a representative from an equivalance class survives.
We employ the notion of generating functions~\cite{graham1989concrete} to obtain the probability of survival.
For every $j$-pattern sequence $\sv \in \mc{P}_j$, we track the distribution of random variable $T(\sv)$, the number of effective parity-check bits for an element in that class.
This is accomplished by evaluating the probability generating function (PGF) of $T(\sv)$, which takes the form
\begin{equation} \label{generatingFunction}
\Phi_{\sv}(x) = \mathbb{E} \left[ x^{T(\sv)} \right]
= \sum_{t=0}^{l_1 + \cdots + l_{j-1}} \Pr (T(\sv)=t)x^t .
\end{equation}
Above, $\Pr (T(\sv)=t)$ is the probability that $t$ parity bits are statistically discriminating, for $j$-pattern sequence $\sv$.
The probability that a particular index sequence with representative class $\sv$ survives at the end of the decoding process is then given by $\Phi_{\sv}(0.5)$.

For example, the $3$-pattern sequence $(1,1,1)$ has generating function $\Phi_{(1,1,1)}(x)=1$ because it corresponds to a valid path in the tree.
The parity bits for a valid sequence match the information bits by construction and, as such, this sequence invariably appears on the output list.
The $3$-pattern sequence $(1,1,3)$ represents states of the form $\{ (i_0, i_1, i_2): i_0=i_1, i_1 \neq i_2 \}$.
From our previous discussion on fragmented codewords, we gather that the first $l_1$ parity bits in this candidate codeword are non-discriminating because the information bits for the first two fragments are drawn from a same message.
On the other hand, the $l_2$ parity check bits in the second sub-block only become non-discriminating when $\wv_{i_0}(0) = \wv_{i_2}(0)$ and $\wv_{i_0}(1) = \wv_{i_2}(1)$.
This conjunction of these events occurs with a probability $\frac{1}{2^{m_0+m_1}}$.
Altogether, the generating function for this 3-pattern sequence is given by $\Phi_{(1,1,3)}(x) = \left( 1 - \frac{1}{2^{m_0+m_1}} \right) x^{l_2} + \frac{1}{2^{m_0+m_1}}$.
It can be verified in an analogous fashion that $\Phi_{(1,2,2)}(x)= \left( 1-\frac{1}{2^{m_0}} \right) x^{l_1+l_2} + \frac{1}{2^{m_0}}$.

Collecting the notions introduced thus far, we can define an aggregate generating function $\boldsymbol{\Phi}(x)$ for the single rooted tree,
\begin{equation} \label{equation:TreeGeneratingFunction}
\boldsymbol{\Phi}(x) = \sum_{\sv \in \mathcal{P}_j} n(\sv) \Phi_{\sv}(x) .
\end{equation}
The expected number of surviving paths within the tree is then given by $\boldsymbol{\Phi}(0.5)$.
In the remainder of this section, we describe the computation of the generating functions for a given $j$-pattern sequence, $\Phi_{\sv}(x)$.
It is pertinent to note that $\Phi_{\sv}(x)$ is sparse;
the coefficient of $x^t$ can only be non-zero when
\begin{equation*}
t \in \left\{ \sum_{q \in S} l_q : \mathcal{S} \subseteq [1:j-1] \right\} . 
\end{equation*}
Thus, it suffices to compute $\Pr (T(\sv)=t)$ for values of $t$ within this set.
We also point out that multiple subsets of $[1:j-1]$ can contribute to a same coefficient.
In other words, there can be distinct subsets, say $\mathcal{S}$ and $\mathcal{S}'$, such that $\sum_{q \in \mathcal{S}} l_q = \sum_{q \in \mathcal{S}'}l_q$.
Altogether, the computation of $\Phi_{\sv}(x)$ mandates inspecting the contribution of all $2^{j-1}$ subsets of $[1:j-1]$.

\subsection{Statistically Discriminating Parity Bits}
\label{subsection:StatisticallyDiscriminatingParityBits}

Again, consider $j$-pattern sequence $\sv$.
We are interested in events where the parity bits associated with column groups in $G$ are statistically discriminating.
Recall that the $l_q$ parity-check bits in $\pv(q)$ are generated via column group
\begin{equation*}
\begin{bmatrix} G_{0,q-1} \\ \vdots \\ G_{q-1, q-1} \\ \mathbf{0} \end{bmatrix} ,
\end{equation*}
and they become collectively discriminating if and only if at least one entry in
\begin{equation*}
\left( \wv_{\sv(0)}(0) + \wv_{\sv(q)}(0), \ldots, \wv_{\sv(q-1)}(q-1) + \wv_{\sv(q)}(q-1), \underline{0} \right)
\end{equation*}
is non-zero.
Define $\mathcal{E}_{\sv,\mathcal{S}}$ to be the event where the parity bits $\pv(q)$ with $q \in \mathcal{S}$ are statistically discriminating, but those with indices in $\mathcal{S}^{\mathrm{c}} = [1:n-1] \setminus \mathcal{S}$ are not.
This somewhat intricate concept warrants an example.
Suppose $\mathcal{S} = \{ 1, 3, 5 \}$; then $\mathcal{E}_{\sv,\mathcal{S}}$ is the event where the parity-check bits $\pv(1)$, $\pv(3)$, and $\pv(5)$ essentially become strings of random Bernoulli trials under the randomness in $G$, which must meet prescribed parity requirements for the candidate message to survive.
Whereas the parity constraints for sub-blocks $[1:n-1] \setminus \{ 1, 3, 5 \}$ are fulfilled solely based on the $j$-pattern and/or message fragments matching one another, regardless of the realization of matrix $G$.
Then, taking into account all the possible subsets of $[1:n-1]$, we can rewrite \eqref{generatingFunction} as
\begin{equation} \label{equation:TreeGeneratingFunctionSingleSet}
\Phi_{\sv}(x) = \sum_{\mathcal{S} \subseteq [1:n-1]}
\Pr \left( \mathcal{E}_{\sv,\mathcal{S}}\right) x^{\sum_{q \in \mathcal{S}}l_q} .
\end{equation}
This expression invites a closer look at the summand $\Pr \left( \mathcal{E}_{\sv, \mathcal{S}} \right)$, which can be computed for every $\mathcal{S} \subseteq [1:n-1]$ based on the randomness in the message sequences. 
 
To better understand such events, we further decompose $\mathcal{E}_{\sv, \mathcal{S}}$ into events that occur within specific blocks.
Let $\mathcal{A}_{q,\sv}$ be the event in which parity-check bits of sub-block $q$ are non-discriminating for the $j$-pattern sequence $\sv$.
Based on our previous discussion on fragmented codewords, we have
\begin{equation} \label{equation:GroupNonDiscriminating}
\mathcal{A}_{q,\sv}
= \bigcap_{\substack{j \in [0:q-1] \\ \sv(q) \neq \sv(j)}} \left\{ \wv_{\sv(q)}(j)=\wv_{\sv(j)}(j) \right\} .
\end{equation}
Then, event $\mathcal{E}_{\sv,\mathcal{S}}$ can be expressed as
\begin{equation} \label{EsS}
\mathcal{E}_{\sv,\mathcal{S}}
= \left( \bigcap_{q \in \mathcal{S}^{\mathrm{c}}} \mathcal{A}_{q,\sv} \right)
\bigcap \left( \bigcap_{q \in \mathcal{S}} \mathcal{A}^{\mathrm{c}}_{q,\sv} \right) .
\end{equation}
We emphasize that the analysis of $\Pr \left( \mathcal{E}_{\sv,\mathcal{S}} \right)$ is cumbersome because the events $\mathcal{A}_{q,\sv}$, $q \in [1:j-1]$ are not independent from each other.
Dependencies arise because of re-entries to previously visited levels, as captured by the $j$-pattern sequence $\sv$.
Nevertheless, before moving forward, it is meaningful to characterize the probability of building-block event $\mathcal{A}_{q,\sv}$.

\begin{lemma} \label{P}
Let $j$-pattern sequence $\sv$ be fixed.
The probability that the parity bits of sub-block $\pv(q)$ are non-discriminating is
\begin{equation*}
\Pr \left( \mathcal{A}_{q,\sv}\right) = g \left( \sum_{\substack{j \in [0:q-1] \\ \sv(q) \neq \sv(j)}}m_{j} \right) ,
\end{equation*}
where $g(t)=2^{-t}$.
\end{lemma}
\begin{IEEEproof}
We already know that, based on equivalence class $\sv$, the contribution of block $G_{j,q-1}$ is lost at every location where $\sv(j) = \sv(q)$.
As such, event $\mathcal{A}_{q,\sv}$ corresponds to a situation where candidate message sequence $(\wv_{\sv(0)}(0), \ldots \wv_{\sv(q-1)}(q-1))$ also matches valid partial message $(\wv_{\sv(q)}(0), \ldots \wv_{\sv(q)}(q-1))$ at every locations where $\sv(j) \neq \sv(q)$.
This renders the parity-check bits in $\pv(q)$ non-discriminating.
The probability of this event occurring, conditioned on $\sv$, is given by
\begin{equation}
\begin{split}
\Pr \left( \mathcal{A}_{q,\sv} \right)
&= \Pr \left(\bigcap_{\substack{j \in [0:q-1] \\ \sv(q) \neq \sv(j)}} \left\{ \wv_{\sv(q)}(j)=\wv_{\sv(j)}(j) \right\} \right) \\
&= \prod_{\substack{j \in [0:q-1] \\ \sv(q) \neq \sv(j)}} \Pr \left( \wv_{\sv(q)}(j)=\wv_{\sv(j)}(j) \right) \\
&= \prod_{\substack{j \in [0:q-1] \\ \sv(q) \neq \sv(j)}}g{(m_{j})}
= g \left( \sum_{\substack{j \in [0:q-1] \\ \sv(q) \neq \sv(j)}}m_{j} \right) ,
\end{split}
\end{equation}
where the second equality is justified through the independence of information bits across distinct messages.
\end{IEEEproof}

In essence, Lemma~\ref{P} reveals how the equivalence class $\sv$ and the randomness in information bits across messages determine the discriminatory power of $\pv(q)$.
Ultimately, the ability to detect an erroneous codeword through $\pv(q)$ also depends on the randomness in $G$.
Under the right conditions, there is a block activation of the $l_q$ bits; otherwise, they collectively become non-discriminatory.
At this point, we turn to the challenge of computing the joint probability that parity-check bits across blocks are discriminating.
To account for dependencies among column groups, we express $\Pr \left(\mathcal{E}_{\sv,\mathcal{S}}\right)$ using conditional probability,
\begin{equation} \label{PEsS}
\Pr \left(\mathcal{E}_{\sv,\mathcal{S}}\right)
= \Pr \left( \bigcap_{q \in \mathcal{S}}\mathcal{A}^{\mathrm{c}}_{q,\sv} \Bigg| \bigcap_{q \in \mathcal{S}^{\mathrm{c}}}\mathcal{A}_{q,\sv} \right)
\Pr \left( \bigcap_{q \in \mathcal{S}^{\mathrm{c}}}\mathcal{A}_{q,\sv} \right) .
\end{equation}

We begin our treatment of \eqref{PEsS} by examining the rightmost term.
We can apply the chain rule of probability to this expression, which yields
\begin{equation} \label{equation:ConditionalChainRule}
\Pr \left( \bigcap_{q \in \mathcal{S}^{\mathrm{c}}} \mathcal{A}_{q,\sv} \right)
= \prod_{q \in \mathcal{S}^{\mathrm{c}}} \Pr \left( \mathcal{A}_{q,\sv} \Bigg| \bigcap_{\ell \in \mathcal{S}^{\mathrm{c}} \cap [1:q-1]}\mathcal{A}_{\ell, \sv}\right) .
\end{equation}
Inspecting \eqref{equation:GroupNonDiscriminating}, we gather that events of the form $\mathcal{A}_{q,\sv}$ are subject to a partial order through set inclusion.
In particular, suppose that $\sv(q) = \sv(\ell)$ with $\ell < q$, then $\mathcal{A}_{q,\sv} \subset \mathcal{A}_{\ell,\sv}$.
This property immediately implies that the most constraining condition within this branch of the partial order is $\mathcal{A}_{p,\sv}$ where $p = \max \left\{ \ell \in [1 : q-1] | \sv(q) = \sv(\ell) \right\}$.
On the other hand, when $\sv(q) \neq \sv(\ell)$, the events $\mathcal{A}_{q,\sv}$ and $\mathcal{A}_{\ell,\sv}$ are independent.
This follows from the fact that, in this latter case, the conditions in \eqref{equation:GroupNonDiscriminating} operate on different pairs of fragments altogether.
The fact that information bits are selected independently across messages is therefore enough to ensure independence.
While the situation is slightly more involved when accounting for $\mathcal{S}$, the concepts are essentially identical.
These observations invite the following definitions,
\begin{align}
\underline{\mathcal{G}}_{q,\mathcal{S},\sv} &= [1:q-1] \cap \mathcal{S}^{\mathrm{c}} \cap \{k \in [1:j-1]: \sv(k) \neq \sv(q)\} 
\label{equation:Gunderline} \\
\underline{\mathcal{Q}}_{q,\mathcal{S},\sv} &= [1:q-1] \cap \mathcal{S}^{\mathrm{c}} \cap \{k \in [1:j-1]: \sv(k) = \sv(q)\} .
\label{equation:Qunderline}
\end{align}
In words, $\underline{\mathcal{G}}_{q,\mathcal{S},\sv}$ contains the numbers of the sub-blocks that precede sub-block $q$ (excluding sub-block $0$), feature a different level than $\sv(q)$, and are not included in $\mathcal{S}$.
Likewise, $\underline{\mathcal{Q}}_{q,\mathcal{S},\sv}$ contains the numbers of the sub-blocks that precede sub-block $q$ (excluding sub-block $0$), possess the same level as $\sv(q)$, and are not included in $\mathcal{S}$.
Based on this notation, we formalize the above discussion into a lemma.

\begin{lemma} \label{P1}
Consider equivalence class $\sv$ and set $\mathcal{S} \subset [1:n-1]$.
Then, when the conditional event has positive probability, we get
\begin{equation} \label{equation:P1}
\Pr \left( \mathcal{A}_{q,\sv} \Bigg| \bigcap_{\ell \in \mathcal{S}^{\mathrm{c}} \cap [1:q-1]}\mathcal{A}_{\ell, \sv}\right)
= g \left( \sum_{\substack{j \in [0:q-1] \\ \sv(q) \neq \sv(j)}} m_{j}
- \sum_{\substack{k \in [0:p-1] \\ \sv(p) \neq \sv(k)}} m_{k} \right),
\end{equation}
where
$p = \max \underline{\mathcal{Q}}_{q,\mathcal{S},\sv}$ and $g(t)=2^{-t}$.
\end{lemma}
\begin{IEEEproof}
Taking into consideration the independence structure identified above, we immediately get
\begin{equation} \label{equation:P1reduced}
\begin{split}
\Pr \left( \mathcal{A}_{q,\sv} \Bigg| \bigcap_{\ell \in \mathcal{S}^{\mathrm{c}} \cap [1:q-1]}\mathcal{A}_{\ell, \sv}\right)
&= \Pr \left( \mathcal{A}_{q,\sv} \Bigg| \bigcap_{\ell \in \underline{\mathcal{Q}}_{q,\mathcal{S},\sv}} \mathcal{A}_{\ell, \sv} ,
\bigcap_{\ell \in \underline{\mathcal{G}}_{q,\mathcal{S},\sv}} \mathcal{A}_{\ell, \sv} \right) \\
&= \Pr \left( \mathcal{A}_{q,\sv} \Bigg| \bigcap_{\ell \in \underline{\mathcal{Q}}_{q,\mathcal{S},\sv}} \mathcal{A}_{\ell, \sv} \right) .
\end{split}
\end{equation}
We know that events of the form $\mathcal{A}_{q,\sv}$ form a partial order through set inclusion.
Moreover, $\mathcal{A}_{k,\sv} \subset \mathcal{A}_{j,\sv}$ whenever j < k and $\sv(j) = \sv(k)$.
This leads to the following two equalities:
\begin{equation*}
\bigcap_{k \in \underline{\mathcal{Q}}_{q,\mathcal{S},\sv}} \mathcal{A}_{k,\sv} = \mathcal{A}_{p, \sv}
\end{equation*}
and $\mathcal{A}_{q,\sv} \cap \mathcal{A}_{p, \sv} = \mathcal{A}_{q,\sv}$, where $p = \max \underline{\mathcal{Q}}_{q,\mathcal{S},\sv}$.
Substituting these expressions into the conditional probability of \eqref{equation:P1reduced}, we get
\begin{equation} \label{equation:P1proof}
\begin{split}
\Pr & \left( \mathcal{A}_{q,\sv} \Bigg| \bigcap_{\ell \in \underline{\mathcal{Q}}_{q,\mathcal{S},\sv}} \mathcal{A}_{\ell,\sv}\right)
= \Pr \left( \mathcal{A}_{q,\sv} \Bigg| \mathcal{A}_{p,\sv} \right) \\
&= \frac{ \Pr \left( \mathcal{A}_{q,\sv} \bigcap \mathcal{A}_{p,\sv} \right) }
{\Pr \left( \mathcal{A}_{p,\sv} \right)}
= \frac{ \Pr \left( \mathcal{A}_{q,\sv} \right) } {\Pr \left(\mathcal{A}_{p,\sv} \right)} \\
&= g \left( \sum_{\substack{j \in [0:q-1] \\ \sv(q) \neq \sv(j)}} m_{j}
- \sum_{\substack{k \in [0:p-1] \\ \sv(p) \neq \sv(k)}} m_{k} \right) .
\end{split}
\end{equation}
The fourth inequality in \eqref{equation:P1proof} is obtained by applying Lemma~\ref{P} and using straightforward properties of the function $g(t) = 2^{-t}$.
\end{IEEEproof}

Together Lemma~\ref{P1} and \eqref{equation:ConditionalChainRule} offer a means to compute the rightmost term in \eqref{PEsS}.
We then turn to the leftmost term on the right-hand-side of \eqref{PEsS}.
To facilitate the description of this slightly more involved expression, we need to expand special cases beyond \eqref{equation:Gunderline} and \eqref{equation:Qunderline}.
Specifically, for every $q \in [1:n-1]$, we introduce the following additional cases:
\begin{align*}
\overline{\mathcal{G}}_{q,\mathcal{S},\sv} &= [q+1:n-1] \cap \mathcal{S}^{\mathrm{c}}\cap \{k \in [1:n-1]:\sv(k) \neq \sv(q)\} \\
\overline{\mathcal{Q}}_{q,\mathcal{S},\sv} &= [q+1:n-1] \cap \mathcal{S}^{\mathrm{c}} \cap \{k \in [1:n-1]:\sv(k) = \sv(q)\} \\
\underline{\tilde{\mathcal{G}}}_{q,\mathcal{S},\sv} &= [1:q-1] \cap \mathcal{S} \cap \{k \in [1:n-1]: \sv(k) \neq \sv(q)\} \\
\underline{\tilde{\mathcal{Q}}}_{q,\mathcal{S},\sv} &= [1:q-1] \cap \mathcal{S} \cap \{k \in [1:n-1]: \sv(k) = \sv(q)\}.
\end{align*}
Above, $\underline{\tilde{\mathcal{G}}}_{q,\mathcal{S},\sv}$ contains the numbers of the sub-blocks that trail sub-block $q$, feature a different level than $\sv(q)$, and are not included in $\mathcal{S}$.
The set $\underline{\tilde{\mathcal{Q}}}_{q,\mathcal{S},\sv}$ contains the numbers of the sub-blocks that trail sub-block $q$, possess the same level as $\sv(q)$, and are not included in $\mathcal{S}$.
The last two conditions, $\underline{\tilde{\mathcal{G}}}_{q,\mathcal{S},\sv}$ and $\underline{\tilde{\mathcal{Q}}}_{q,\mathcal{S},\sv}$ are analogous to $\underline{\mathcal{G}}_{q,\mathcal{S},\sv}$ and $\underline{\mathcal{Q}}_{q,\mathcal{S},\sv}$, albeit they contain numbers belonging to $\mathcal{S}$ rather than its complement.

The strategy to tackle
\begin{equation} \label{equation:ConditonalProb}
\Pr \left( \bigcap_{q \in \mathcal{S}} \mathcal{A}^{\mathrm{c}}_{q,\sv} \Bigg| \bigcap_{\ell \in \mathcal{S}^{\mathrm{c}}} \mathcal{A}_{\ell,\sv} \right)
\end{equation}
is similar to our previous step.
We apply the chain rule of probability and analyze the relation between events.
The expanded version of \eqref{equation:ConditonalProb} can be written as
\begin{equation} \label{equation:ConditonalProbChainRule}
\prod_{q \in \mathcal{S}} \Pr \left( \mathcal{A}^{\mathrm{c}}_{q,\sv} \Bigg| \bigcap_{\ell \in \mathcal{S}^{\mathrm{c}}} \mathcal{A}_{\ell,\sv},
\bigcap_{k \in \mathcal{S} \cap [1:q-1]} \mathcal{A}^{\mathrm{c}}_{k,\sv} \right) .
\end{equation}
We can further partition the conditional events via
\begin{gather}
\mathcal{S}^{\mathrm{c}} = \underline{\mathcal{G}}_{q,\mathcal{S},\sv} \uplus \underline{\mathcal{Q}}_{q,\mathcal{S},\sv}
\uplus \overline{\mathcal{G}}_{q,\mathcal{S},\sv} \uplus \overline{\mathcal{Q}}_{q,\mathcal{S},\sv} \\
\mathcal{S} \cap [1:q-1] = \underline{\tilde{\mathcal{G}}}_{q,\mathcal{S},\sv} \uplus \underline{\tilde{\mathcal{Q}}}_{q,\mathcal{S},\sv} .
\end{gather}
The event $\mathcal{A}^{\mathrm{c}}_{q,\sv}$ in which parity-check bits of sub-block $q$ are discriminating; together with what occurs at level $\sv(q)$, namely $\bigcap_{\ell \in \underline{\mathcal{Q}}_{q,\mathcal{S},\sv} \uplus \overline{\mathcal{Q}}_{q,\mathcal{S},\sv}} \mathcal{A}_{\ell,\sv}$ and $\bigcap_{k \in \underline{\tilde{\mathcal{Q}}}_{q,\mathcal{S},\sv}} \mathcal{A}^{\mathrm{c}}_{k,\sv}$, is independent of sub-blocks associated with other levels, namely $\bigcap_{\ell \in \underline{\mathcal{G}}_{q,\mathcal{S},\sv} \uplus \overline{\mathcal{G}}_{q,\mathcal{S},\sv}} \mathcal{A}_{\ell,\sv}$ and $\bigcap_{k \in \underline{\tilde{\mathcal{G}}}_{q,\mathcal{S},\sv}} \mathcal{A}^{\mathrm{c}}_{k,\sv}$.
As before, this property follows from the fact that, at levels other than $\sv(q)$, the conditions in \eqref{equation:GroupNonDiscriminating} operate on different pairs of fragments altogether.
Since information bits are selected independently across messages, this is enough to ensure independence between the groupings above.
Thus, \eqref{equation:ConditonalProbChainRule} reduces to
\begin{equation} \label{equation:ReducedConditonalProbChainRule}
\begin{split}
&\prod_{q \in \mathcal{S}} \Pr \left( \mathcal{A}^{\mathrm{c}}_{q,\sv} \Bigg|
\bigcap_{\ell \in \underline{\mathcal{Q}}_{q,\mathcal{S},\sv}} \mathcal{A}_{\ell,\sv},
\bigcap_{\ell \in \overline{\mathcal{Q}}_{q,\mathcal{S},\sv}} \mathcal{A}_{\ell,\sv},
\bigcap_{k \in \underline{\tilde{\mathcal{Q}}}_{q,\mathcal{S},\sv}} \mathcal{A}^{\mathrm{c}}_{k,\sv} \right) \\
&= \prod_{q \in \mathcal{S}} \left( 1 - \Pr \left( \mathcal{A}_{q,\sv} \Bigg|
\bigcap_{\ell \in \underline{\mathcal{Q}}_{q,\mathcal{S},\sv}} \mathcal{A}_{\ell,\sv},
\bigcap_{\ell \in \overline{\mathcal{Q}}_{q,\mathcal{S},\sv}} \mathcal{A}_{\ell,\sv},
\bigcap_{k \in \underline{\tilde{\mathcal{Q}}}_{q,\mathcal{S},\sv}} \mathcal{A}^{\mathrm{c}}_{k,\sv} \right) \right) .
\end{split}
\end{equation}
The probability expression in \eqref{equation:ReducedConditonalProbChainRule} is in a form reminiscent of Lemma~\ref{P1}.
At this stage, we proceed by breaking the conditional expression above into special cases.

\paragraph{Case I -- $( \overline{\mathcal{Q}}_{q,\mathcal{S},\sv} \neq \emptyset )$}
If this set is not empty, then there exists an element $\ell \in \overline{\mathcal{Q}}_{q,\mathcal{S},\sv}$, $\ell > q$, such that the parity bits generated by column group~$\ell$ are non-discriminating.
This implies that vector $( \wv_{\sv(0)}(0) + \wv_{\sv(\ell)}(0), \ldots, \wv_{\sv(\ell-1)}(\ell-1) + \wv_{\sv(\ell)}(\ell-1)) = \underline{0}$.
Since $\sv(\ell) = \sv(q)$, we deduce that sub-vector $( \wv_{\sv(0)}(0) + \wv_{\sv(q)}(0), \ldots, \wv_{\sv(q-1)}(q-1) + \wv_{\sv(q)}(q-1)) = \underline{0}$.
Thus, the parity-check bits generated by column group~$q$ are non-discriminating and
\begin{equation} \label{caseI}
\Pr \left( \mathcal{A}_{q,\sv} \Bigg|
\bigcap_{\ell \in \underline{\mathcal{Q}}_{q,\mathcal{S},\sv}} \mathcal{A}_{\ell,\sv},
\bigcap_{\ell \in \overline{\mathcal{Q}}_{q,\mathcal{S},\sv}} \mathcal{A}_{\ell,\sv},
\bigcap_{k \in \underline{\tilde{\mathcal{Q}}}_{q,\mathcal{S},\sv}} \mathcal{A}^{\mathrm{c}}_{k,\sv} \right) = 1 .
\end{equation}

\paragraph{Case II -- $( \underline{\tilde{\mathcal{Q}}}_{q,\mathcal{S},\sv} \neq \emptyset )$}
If this set in not empty, then there exists an element $k \in \underline{\tilde{\mathcal{Q}}}_{q,\mathcal{S},\sv}$, $k < q$, such that the parity bits generated by column group~$k$ are discriminating.
In other words, vector $( \wv_{\sv(0)}(0) + \wv_{\sv(k)}(0), \ldots, \wv_{\sv(k-1)}(k-1) + \wv_{\sv(k)}(k-1) )$ is non-zero.
Yet, this vector appears as a sub-component of $( \wv_{\sv(0)}(0) + \wv_{\sv(q)}(0), \ldots, \wv_{\sv(q-1)}(q-1) + \wv_{\sv(q)}(q-1))$ and, consequently, the latter vector must also be a non-zero vector.
By Lemma~\ref{lemma:RandomGenerator}, this implies that the parity-check bits associated with column group~$q$ are discriminatory and
\begin{equation} \label{caseII}
\Pr \left( \mathcal{A}_{q,\sv} \Bigg|
\bigcap_{\ell \in \underline{\mathcal{Q}}_{q,\mathcal{S},\sv}} \mathcal{A}_{\ell,\sv},
\bigcap_{\ell \in \overline{\mathcal{Q}}_{q,\mathcal{S},\sv}} \mathcal{A}_{\ell,\sv},
\bigcap_{k \in \underline{\tilde{\mathcal{Q}}}_{q,\mathcal{S},\sv}} \mathcal{A}^{\mathrm{c}}_{k,\sv} \right) = 0 .
\end{equation}

\paragraph{Case III -- $( \overline{\mathcal{Q}}_{q,\mathcal{S},\sv} = \emptyset$ and $\underline{\tilde{\mathcal{Q}}}_{q,\mathcal{S},\sv} = \emptyset )$}
In this case, we get
\begin{equation} \label{caseIII}
\begin{split}
\Pr & \left( \mathcal{A}_{q,\sv} \Bigg|
\bigcap_{\ell \in \underline{\mathcal{Q}}_{q,\mathcal{S},\sv}} \mathcal{A}_{\ell,\sv},
\bigcap_{\ell \in \overline{\mathcal{Q}}_{q,\mathcal{S},\sv}} \mathcal{A}_{\ell,\sv},
\bigcap_{k \in \underline{\tilde{\mathcal{Q}}}_{q,\mathcal{S},\sv}} \mathcal{A}^{\mathrm{c}}_{k,\sv} \right) \\
&= \Pr \left( \mathcal{A}_{q,\sv} \Bigg|
\bigcap_{\ell \in \underline{\mathcal{Q}}_{q,\mathcal{S},\sv}} \mathcal{A}_{\ell,\sv} \right)
\end{split}
\end{equation}
and we can apply Lemma~\ref{P1}, which gives us a means to compute the desired probability.

\begin{remark}
The fourth case where both $\overline{\mathcal{Q}}_{q,\mathcal{S},\sv} \neq \emptyset$ and $\underline{\tilde{\mathcal{Q}}}_{q,\mathcal{S},\sv} \neq \emptyset$ cannot occur and need not be considered.
The first condition prevents $\pv(q)$ from being statistically discriminating, whereas the second condition forces these parity-check bits to be discriminating.
In such scenarios, $\Pr \left( \mathcal{E}_{\sv,\mathcal{S}} \right)$ does not admit the conditional form of \eqref{PEsS}, but it can readily be identified as having probability zero.
\end{remark}

We collect the results discussed above and summarize our findings in the form of a proposition.

\begin{proposition} \label{proposition:ProbabilityPatternBasedEvent}
Consider equivalence class $\sv$ and set $\mathcal{S} \subset [1:n-1]$.
The probability that the parity-check bits $\pv(q)$ with $q \in \mathcal{S}$ are statistically discriminating, and those with indices in $\mathcal{S}^{\mathrm{c}} = [1:n-1] \setminus \mathcal{S}$ are non-discriminating is given by
\begin{equation}
\begin{split}
\Pr \left( \mathcal{E}_{\sv,\mathcal{S}} \right)
&= \prod_{q \in \mathcal{S}^{\mathrm{c}}} g \left( \sum_{\substack{j \in [0:q-1] \\ \sv(q) \neq \sv(j)}} m_{j}
- \sum_{\substack{k \in [0:p-1] \\ \sv(p) \neq \sv(k)}} m_{k} \right) \\
&\times \prod_{q \in \mathcal{S}} \left[ 1 - \left[
\mathbbm{1}_{ \left\{ \overline{\mathcal{Q}}_{q,\mathcal{S},\sv} \neq \emptyset \right\} }
+ \mathbbm{1}_{ \left\{ \overline{\mathcal{Q}}_{q,\mathcal{S},\sv} = \emptyset \right\} }
\mathbbm{1}_{ \left\{ \underline{\tilde{\mathcal{Q}}}_{q,\mathcal{S},\sv} = \emptyset \right\} }
g \left( \sum_{\substack{j \in [0:q-1] \\ \sv(q) \neq \sv(j)}} m_{j}
- \sum_{\substack{k \in [0:p-1] \\ \sv(p) \neq \sv(k)}} m_{k} \right)
\right] \right]
\end{split}
\end{equation}
where $\mathbbm{1}_{ \{ \cdot \} }$ is the set indicator function, $p = \max \underline{\mathcal{Q}}_{q,\mathcal{S},\sv}$ and $g(t)=2^{-t}$.
\end{proposition}

This proposition, together with \eqref{equation:TreeGeneratingFunction} and \eqref{equation:TreeGeneratingFunctionSingleSet}, provides an algorithmic procedure to compute the expected number of erroneous paths that survive the decoding process.
While the proposition is stated for the full length of the transmission process, the reader will notice that the result can also be applied at any stage of the decoding process.
Hence, the expected number of erroneous surviving paths at level $j-1$ is given by,
\begin{equation*}
\mathbb{E}[L_{j-1}] = \sum_{\sv \in \mc{P}_j} n(\sv) \Phi_{\sv} \left( \frac{1}{2} \right) - 1,
\quad j \in [1:n].
\end{equation*}

Our expression for the expected number of surviving candidate codewords at the end of the tree decoding process is convoluted.
Unfortunately, there is no apparent structure, beyond the presence of equivalence classes, that enable a reduction in computational complexity.
It is pertinent to mention that the Bell numbers grow rapidly and, as such, the computation of the expected number of surviving paths through the tree can be challenging.
Nevertheless,  exact numbers can be obtained for the parameters we are interested in.

\section{Approximate Performance of Tree Code}
\label{Section:ApproximatePerformanceTreeCodeAppendix}

This appendix contains proofs for some of the results contained in Section~\ref{subsection:ApproximateTreeCodeAnalysis}.
Recall that the approximate analysis for tree coding is performed under Assumption~\ref{assumption:DistinctFragments}.
That is, the collection of messages $\mathcal{M}$ is such that, whenever $i \neq k$,
\begin{equation*}
\wv_i(j) \neq \wv_k(j)
\end{equation*}
for all $j \in [0:n-1]$ with $m_j > 0$.

\subsection{Proof of Proposition~\ref{proposition:ExpectedValuesDistinctFragments}}
\label{subsection:ExpectedValuesDistinctFragmentsProof}

We begin the proof with the simplest scenario.
Consider erroneous partial vector $\wv_{i_0}(0) \wv_{i_1}(1)$ where $i_0 \neq i_1$.
Under Assumption~\ref{assumption:DistinctFragments}, we know that $\wv_{i_0} (0) \neq \wv_{i_1}(0)$.
Then, by Lemma~\ref{lemma:RandomGenerator}, the probability that $p_{i_0}(1) = p_{i_1}(1)$ is given by
\begin{equation*}
\Pr \left( p_{i_0}(1) = p_{i_1}(1) \right) = 2^{-l_1} .
\end{equation*}
That is, the probability that an erroneous vector survives stage~$1$ is equal to $2^{-l_1}$.
Moving forward, we can extend this analysis to subsequent stages.
Consider erroneous partial vector $\wv_{i_0}(0) \wv_{i_1}(1) \cdots \wv_{i_j}(j)$.
Since this is not a valid message, there exists $q < j$ such that $i_q \neq i_j$.
Under Assumption~\ref{assumption:DistinctFragments}, we then have $\wv_{i_q} (q) \neq \wv_{i_j}(q)$.
Again, by Lemma~\ref{lemma:RandomGenerator}, we gather that
\begin{equation} \label{equation:ProbabilitySurvival}
\Pr \left( \sum_{\ell = 0}^{j-1} \wv_{i_{\ell}}(\ell) G_{\ell,j-1} = p_{i_j}(j) \right) = 2^{-l_j} ,
\end{equation}
where $p_j=2^{-l_j}$ and $q_j= 1-p_j$.
We note, briefly, that the event in \eqref{equation:ProbabilitySurvival} is equivalent to the parity condition found in \eqref{equation:GeneralG}, which is the requirement for a candidate vector to pass stage~$j$.

Suppose $j \geq 2$ and assume $\tilde{L}_{j-1}$ is given.
In view of \eqref{equation:ProbabilitySurvival}, $\tilde{L}_j$ can be interpreted as the sum of $K (\tilde{L}_{j-1}+1) - 1$ Bernoulli random variables, each with parameter $p_j$.
Consequently, the expected value $\mathbb{E} [ \tilde{L}_j ]$ admits the recursive form
\begin{equation} \label{exprec1}
\begin{split}
\mathbb{E} \left[ \tilde{L}_j \right]
&= \mathbb{E} \left[ \mathbb{E} \left[ \tilde{L}_j \mid \tilde{L}_{j-1} \right] \right] \\
&= \mathbb{E} \left[ \left( \left( \tilde{L}_{j-1}+1 \right) K-1 \right) p_j \right] \\
&= p_j K \mathbb{E} \left[ \tilde{L}_{j-1} \right] + p_j(K-1) .
\end{split}
\end{equation}
The above equation can be solved recursively using initial condition $\mathbb{E}[\tilde{L}_1 ]= (K-1)p_1$.
This yields the closed-form expression that appears in \eqref{expLi}.

\section{Asymptotic Performance Analysis}
\label{section:AsymptoticPerformanceAppendix}

This section contains demonstrations for the two main results contained in Section~\ref{section:AsymptoticAnalysis}, which focuses on the asymptotic analysis of CCS.

\subsection{Proof of Theorem~\ref{theorem:AsymptoticBehaviorParityTrailing}}

For mathematical convenience, we assume that the total number of parity bits $P$ is an integer multiple of the length of coded sub-blocks $J$.
Since all the parity bits are located towards the end of the codewords, the final $P/J$ sub-blocks only contain parity bits, and the initial $n-P/J$ sub-blocks are composed of information bits.
In other words, the number of parity bits in sub-block~$k$ is given by
\begin{equation} \label{li}
l_j = \begin{cases}
0 & j < n - P/J \\
J & j \geq n - P/J .
\end{cases}
\end{equation}
We note that with this parity-check profile, all the parity bits are statistically discriminating.
This stems from the fact that for erroneous codeword $\wv_{\mathrm{e}}$, all the parity equations act on $\wv_{\mathrm{e}} \neq \wv$.
This situation is akin to the conditions of Lemma~\ref{lemma:RandomGenerator} in Appendix~\ref{sectionRandomLInearCodes}.
Thus, applying tree decoding in parallel, Proposition~\ref{proposition:ExpectedValuesDistinctFragments} and Proposition~\ref{proposition:ExpectedParityChecks} become exact rather than approximate.
Leveraging these results, we find conditions on $P$ for which $\mathbb{E} [L_{n-1}]$ decays to zero in the limit as $\Ka \rightarrow \infty$.
This immediately results in a vanishing tree decoding error probability because
\begin{equation*}
\ptree = \Pr (L_{n-1} \ge 1) \le \mathbb{E} [L_{n-1}]
\end{equation*}
by the Markov inequality.
The quantity $\prod_{\ell=q}^{n-1}p_\ell$ appearing in \eqref{expLi} can be computed using \eqref{li} as
\begin{equation*}
\prod_{\ell=q}^{n-1}p_{\ell} =
\begin{cases}
\left( \frac{1}{2^J} \right)^{P/J} = \frac{1}{2^P}
& q \leq n-{P}/{J} \\
\left( \frac{1}{2^J} \right)^{n-q} = \frac{1}{2^{J(n-q)}}
& q > n - {P}/{J} .
\end{cases}
\end{equation*}
Substituting the above expression into \eqref{expLi}, we can upper bound the expected value $\mathbb{E} [L_{n-1}]$ by
\begin{equation*}
\begin{split}
\mathbb{E}[L_{n-1}]
&\le \sum_{j=1}^{n-P/J} \frac{\Ka^{n-1-j}(\Ka-1)}{2^P}
+ \sum_{j=n-P/J+1}^{n-1} \frac{\Ka^{n-1-j}(\Ka-1)}{2^{J(n-j)}} \\
&= \frac{\Ka^{P/J-1}}{2^P}\left[ K^{n-P/J}-1 \right]
+ \frac{\Ka-1}{2^J} \left[ \frac{1-\left(\frac{\Ka}{2^J}\right)^{P/J-1}}{1-\frac{\Ka}{2^J}} \right] \\
& \le \frac{\Ka^{n-1}}{2^P} + \frac{\Ka}{2^J-\Ka} \\
& \le \frac{2^{(n-1)\log_2 \Ka}}{2^P} + \frac{1}{\Ka^{c_1-1}-1} .
\end{split}
\end{equation*}
The quantity $\frac{1}{\Ka^{c_1-1}-1} \rightarrow 0 $ as $\Ka \rightarrow \infty$ since $c_1>1$.
The quantity $\frac{2^{(n-1)\log_2 \Ka}}{2^P} \rightarrow 0$  if $P=(n-1)\log_2 \Ka + h(\Ka)$, where $h(\cdot)$ is any function that grows unbounded with its argument.
For instance, if we choose $h(\Ka) = \delta \log_2 \Ka$ for some fixed constant $\delta>0$, we get $\ptree \le \frac{1}{\Ka^{c_1-1}-1} + \frac{1}{\Ka^\delta}$, which implies $\ptree \rightarrow 0$ as $\Ka \rightarrow \infty$.

When the total number of parity-check bits is $P=(n-1+\delta)\log_2 \Ka$, we have $M=nJ=B+(n-1+\delta)\log_2 \Ka$. Substituting $J=c_1\log_2 \Ka$ into the above equation, we obtain $n=\frac{B+(\delta-1)\log_2\Ka}{(c_1-1)\log_2\Ka}$. Hence, in regime (i), we have $n=c_2=\frac{\alpha+\delta-1}{c_1-1}$. In regime (ii), this leads to $n=c_2\frac{\Ka}{\log_2 \Ka}$, where $c_2 \approx \frac{1}{c_1-1}$ for large $\Ka$.
These results imply that the number of sub-blocks needed is $\mathcal{O}(1)$ in regime (i); and it scales nearly linearly with $\Ka$ in regime (ii).

Let $N_{\mathrm{s}}$ denote the number of channel uses allocated to each CS sub-problem. Standard results in the compressed sensing literature offer sufficient conditions on $N_{\mathrm{s}}$ for exact support recovery using conventional CS algorithms such as LASSO and OMP~\cite{wainwright2009sharp, cai2011orthogonal}.
For instance, if the measurement matrix is from the uniform Gaussian ensemble, then
\begin{equation*}
N_{\mathrm{s}}=\mathcal{O}(\Ka\log_2(2^J-\Ka))=\mathcal{O}(\Ka J)
\end{equation*}
channel uses are sufficient to ensure that the error probability of the support recovery process is $\mathcal{O}(\beta_1\exp(-\beta_2\min(2^J,\Ka)))$ when LASSO is employed with a proper choice of the regularization parameter.
Parameters $\beta_1$ and $\beta_2$ are fixed constants in the expression above~\cite{wainwright2009sharp}.
Altogether, having $N_{\mathrm{s}}=\lambda \Ka J$ for some fixed constant $\lambda$ ensures that $n \pcs \rightarrow 0$ for both regimes (i) and (ii).

Since there are $n$ CS sub-problems, the relation between $N_{\mathrm{s}}$ and the total number of channel uses is $N=nN_{\mathrm{s}}=n\lambda \Ka J$.
This yields $N=\lambda \frac{(\alpha+\delta-1)c_1}{c_1-1} \Ka\log_2\Ka$ for regime (i), and $N \approx \lambda \frac{c_1}{c_1-1} \Ka^2$ for regime (ii), both of which are consistent with the statement of the theorem.
The computational complexity of the overall scheme is dictated by its two components: the complexity of solving the CS sub-problems, and the expected complexity associated with the tree decoding process.

\paragraph{CS Sub-Problems}
The computational complexity of conventional CS solvers (e.g., LASSO, OMP, nnLS) scales polynomially in the length of the unknown vector.
Since we have $n$ sub-problems, the computational complexity of the CS sub-problems is given by
\begin{equation*}
\begin{split}
C_{\mathrm{cs}} = \mathcal{O}\left( n \left( 2^J \right)^\rho \right) 
= \mathcal{O} \left( n \Ka^{\rho c_1} \right)
\end{split}
\end{equation*}
where $\rho \geq 1$.
The above expression simplifies to $C_{\mathrm{cs}} = \mathcal{O} \left(\Ka^{\rho c_1}\right)$ for regime (i) and  $C_{\mathrm{cs}} = \mathcal{O} \left(\frac{\Ka^{\rho c_1+1}}{\log_2 \Ka}\right)$ for regime (ii). Hence, the computational cost of solving the CS sub-problems is bounded by a polynomial in $\Ka$ for both regimes. If CS solvers with linear complexity in the length of the unknown vector are employed for the sub-problems (i.e., $\rho=1$), then $C_{\mathrm{cs}}$ scales as $\Ka^{c_1}$ and $\frac{\Ka^{c_1+1}}{\log_2\Ka}$ in regimes (i) and (ii), respectively.

\paragraph{Tree Decoder}
Under regime~(i), we can leverage the identity $n=c_2=\frac{\alpha+\delta-1}{c_1-1}$.
With parameters $J=c_1 \log_2 \Ka$ and  $P=(n-1+\delta) \log_2\Ka$, we can rewrite the boundary of \eqref{li} as
\begin{equation*}
n-\frac{P}{J}
= n - \frac{n-1+\delta}{c_1}
= \frac{\alpha}{c_1} .
\end{equation*}
This yields the compact form
\begin{equation} \label{li1}
l_j = \begin{cases}
0 & j < \alpha/c_1 \\
c_1 \log_2 \Ka & j \geq \alpha/c_1 .
\end{cases}
\end{equation}
The product $\prod_{\ell=q}^j p_{\ell}$ appearing in \eqref{expcomp} can then be computed using \eqref{li1},
\begin{equation} \label{equation:RunningProduct}
\prod_{\ell=q}^j p_{\ell} = \begin{cases}
1 & q \leq j < \alpha/c_1 \\
\Ka^{- c_1(j+1) + \alpha} & q < \alpha/c_1 \leq j \\
\Ka^{- c_1(j+1-q)} & \alpha/c_1 \leq q \leq j .
\end{cases}
\end{equation}

We get an upper bound on the expected complexity for this scenario by substituting \eqref{li1}, \eqref{equation:RunningProduct}, $J=c_1\log_2\Ka$, and $P=(n-1+\delta) \log_2\Ka$ into \eqref{expcomp}.
Recall the form of \eqref{expcomp}, under $K = \Ka$,
\begin{equation} \label{equation:ComplexityTreeSum}
\mathbb{E}[C_{\mathrm{tree}}]
= \Ka \sum_{j=1}^{n-1} l_j
+ \Ka \sum_{j=1}^{n-2} l_{j+1} \sum_{q=1}^j K^{j-q}(K-1) \prod_{\ell=q}^j p_{\ell} .
\end{equation}
We recognize the first summand as $\Ka P$ and, therefore,
\begin{equation}
\Ka \sum_{j=1}^{n-1} l_j
= \Ka (n-1+\delta) \log_2 \Ka
= \mathcal{O} (\Ka \log_2\Ka ) .
\end{equation}
We break down the second summand along the cases of \eqref{equation:RunningProduct}.
The first portion of the triplet is
\begin{equation} \label{equation:TripletOne}
\begin{split}
&\Ka \sum_{j=1}^{\alpha/c_1-1} l_{j+1} \sum_{q=1}^j \Ka^{j-q}(\Ka-1) \prod_{\ell=q}^j p_{\ell} \\
&= \Ka c_1 \log_2 (\Ka) \sum_{q=1}^{{\alpha/c_1 -1}} \Ka^{\alpha/c_1 - 1 - q} (\Ka-1) \\
&\leq c_1 \Ka^{\alpha/c_1} \log_2 \Ka
= \mathcal{O}\left(\Ka^{\alpha/c_1} \log_2 \Ka \right) .
\end{split}
\end{equation}
We note that, among the indices of the first sum in \eqref{equation:TripletOne}, $l_{j+1}$ is non-zero only when $j = \alpha/c_1-1$.
This greatly simplifies the expression and leads to the succinct characterization.
The second portion of the triplet is
\begin{equation} \label{equation:TripletTwo}
\begin{split}
&\Ka \sum_{j=\alpha/c_1}^{n-2} l_{j+1} \sum_{q=1}^{\alpha/c_1-1} \Ka^{j-q}(\Ka-1) \prod_{\ell=q}^j p_{\ell} \\
&= \Ka \sum_{j=\alpha/c_1}^{n-2} c_1 \log_2 (\Ka) \sum_{q=1}^{\alpha/c_1-1} \Ka^{j-q}(\Ka-1) \Ka^{- c_1(j+1) + \alpha} \\
&\leq c_1 \frac{\Ka^{\alpha/c_1} \log_2 \Ka}{\Ka^{c_1 - 1}-1}
= \mathcal{O} \left( \frac{\Ka^{\alpha/c_1} \log_2 \Ka}{\Ka^{c_1-1}-1} \right) .
\end{split}
\end{equation}
While tedious, the manipulations in \eqref{equation:TripletTwo} revolve around the evaluation of geometric sums and the application of straightforward upper bounds.
This same strategy is applied to the last portion of the sum, with
\begin{equation} \label{equation:TripletThree}
\begin{split}
&\Ka \sum_{j=\alpha/c_1}^{n-2} l_{j+1} \sum_{q=\alpha/c_1}^j \Ka^{j-q}(\Ka-1) \prod_{\ell=q}^j p_{\ell} \\
&= \Ka (\Ka-1) \sum_{j=\alpha/c_1}^{n-2} c_1 \log_2 (\Ka) \sum_{q=\alpha/c_1}^j \Ka^{j-q} \Ka^{- c_1(j+1-q)} \\
&\leq c_1 (n-1 - \alpha/c_1) \frac{\Ka}{\Ka^{c_1 - 1} - 1} \log_2 \Ka
= \mathcal{O} \left( \frac{\Ka \log_2 \Ka}{\Ka^{c_1-1}-1} \right) .
\end{split}
\end{equation}
Collecting the first summand and the three partial sums of the second summand, we gather that
\begin{equation} \label{equation:TreeComplexityTrailRegime1}
\begin{split}
\mathbb{E}[C_\mathrm{tree}]
&= \mathcal{O} (\Ka \log_2\Ka )
+ \mathcal{O}\left(\Ka^{\alpha/c_1} \log_2 \Ka \right) \\
&+ \mathcal{O} \left( \frac{\Ka^{\alpha/c_1} \log_2 \Ka}{\Ka^{c_1-1}-1} \right)
+ \mathcal{O} \left( \frac{\Ka \log_2 \Ka}{\Ka^{c_1-1}-1} \right) .
\end{split}
\end{equation}
The above expression is dominated by the contribution of its second term.
Consequently, the average complexity of tree decoding in regime (i) has an order complexity of $\mathcal{O}\left(\Ka^{\alpha/c_1} \log_2 \Ka \right)$.

Under regime~(ii), we inherit the identity $n= \frac{1}{c_1-1}\frac{\Ka}{\log_2\Ka} + \frac{\delta-1}{c_1-1}$.
Paralleling the steps above, we can rewrite the boundary of \eqref{li} as
\begin{equation*}
n-\frac{P}{J}
= \frac{(c_1 - 1)n}{c_1} - \frac{\delta - 1}{c_1}
= \frac{\Ka}{c_1 \log_2\Ka} .
\end{equation*}
We can avoid duplicating most of the derivations performed under regime~(i) by recognizing that the boundary condition above can be obtained from \eqref{li1} by substituting $\alpha = \frac{\Ka}{\log_2 \Ka}$.
We can discern where the modified formulation has an impact beyond substitution by carefully tracking steps.
For the first summand in the expansion of \eqref{equation:ComplexityTreeSum}, we get
\begin{equation}
\begin{split}
&\Ka \sum_{j=1}^{n-1} l_j = \Ka (n-1+\delta) \log_2 \Ka \\
&= \Ka \left( c_2 \frac{\Ka}{\log_2 \Ka} - 1 + \delta \right) \log_2 \Ka
= \mathcal{O} \left( \Ka^2 \right) .
\end{split}
\end{equation}
The first triplet of the second summand in \eqref{equation:TripletOne} is unaltered beyond the aforementioned substitution.
This yields
\begin{equation}
\mathcal{O} \left( \Ka^{\frac{\Ka}{c_1 \log_2 \Ka}} \log_2 \Ka \right)
= \mathcal{O} \left( 2^{\Ka/c_1} \log_2 \Ka \right) ,
\end{equation}
where we have leveraged the identity $x^{\frac{x}{\log_2x}}=2^x$ for all $x>0$.
The second portion of the triplet is also unaltered beyond substitution.
Leveraging the same identity, \eqref{equation:TripletTwo} becomes
\begin{equation}
\mathcal{O} \left( \frac{\Ka^{\frac{\Ka}{c_1 \log_2 \Ka}} \log_2 \Ka}{\Ka^{c_1-1}-1} \right)
= \mathcal{O} \left( \frac{2^{\Ka/c_1} \log_2 \Ka}{\Ka^{c_1-1}-1} \right)
\end{equation}
under regime~(ii).
The final component comes from the upper bound in \eqref{equation:TripletThree}.
After proper substitution, it changes to
\begin{equation}
\begin{split}
&c_1 \left( n - 1 - \frac{\Ka}{c_1 \log_2 \Ka} \right) \frac{\Ka}{\Ka^{c_1 - 1} - 1} \log_2 \Ka \\
&= c_1 \left( c_2 \frac{\Ka}{\log_2 \Ka} - 1 - \frac{\Ka}{c_1 \log_2 \Ka} \right) \frac{\Ka}{\Ka^{c_1 - 1} - 1} \log_2 \Ka \\
&= \mathcal{O} \left( \frac{\Ka^2}{\Ka^{c_1 - 1} - 1} \right) .
\end{split}
\end{equation}
Aggregating the contributions from these four parts, we get
\begin{equation} \label{equation:TreeComplexityTrailRegime2}
\begin{split}
\mathbb{E}[C_\mathrm{tree}]
&= \mathcal{O} \left( \Ka^2 \right) + \mathcal{O} \left( 2^{\Ka/c_1} \log_2 \Ka \right) \\
&+ \mathcal{O} \left( \frac{2^{\Ka/c_1} \log_2 \Ka}{\Ka^{c_1-1}-1} \right) + \mathcal{O} \left( \frac{\Ka^2}{\Ka^{c_1 - 1} - 1} \right) .
\end{split}
\end{equation}
The above expression is dominated by the second term
Thus, the average complexity of tree decoding under regime (ii) is on the order of $\mathcal{O} \left( 2^{\Ka/c_1} \log_2 \Ka \right)$.
This complete the proof of Theorem~\ref{theorem:AsymptoticBehaviorParityTrailing}.

\subsection{Proof of Theorem~\ref{theorem:AsymptoticBehaviorParityUniform}}

In this section, we provide a proof for Theorem~\ref{theorem:AsymptoticBehaviorParityUniform}.
We wish to study the scenario where parity bits are assigned uniformly across sub-blocks under the simplifying conditions of Assumption~\ref{assumption:DistinctFragments}.
For uniform allocation, these conditions reduce to $\wv_i(j) \neq \wv_k(j)$ for all $j \in [0:n-1]$.
To do so, we find circumstances under which the probability of Assumption~\ref{assumption:DistinctFragments} being violated goes to zero as $\Ka$ grows unbounded.
We begin by establishing an upper bound on the expected number of sub-blocks for which the uniqueness condition is violated.
The expected number of fragments that are repeated more than once can the be bounded by
\begin{equation*}
\begin{split}
&\mathbb{E} [R] = \mathbb{E} \left[ \sum_{i=1}^{\Ka} \sum_{k=1}^{i-1} \sum_{j=0}^{n-1} \mathbbm{1}_{\left\{ \wv_i(j) = \wv_k(j) \right\}} \right] \\
&= \sum_{i=1}^{\Ka} \sum_{k=1}^{i-1} \sum_{j=0}^{n-1} \Pr \left( \wv_i(j) = \wv_k(j) \right)
\leq \frac{n \Ka^2}{2^m} .
\end{split}
\end{equation*}
Then, by the Markov inequality, we get
\begin{equation} \label{equation:ProbAssumptionViolated}
\Pr (R \geq 1) \leq \mathbb{E} [R] \leq \frac{n \Ka^2}{2^m} .
\end{equation}
It remains to choose suitable parameters for $\alpha$, $c_1$, and $c_2$ to ensure that \eqref{equation:ProbAssumptionViolated} vanishes as $\Ka \rightarrow \infty$.
The selection of suitable parameters depends on the asymptotic regime under consideration.
Thus, we treat the cases associated with regime~(i) and regime~(ii) separately.

\paragraph{Regime~(i)}
In this case, the number of information bits per non-root sub-block is dictated by the first column in Table~\ref{table:AsymptoticParameters}, with
\begin{equation} \label{equation:mAsymptoticUniformRegime1}
m = \frac{B-J}{n-1} = \frac{\alpha-c_1}{c_2-1} \log_2 \Ka .
\end{equation}
The total number of parity bits for this setting becomes $P = nJ - B = (c_1 c_2 - \alpha) \log_2 \Ka$ and, consequently, the number of parity bits per non-root sub-block becomes
\begin{equation} \label{equation:lAsymptoticUniformRegime1}
l = \frac{nJ - B}{n-1} = \frac{c_1 c_2 - \alpha}{c_2 - 1} \log_2 \Ka .
\end{equation}
Substituting the expression for $m$ in \eqref{equation:ProbAssumptionViolated}, we get
\begin{equation}
\Pr (R \geq 1) \leq \frac{c_2 \Ka^2}{2^{\frac{\alpha-c_1}{c_2-1} \log_2 \Ka}}
= c_2 \Ka^{2 - \frac{\alpha-c_1}{c_2-1}} .
\end{equation}
Picking $c_1$ and $c_2$ such that $\frac{\alpha-c_1}{c_2-1} > 2$ meets our requirements.
In words, limiting the number of sub-blocks $n = c_2$ simultaneously ensures that the number of information bits per fragment $m$ is large and, hence, the probability of a match $2^{-m}$ is low.
At the same time, the number of sub-blocks must be large enough to account for all the information bits while leaving room for parity bits, $c_1 c_2 - \alpha > 0$.

\paragraph{Regime~(ii)}
The number of information bits for this second case is characterized by
\begin{equation} \label{equation:mAsymptoticUniformRegime2}
\begin{split}
m &= \frac{B-J}{n-1}
= \frac{\Ka-c_1 \log_2 \Ka}{c_2 \frac{\Ka}{\log_2 \Ka} - 1} \\
&\geq \frac{1}{c_2} \log_2 \Ka - \frac{c_1}{c_2} \frac{(\log_2 \Ka)^2}{\Ka} \\
&= \frac{1}{c_2} \log_2 \Ka - o(1) .
\end{split}
\end{equation}
The number of parity bits in this setting is $P = nJ - B = (c_1 c_2 - 1) \Ka$, which yields
\begin{equation} \label{equation:lAsymptoticUniformRegime2}
\begin{split}
l &= \frac{nJ - B}{n-1} = \frac{c_1 c_2 - 1}{c_2 \frac{\Ka}{\log_2 \Ka} - 1} \Ka \\
&\geq \frac{c_1 c_2 - 1}{c_2} \log_2 \Ka
\end{split}
\end{equation}
After incorporating this admittedly intricate inequality governing $m$ in \eqref{equation:ProbAssumptionViolated}, we obtain
\begin{equation}
\begin{split}
\Pr (R \geq 1)
&\leq \frac{c_2 \frac{\Ka}{\log_2 \Ka} \Ka^2}{2^{\frac{1}{c_2} \log_2 \Ka - o(1)}} \\
&= c_2 \frac{\Ka^{3-\frac{1}{c_2}}}{\log_2 \Ka} 2^{o(1)} .
\end{split}
\end{equation}
Choosing $c_1$ and $c_2$ such that $c_2 < \frac{1}{3}$ implies that the right-hand side of this expression goes to zero as $\Ka \rightarrow \infty$.
At the same time, the number of sub-blocks cannot grow too fast.
Furthermore, the number of sub-blocks must be large enough to account for all the information bits and leave room for parity bits.
This is accomplished with $c_1 c_2 > 1$.

\subsubsection{Asymptotic Reliability}

The conditions outlined above control the number of possible matches, and it pushes $\Pr (R \geq 1)$ to zero.
In other words, given the stipulations included in the statement of Theorem~\ref{theorem:AsymptoticBehaviorParityUniform}, we know that the conditions of Assumption~\ref{assumption:DistinctFragments} are met with probability approaching one as $\Ka \rightarrow \infty$.
Through conditioning, we can therefore focus on cases where Assumption~\ref{assumption:DistinctFragments} holds.
A decoding error can be declared when the uniqueness conditions are not met at the output of the CS problems.
This idiosyncratic situation has no impact on asymptotic performance when system parameters are selected judiciously.
Pragmatically, this enables us to use the simplified expressions of Section~\ref{subsection:ApproximateTreeCodeAnalysis} in characterizing true asymptotic behavior.

For ease of exposition, we take the total number of parity bits $P$ to be an integer multiple of $n-1$.
With careful accounting, it is possible to relax this assumption and retain the character of our findings.
Still, this structure simplifies the proof considerably.
For the current scenarios, the number of parity bits per non-root sub-block is given by
\begin{equation} \label{liep}
l_j = \frac{P}{n-1} \quad j \in [1:n-1] .
\end{equation}
The quantity $\prod_{\ell=q}^{n-1}p_{\ell}$ appearing in \eqref{expLi} can then be computed using \eqref{liep},
\begin{equation*}
\prod_{\ell=q}^{n-1}p_{\ell} = \left(\frac{1}{2^{ \frac{P}{n-1}}}\right)^{n-q} = 2^{- \frac{P(n-q)}{n-1}} .
\end{equation*}
Substituting this expression into \eqref{expLi}, we get an upper bound for the expected value $\mathbb{E} [L_{n-1}]$,
\begin{equation} \label{equation:LAsymptoticUpperBound}
\begin{split}
\mathbb{E} [ L_{n-1} ]
&= \sum_{q=1}^{n-1} \Ka^{n-1-q}(\Ka-1) \frac{1}{2^{ \frac{P(n-q)}{n-1}}} \\
&\le \frac{\Ka}{2^l-\Ka} .
\end{split}
\end{equation}
For asymptotic reliability of the CCS algorithm, we need the expected number of erroneous paths that survive at the end of the decoding process to go to zero.
Again, we find suitable conditions for the two regimes separately.


\paragraph{Regime~(i)}
In this regime, $l$ is given by \eqref{equation:lAsymptoticUniformRegime1}.
Substituting this expression in \eqref{equation:LAsymptoticUpperBound}, we get the upper bound
\begin{equation*}
\begin{split}
\mathbb{E}\left[L_{n-1}\right]
&\leq \frac{1}{\Ka^{\frac{c_1 c_2 - \alpha}{c_2 -1} - 1} - 1} .
\end{split}
\end{equation*}
This quantity and $\ptree$ go to zero as $\Ka \rightarrow \infty$ provided that $\frac{\alpha - c_1}{c_2 - 1} < c_1 - 1$.
In words, the discriminating power of parity bits has to counteract the per stage growth of the tree.

\paragraph{Regime~(ii)}
In this case, $l$ is governed by the inequality in \eqref{equation:lAsymptoticUniformRegime2} and, after substitution and manipulations, we obtain
\begin{equation*}
\begin{split}
\mathbb{E}\left[L_{n-1}\right]
&\leq \frac{\Ka}{\Ka^{\frac{c_1 c_2 - 1}{c_2}}-\Ka} .
\end{split}
\end{equation*}
The expected number of extraneous paths and $\ptree$ tend to zero as $\Ka \rightarrow \infty$ whenever $c_1 - 1 > \frac{1}{c_2}$.
This condition again arises from the tension between between the growth at every stage in the tree and the ability of parity bits to identify eroneous candidate paths.

\subsubsection{Complexity Analysis}

In a manner akin to the proof of Theorem~\ref{theorem:AsymptoticBehaviorParityTrailing}, the number of channel uses for the CS sub-problem is chosen as $N_{\mathrm{s}}= \lambda \Ka J$.
This ensures that the probability of failure for one CS sub-component goes to zero exponentially fast and, hence, $n \pcs \rightarrow 0$ for both the regimes.
The total number of channel uses $N$ is then given by $N = n N_{\mathrm{s}} = c_1 n \lambda \Ka \log_2 \Ka$.
This expression simplifies to $N = c_1 c_2 \lambda \Ka \log_2 \Ka$ under regime~(i); and it becomes $N = c_1 c_2 \lambda \Ka^2$ under regime~(ii).
Both cases are consistent with the statement of the theorem.
As before, the computational complexity for the CCS algorithm arises from two components: the complexity of solving the CS sub-problems, and the average complexity associated with the tree decoder.

\paragraph{CS Sub-Problems}
The analysis of this component is similar to that contained in the proof of Theorem~\ref{theorem:AsymptoticBehaviorParityTrailing}.
The CS sub-problems have polynomial complexity in the length of the unknown vector.
The complexity of the CS sub-problems $C_{\mathrm{cs}}$ can then be expressed as
\begin{equation*}
C_{\mathrm{cs}} = \mathcal{O}\left( n \left( 2^J \right)^\rho \right) 
= \mathcal{O} \left( n \Ka^{\rho c_1} \right)
\end{equation*}
where $\rho \geq 1$.
Thus, $C_{\mathrm{cs}}$ scales as $\mathcal{O} \left( \Ka^{c_1} \right)$ for regime~(i) and $\mathcal{O} \left( \frac{\Ka^{c_1+1}}{\log_2\Ka} \right)$ for regime~(ii) when a CS algorithm of linear complexity is employed to solve the CS sub-problems.

\paragraph{Tree Decoder}
For uniformly distributed parity bits, the expected computational complexity of \eqref{expcomp} can be simplified to
\begin{equation}
\begin{split}
\mathbb{E}[C_{\mathrm{tree}}]
&= \Ka \left( \sum_{j=1}^{n-1} l + \sum_{j=1}^{n-2} l \sum_{q=1}^j \Ka^{j-q}(\Ka-1) \prod_{\ell=q}^j 2^{-l} \right) \\
&\leq \Ka P + \frac{\Ka (\Ka-1)}{2^l - \Ka} P .
\end{split}
\end{equation}
From our characterization of $\mathbb{E} [L_{n-1}]$ and its upper bound in \eqref{equation:LAsymptoticUpperBound}, we know that
\begin{equation*}
\frac{\Ka}{2^l-\Ka} = o(1)
\end{equation*}
whenever parameters are selected judiciously.
In regime~(i), $P = \mathcal{O}(\log_2 \Ka)$ and, hence, $\mathbb{E}[C_{\mathrm{tree}}] = \mathcal{O}(\Ka\log_2\Ka)$.
Likewise, in regime~(ii), $P = \mathcal{O}(\Ka)$ and $\mathbb{E}[C_{\mathrm{tree}}] = \mathcal{O} \left( \Ka^2 \right)$.
These results are those reported in the statement of Theorem~\ref{theorem:AsymptoticBehaviorParityUniform}.

\bibliographystyle{IEEEbib}
\bibliography{IEEEabrv,MACcollision}

\end{document}